\documentclass[twocolumn]{aastex631}

\usepackage{booktabs}
\usepackage{graphicx}	

\usepackage{amsmath}	
\usepackage{amssymb}	
\usepackage{multirow}
\usepackage{longtable}
\usepackage{subfigure}
\usepackage{newtxtext,newtxmath}

\newcommand{\photoz}{photo-$z$}
\newcommand{\Photoz}{Photo-$z$}
\newcommand{\photozs}{photo-$z$'s}

\newcommand{\specz}{spec-$z$}

\newcommand{\speczs}{spec-$z$'s}
\newcommand{\Speczs}{Spec-$z$'s}

\newcommand{\grismzs}{grism-$z$'s}

\newcommand{\fdiv}{$f$-divergence}
\newcommand{\fdivs}{$f$-divergences}

\newcommand{\fopco}{$f_\mathrm{op \, \& \, co}$}
\newcommand{\catalogurl}{\url{https://archive.stsci.edu/hlsp/candels}}

\shorttitle{CANDELS Photometric Redshifts}
\shortauthors{Kodra et al.}
\graphicspath{{./}{img/}}

\begin{document}

\title{Optimized Photometric Redshifts for the Cosmic Assembly Near-infrared Deep Extragalactic Legacy Survey (CANDELS)}

\correspondingauthor{Brett H.~Andrews}
\email{andrewsb@pitt.edu}

\author{Dritan Kodra}
\affiliation{Department of Physics and Astronomy, University of Pittsburgh, Pittsburgh, PA 15260, USA}
\affiliation{Pittsburgh Particle Physics, Astrophysics, and Cosmology Center (PITT PACC), University of Pittsburgh, Pittsburgh, PA 15260, USA}

\author[0000-0001-8085-5890]{Brett H.~Andrews}
\affiliation{Department of Physics and Astronomy, University of Pittsburgh, Pittsburgh, PA 15260, USA}
\affiliation{Pittsburgh Particle Physics, Astrophysics, and Cosmology Center (PITT PACC), University of Pittsburgh, Pittsburgh, PA 15260, USA}

\author{Jeffrey A. Newman}
\affiliation{Department of Physics and Astronomy, University of Pittsburgh, Pittsburgh, PA 15260, USA}
\affiliation{Pittsburgh Particle Physics, Astrophysics, and Cosmology Center (PITT PACC), University of Pittsburgh, Pittsburgh, PA 15260, USA}


\author[0000-0001-8519-1130]{Steven L.~Finkelstein}
\affiliation{Department of Astronomy, The University of Texas at Austin, 2515 Speedway, Austin, TX 78712, USA}

\author[0000-0003-3820-2823]{Adriano Fontana}
\affiliation{Osservatorio Astronomico di Roma, via di Frascati 33, I-00078 Monte Porzio Catone (Roma), Italy}

\author{Nimish Hathi}
\affiliation{Space Telescope Science Institute, 3700 San Martin Drive, Baltimore, MD 21218, USA}

\author{Mara Salvato}
\affiliation{Max-Planck-Institut f\"ur extraterrestrische Physik, Giessenbachstr. 1, 85748 Garching, Germany}

\author{Tommy Wiklind}
\affiliation{Department of Physics, Catholic University of America, Washington D.C.\ 20064, USA}

\author{Stijn Wuyts}
\affiliation{Department of Physics, University of Bath, Claverton Down, Bath BA2 7AY, UK}


\author[0000-0002-7767-5044]{Adam Broussard}
\affiliation{Department of Physics and Astronomy, Rutgers, the State University of New Jersey, Piscataway, NJ 08854, USA}

\author[0000-0003-3691-937X]{Nima Chartab}
\affiliation{Department of Physics and Astronomy, University of California, Irvine, CA 92697, USA}

\author{Christopher Conselice}
\affiliation{Jodrell Bank Centre for Astrophysics, University of Manchester, Oxford Road, Manchester, UK}

\author[0000-0003-1371-6019]{M.\ C.\ Cooper}
\affiliation{Center for Cosmology, Department of Physics \& Astronomy, University of California, Irvine, 4129 Reines Hall, Irvine, CA 92697, USA}

\author{Avishai Dekel}
\affiliation{Racah Institute of Physics, The Hebrew University, Jerusalem 91904 Israel}

\author[0000-0001-5414-5131]{Mark Dickinson}
\affiliation{NSF's NOIRLab, 950 N.\ Cherry Ave., Tucson, AZ 85719, USA}

\author{Henry C.~Ferguson}
\affiliation{Space Telescope Science Institute, 3700 San Martin Drive, Baltimore, MD 21218, USA}

\author[0000-0003-1530-8713]{Eric Gawiser}
\affiliation{Department of Physics and Astronomy, Rutgers, the State University of New Jersey, Piscataway, NJ 08854, USA}

\author{Norman A.~Grogin}
\affiliation{Space Telescope Science Institute, 3700 San Martin Drive, Baltimore, MD 21218, USA}

\author[0000-0001-9298-3523]{Kartheik Iyer}
\affiliation{Dunlap Institute for Astronomy and Astrophysics, University of Toronto, 50 St.\ George Street, Toronto, ON M5S 3H4, Canada}

\author{Jeyhan Kartaltepe}
\affiliation{School of Physics and Astronomy, Rochester Institute of Technology, Rochester, NY 14623, USA}

\author{Susan Kassin}
\affiliation{Space Telescope Science Institute, 3700 San Martin Drive, Baltimore, MD 21218, USA}
\affiliation{The William H.\ Miller III Department of Physics \& Astronomy, Johns Hopkins University, 3400 N.\ Charles Street, Baltimore, MD 21218, USA}

\author[0000-0002-6610-2048]{Anton M.~Koekemoer}
\affiliation{Space Telescope Science Institute, 3700 San Martin Drive, Baltimore, MD 21218, USA}

\author{David C.~Koo}
\affiliation{Department of Astronomy and Astrophysics, University of California, Santa Cruz, Santa Cruz, CA 95064, USA}

\author[0000-0003-1581-7825]{Ray A.~Lucas}
\affiliation{Space Telescope Science Institute, 3700 San Martin Drive, Baltimore, MD 21218, USA}

\author{Kameswara Bharadwaj Mantha}
\affiliation{Minnesota Institute for Astrophysics, University of Minnesota, 116 Church St SE, Minneapolis, MN 55455, USA}
\affiliation{School of Physics and Astronomy, University of Minnesota, 116 Church St SE, Minneapolis, MN 55455, USA}

\author{Daniel H.\ McIntosh}
\affiliation{Department of Physics \& Astronomy, University of Missouri, Kansas City, Kansas City, MO 64110, USA}

\author{Bahram Mobasher}
\affiliation{Department of Physics and Astronomy, University of California, Riverside, 900 University Ave, Riverside, CA 92521, USA}

\author{Camilla Pacifici}
\affiliation{Space Telescope Science Institute, 3700 San Martin Drive, Baltimore, MD 21218, USA}

\author{Pablo G.\ P\'erez-Gonz\'alez}
\affiliation{Centro de Astrobiolog\'{\i}a, CAB/CSIC-INTA, Ctra.\ de Torrej\'on a Ajalvir, km 4, 28850, Torrej\'on de Ardoz, Madrid, Spain}

\author{Paola Santini}
\affiliation{Osservatorio Astronomico di Roma, via di Frascati 33, I-00078 Monte Porzio Catone (Roma), Italy}



\begin{abstract}

We present the first comprehensive release of photometric redshifts (photo-$z$'s) from the Cosmic Assembly Near-Infrared Deep Extragalactic Legacy Survey (CANDELS) team.  We use statistics based upon the Quantile--Quantile ($Q$--$Q$) plot to identify biases and signatures of underestimated or overestimated errors in \photoz\ probability density functions (PDFs) produced by six groups in the collaboration; correcting for these effects makes the resulting PDFs better match the statistical definition of a PDF.  After correcting each group's PDF, we explore three methods of combining the different groups' PDFs for a given object into a consensus curve. Two of these methods are based on identifying the minimum \fdiv\ curve, i.e., the PDF that is closest in aggregate to the other PDFs in a set (analogous to the median of an array of numbers).  We demonstrate that these techniques yield improved results using sets of spectroscopic redshifts independent of those used to optimize PDF modifications.  The best photo-$z$ PDFs and point estimates are achieved with the minimum \fdiv\ using the best 4 PDFs for each object (mFDa4) and the hierarchical Bayesian (HB4) methods, respectively.  The HB4 photo-$z$ point estimates produced $\sigma_{\rm NMAD} = 0.0227/0.0189$ and $|\Delta z/(1+z)| > 0.15$ outlier fraction = 0.067/0.019 for spectroscopic and 3D-HST redshifts, respectively.  Finally, we describe the structure and provide guidance for the use of the CANDELS photo-$z$ catalogs, which are available at \catalogurl.

\end{abstract}



\section{Introduction} \label{sec:intro}

Cosmic Noon ($z \sim 2$) saw the most intense star formation in the history of the universe; the rapid buildup of galactic disks and bulges, and the end of star formation in the most massive galaxies.  Deep galaxy surveys investigate this critical epoch of galaxy formation by tracking the evolving galaxy demographics via the galaxy luminosity function \citep[e.g.,][]{Cole_2001, Reddy_2009, Finkelstein_2015}, stellar mass function \citep[e.g.,][]{Marchesini_2009, Muzzin_2013, Duncan_2014, Tomczak_2014, Davidzon_2017}, and the stellar mass--star formation rate relation \citep[e.g.,][]{Noeske_2007, Whitaker_2012, Speagle_2014, Salmon_2015, Sandles_2022}.  The Cosmic Assembly Near-infrared Deep Extragalactic Legacy Survey \citep[CANDELS;][]{CANDELS_2011_Grogin, CANDELS_2011_Koekemoer} combines high spatial resolution Hubble Space Telescope (HST) Wide Field Camera 3 (WFC3)/IR and Advanced Camera for Surveys (ACS) imaging with existing intermediate-resolution optical and IR imaging down to faint limits (5$\sigma$ point-source limit of $H \sim 27$ for wide and $H \sim 27.7$ for deep) to extend our knowledge of these important galaxy evolution probes.  However, the measurements required for these probes rely on accurate redshifts.

Consequently, there have been multiple concerted efforts on 8--10 m class telescopes to obtain spectroscopic redshifts (\speczs) for CANDELS galaxies (e.g., VVDS, \citealt{LeFevre_2005}; zCOSMOS, \citealt{zCOSMOS}; VUDS, \citealt{VUDS_2015}; and VANDELS, \citealt{McLure_2018}).  \Speczs\ are the gold standard for accurate redshifts, but the galaxies with successfully measured \speczs\ tend to be brighter, at lower redshifts, and preferentially biased toward star-forming galaxies with strong emission lines.  Given the resource-intensive nature of measuring \speczs, spectroscopic follow-up of a large and representative sample, especially of the faintest CANDELS objects, is impossible with current and near-future facilities \citep{Newman_2015}.  Grism-based redshifts (\grismzs) from HST (e.g., 3D-HST, \citealt{3DHST}) can help alleviate some of the issues with ground-based \speczs\ because they can cover many more galaxies and probe a more representative population of galaxies \citep{Bezanson_2016}.  Nevertheless, grism spectroscopy is technically challenging, does not cover the full depth of the CANDELS imaging, and is not available for the whole CANDELS footprint.  Thus, we must rely on photometric redshifts (\photozs) for distance information \citep{Baum_1957} across the full range of galaxies, albeit at lower accuracy and precision than \speczs\ or \grismzs.

\Photoz\ estimation approaches mostly fall into two main groups---machine learning (ML)--based methods and template-based methods---with the optimal approach depending on the amount and quality of training data and science goals (see \citealt{Salvato_2019} and references therein).  ML methods estimate redshift from galaxy photometry by training an ML model on galaxies with \speczs.  ML methods outperform template-based methods in the regime of complete and representative coverage of color--$z$ space, such as Sloan Digital Sky Survey \photoz\ efforts (e.g., \citealt{Beck_2016}; \citealt{Pasquet_2019}; \citealt{Hayat_2021}; \citealt{dey2022_capsnet}), where data-driven interpolation is very accurate.  Template-based methods involve redshifting spectral energy distribution (SED) templates (or combinations thereof) to find the best fit to a galaxy's SED.  Template-based methods are superior to ML methods when spectroscopic coverage is incomplete and/or biased (e.g., in deep fields) because they leverage our knowledge of galaxy physics.   Hence, only template-based methods have been employed for CANDELS due to the biased and incomplete \specz\ training sets.  However, shortcomings in our knowledge of galaxy physics can limit the success of template-based methods; issues include template mismatch (i.e., when the SED templates differ from the observed SEDs), incorrect treatment of dust attenuation, and inappropriate priors \citep{Salvato_2019, newman2022}.

Template libraries fall intro two broad classes, empirical and theoretical, each with its own relative advantages and disadvantages.  Empirical templates (e.g., \citealt{coleman1980} and \citealt{kinney1996}) are principally derived from spectroscopic observations of nearby galaxies (and supplemented with model spectra to increase wavelength coverage), so they can accurately capture spectral features in real galaxies.  However, they are based on a relatively small set of observations at very low redshift, and do not span the full range of potential ages, metallicities, or galaxy types in CANDELS.  Empirical templates can be optimized for high-redshift galaxies through calibration (e.g., \citealt{Ilbert_2006}), but this process requires a large spectroscopic redshift sample that is usually not available and not representative of the photometric galaxy sample.

Theoretical template sets are constructed from stellar population synthesis models (e.g., \citealt{Fioc}, \citealt{BC03}, and \citealt{Maraston}), so they cover more of color--redshift space and perform better at high redshift \citep{EAZY}.  All of the groups in this work used theoretical template libraries.  However, theoretical template libraries cannot fully cover color--redshift space, and the mismatch between template and observed SEDs remains one of the biggest sources of systematic uncertainty in \photoz\ estimation using template-based methods \citep{newman2022}.   Furthermore, theoretical template libraries must assume star formation histories (SFHs) for their template SEDs, include (or not) emission lines, and adopt a dust attenuation law.  These choices represent a simplified version of the properties of real galaxies, especially given the wide range in redshift, luminosity, color, and galaxy type of CANDELS galaxies.  Priors can help break degeneracies between multiple redshift solutions by incorporating knowledge about the galaxy population as a whole (e.g., the abundance of galaxies as a function of luminosity and redshift), but they can introduce biases.  For instance, in Figure \label{fig:All6_summed_egs} we show that the stacked redshift probability distribution functions (PDFs) from different codes produce different redshift peaks because of differences in template sets and assumed priors even though they all perform well according to the point estimate metrics used.

\Photoz\ methods can be evaluated using spectroscopic training sets \citep[e.g.,][]{Hogg_1998, Hildebrandt_2008, Hildebrandt_2010}, clustering measurements \citep{Newman_2008}, and mock catalogs generated from simulations \citep{Schmidt_2020}, but for CANDELS, only the first is a viable option.  Most relevant to this work is the comparison by \citet{Dahlen_2013} that studied 11 \photoz\ analyses of the same CANDELS photometry and \specz\ data set.  These analyses utilized a diversity of codes, template sets, and priors---all of which impact the resulting \photoz\ PDFs.  \citet{Dahlen_2013} found that none of the \photoz\ codes significantly outperformed the others.  However, the analyses that used \speczs\ to calibrate the \photoz\ PDFs (e.g., by implementing zero-point offsets to the photometry or modifying the templates) achieved better results.  They found that \photoz\ PDFs from individual analyses were generally too narrow to be consistent with the statistical definition of a PDF (as did \citealt{Bezanson_2016}), for which they implemented a correction.  Finally, they were able to improve the \photoz\ PDFs by combining the PDFs from individual analyses via several methods: taking the median, summing PDFs, and a hierarchical Bayesian approach.

This paper builds off of \citet{Dahlen_2013} by comparing a more recent effort by six groups running five different \photoz\ codes with a larger \specz\ data set and \grismzs\ to better study performance at fainter magnitudes where the low signal-to-noise ratio of the photometry can drive discrepant results between analyses.  We calibrate the PDFs from individual codes and introduce a better method for combining PDFs, the minimum \fdiv\ \citep{renyi1961}.  We also test the performance of each group's PDFs using point estimates (specifically, the weighted-mean redshift $z_{\rm weight}$).  Our results are presented as the final CANDELS \photoz\ catalog.

This paper is organized as follows. Section \ref{sec:data} describes the data sets employed in this paper.  Section \ref{sec:optimization} explains the methods used to improve PDFs from individual \photoz\ codes and assesses the results with a variety of summary statistics.  In Section \ref{sec:combining_pdfs} we test several methods of combining PDF results from multiple \photoz\ codes.  Section \ref{sec:catalogs} presents the final \photoz\ catalog for the CANDELS fields.  Finally, Section \ref{sec:summary} summarizes the major findings of this study and provides guidance on the use of these catalogs.

Throughout the paper we assume AB magnitudes unless differently stated. In order to allow direct comparison with existing works from the literature of X-ray surveys, we adopt a flat $\Lambda$ cold dark matter cosmology with $h=H_0/[100\,\mathrm{km\, s}^{-1} \mathrm{Mpc}^{-1}]=0.7$, $\Omega_M$=0.3, and $\Omega_\Lambda$=0.7.


\section{Data}
\label{sec:data}

\subsection{Photometry}
\label{sec:photometry}

Throughout this study we use data from the CANDELS collaboration.  CANDELS obtained observations in HST/WFC3 F105W, F125W, and F160W as the primary exposures and HST/ACS F606W and F814W as parallel exposures.  These observations covered five fields spanning a total area of $\mathrm{\sim 800 \, arcmin^2}$ on the sky with $\mathrm{\sim 125 \, arcmin^2}$ in the deep portion and $\mathrm{\sim 675 \, arcmin^2}$ in the wide portion. The deep portion has a depth of $\sim 13$ orbits per pointing (spread across three filters) and includes data in two of the CANDELS fields, GOODS{-}North and  GOODS{-}South. The wide portion consists of $\sim 2$ orbits per pointing and includes buffer regions around the deep fields as well as three additional fields, COSMOS, the Extended Groth Strip (EGS), and the UKIDSS Ultra-Deep Survey field (UDS). Detailed descriptions of the sky coverage and observing strategy of $\mathrm{CANDELS}$ can be found in \cite{CANDELS_2011_Grogin} and \cite{CANDELS_2011_Koekemoer}.

One of the most powerful aspects of the CANDELS data set is the wealth of complementary photometry for these heavily studied extragalactic fields.  We utilized matched-model TFIT \citep{laidler2007} photometry (from the \texttt{v1} photometric catalogs) from ground- and space-based UV to IR imaging as the input for the photometric redshift codes.  The sources and bands vary somewhat between fields but are typically $u$-band through Spitzer/IRAC 8 $\mu$m with F160W serving as the selection band for the photometric catalogs.  Full descriptions of the catalogs for each CANDELS field are given in \citet[COSMOS]{CANDELS_COSMOS}, \citet[EGS]{CANDELS_EGS}, \citet[GOODS-S]{CANDELS_GOODSS}, \citet[UDS]{CANDELS_UDS}, and \citet[GOODS-N]{Barro_2019}.  In the interest of uniformity across fields, we only used broadband photometry.   For instance, we did not utilize the medium-band photometry from the SHARDS survey \citep{PerezGonzalez2013} for the GOODS-N field, so for certain use cases (e.g., environment) the higher-precision \photozs\ from \citet{Barro_2019} that use those bands will be preferable.


\subsection{Individual Photometric Redshifts}
\label{sec:indiv_photoz}

Six groups within the CANDELS collaboration have used the TFIT photometric catalogs to estimate \photoz\ PDFs for each galaxy observed by CANDELS. For each galaxy, the codes fit a set of SED templates in redshift space and minimized $\chi^2$ between the observed and template SEDs.  To calculate a redshift posterior PDF, most codes use a magnitude-dependent redshift prior for each object combined with a likelihood proportional to $\exp(-\chi^2/2)$.  To measure \photozs, the groups used different codes for calculation, different template SEDs, and/or different criteria when using a code (e.g., different priors).  In the following, we briefly explain each of the codes used to estimate \photoz\ PDFs in this paper.

\bigskip

\noindent \textbf{EAZY} (\citealt{EAZY})\footnote{\url{http://www.astro.yale.edu/eazy/}} fits a linear combination of stellar population templates to the observed $U$-to-8~$\mu$m SEDs (in flux space) by minimizing the $\chi^2$ statistics.  Corrections for absorption by intervening HI clouds are applied following \citet{Madau_1995}.  EAZY allows down-weighting data at rest-frame wavelengths $\geq 2 \, \mu$m via a template error function.  It incorporates a ``luminosity-function $\times$ volume'' prior that reduces the probability of low redshifts (due to the small volume) and of high redshifts for objects with bright apparent magnitudes.  In addition, an iterative application of photometric zero point offsets improves the match to the available spectroscopic redshifts.

EAZY was run by two separate groups, led by S.~Finkelstein and S.~Wuyts, respectively.  The two groups differed in their choices of parameters and priors.  The Wuyts group used the ``EAZY\_v1.0\_lines'' templates, which are identical to the original ``EAZY\_v1.0'' templates (based on the P\'{E}GASE stellar population synthesis models; \citealt{Fioc}) but with additional emission lines added following the prescription by \citet{Ilbert_2008}.  The six templates contain a subset of five ``principal component'' templates constructed as described by \citet{EAZY}, complemented by an additional template to account for dusty galaxies that do not appear in the semianalytical models that the principal component set was based on.  The Finkelstein group used the ``EAZY\_v1.1\_lines'' templates, whose first six templates are the same as the templates from the ``EAZY\_v1.0\_lines'' set except that all of the P\'{E}GASE emission lines were replaced using the \citet{Ilbert_2008} prescription.  It also has an additional template to account for massive old red galaxies.

\bigskip

\defcitealias{BC03}{BC03}

\noindent \textbf{zphot} \citep{Giallongo_1998, Fontana_2000} is a code that minimizes $\chi^2$ statistics when fitting template $\mathrm{SEDs}$.
It returns the best-fitting values and estimated \photoz\ uncertainties as well as other parameters of the galaxy template (e.g., age, metallicity, and dust extinction).
zphot accepts both fluxes and magnitudes as input with a rigorous treatment of nondetections in the latter case.  The code allows for a minimum photometric error to be set in each photometric band and optionally applies a cut to avoid unrealistically large negative fluxes.  For the templates, we used the P\'{E}GASE 2.0 models  \citep{Fioc}, which assumed the stellar initial mass function from \citet{Rana_1992} because they yielded the most accurate photo-$z$'s with the set of parameters explored.  Emission lines were not included in the templates.  Our adopted template library was constructed with a combination of SFHs.  Specifically, we used constant and truncated SFHs as well as an SFH with the star formation rate proportional to the amount of gas.  Metallicity evolution is measured self-consistently during the galaxy evolution.  We implemented dust extinction with a \cite{Calzetti_2000} extinction law with E(B-V) ranging from 0 to 1.  Finally, to avoid contamination from nonstellar emission it excludes $\mathrm{IRAC}$ bands that probe rest frame $> 5.5\, \mu$m (ch3 at $z \leq 0.15$ and ch4 at $z \leq 0.6$).  This code was run by the group led by A.~Fontana.

\bigskip

\noindent \textbf{HyperZ} (\citealt{HyperZ_2000})\footnote{\url{http://webast.ast.obs-mip.fr/hyperz/}} performs fits by minimizing the $\chi^2$ statistics in flux space while excluding negative fluxes. The user can specify shifts to magnitudes. To avoid unrealistic solutions, a prior for the $\mathrm{F160W}$ band absolute magnitude in the range $-30 < M < -9$ (Vega mag) was used. Reddening was included in the form of the \cite{Calzetti_2000} reddening law varying $A_V$ from 0 to 3 in steps of 0.2. SED templates are based on the \cite{Maraston} models with exponentially declining SFHs (with e-folding times $\tau=0.1$, 0.3, 1, and 2 Gyr) at four metallicity values (1/5, 1/2, 1, and 2 $\times$ solar metallicity).  The allowed age range of the templates was restricted between 0.1 Gyr and the age of the universe at the given redshift of the galaxy in question. Finally, the option for a minimum photometric error was set to 0.05 magnitudes. This recipe was introduced in \cite{Pforr_2013}. This code was run by the group led by J.~Pforr.

\bigskip

\noindent \textbf{LePhare} (\citealt{LePHARE_2011})\footnote{\url{http://www.cfht.hawaii.edu/~arnouts/LEPHARE/lephare.html}} minimizes $\chi^2$ statistics when fitting template SEDs, both in magnitude and in flux space. It has the capability to include luminosity priors, to add extra photometry errors, to use a training sample to optimize the template SEDs and to derive zero-point offsets, and to account for contributions from emission lines \citep{Ilbert_2006, Ilbert_2009}.  The median values from the photo-$z$ PDFs are also provided. Here, a prior on the optical absolute magnitude in the range $-24 < M < -8$ is used, while IRAC ch3 and ch4 are excluded.  The SED template set was a combination of the  \citetalias{BC03} and \citet{Polletta_2007} templates.  This code was run by the group led by M.~Salvato.

\bigskip

\noindent \textbf{WikZ} (\citealt{WikZ_2008}) fits SED templates to observed photometry in flux space by minimizing $\chi^2$ statistics.  The code can be run with two different parameterized SFHs, an exponentially declining star formation rate or a delayed-$\tau$ SFH. In the current application, the code was run using a delayed-$\tau$ SFH with the \citetalias{BC03} SED templates. Negative fluxes are not completely excluded, but they add to the $\chi^2$ when the template flux is brighter than the $1\sigma$ upper limit, while they are excluded when the flux is lower than the $1\sigma$ upper limit. Additionally, it excludes IRAC ch3 and ch4 for $z < 0.5$ and $z < 0.7$, respectively. Finally, it has the capability to add extra smoothing errors to the photometry.  This code was run by the group led by T.~Wiklind.


\subsection{Spectroscopic and 3D-HST Grism Redshifts}
\label{s:speczs_grismzs}

Spectroscopic redshift are used to train photometric redshift codes (e.g., by identifying zero-point offsets in observed-frame photometry that, if removed, will improve fits) and to test the accuracy of the photo-$z$ estimates. The six \photoz\ participant groups used a training set of 5807 high-quality spectroscopic redshifts spanning all five CANDELS fields.  The same set of redshifts is used to recalibrate the PDFs in Section \ref{sec:optimization}.

Our primary testing set consists of 4089 high quality spectroscopic redshifts drawn from a variety of sources completely independent of those included in the training set with any overlapping objects removed. We use this set to assess the performance of point statistics (e.g., the redshift of maximum probability) and to test the quality of photo-$z$ $\mathrm{PDFs}$ from each code, both before and after the optimization procedure described below.

We also use 3D-HST grism redshifts (grism-$z$'s; \citealt{3DHST}) to test the \photoz\ point estimate and PDF quality. This set consists of 3367 of the highest-quality redshifts spanning all of the CANDELS fields. The grism-$z$'s were determined spectrophotometrically with EAZY, but run by other investigators with different priors and using different photometry. To mitigate the impact of incorporating photo-$z$ information in the grism redshift fits, we restricted to objects where the spectrum, not the photo-$z$, drove the fit.  We also include only objects with redshifts larger than $\mathrm{\mathit{z}_{grism}} > 0.6$.

\begin{figure*}
  \begin{center}
      \includegraphics[width=0.7\textwidth]{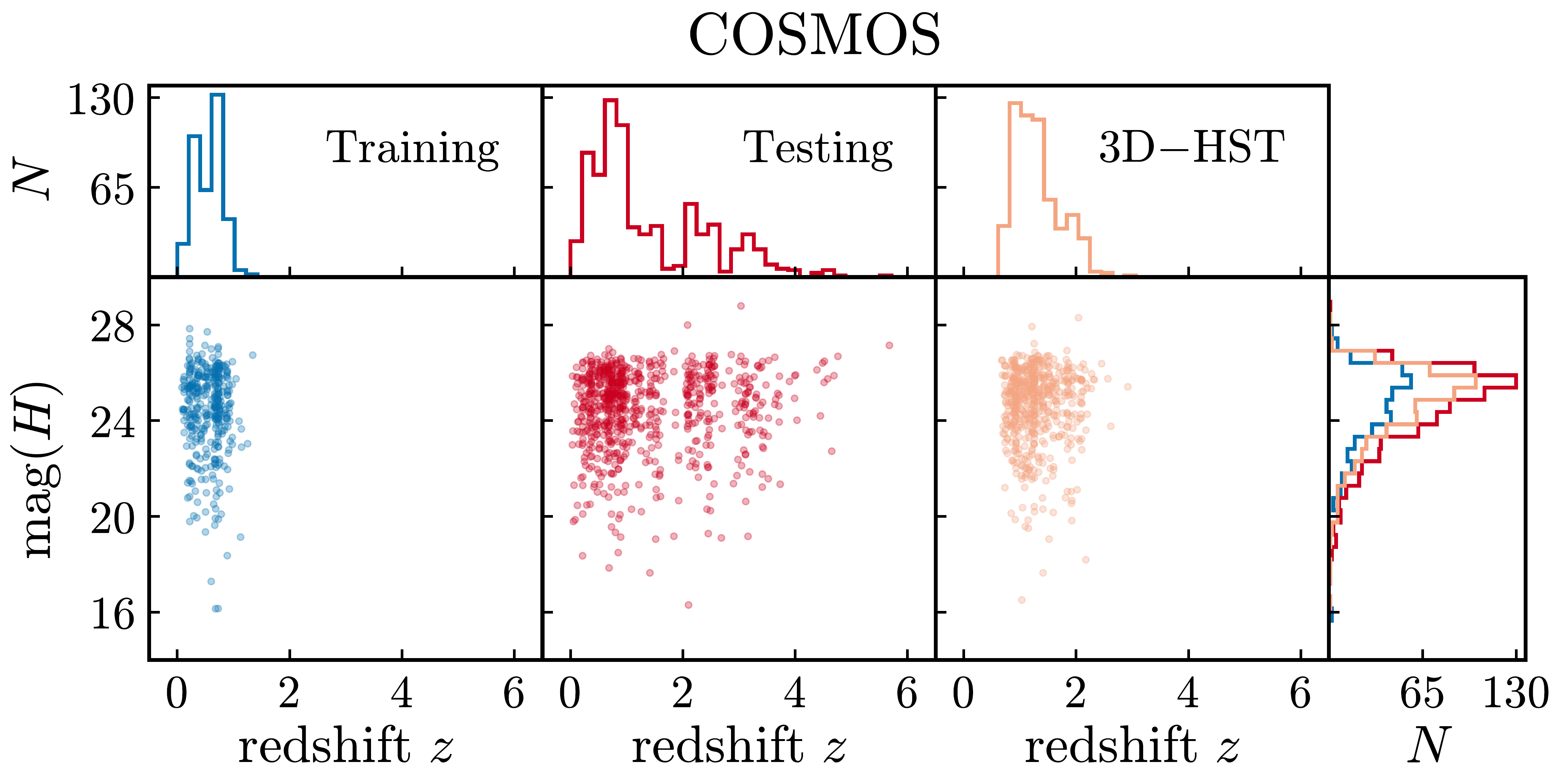}
      \\[0.5cm]
      \includegraphics[width=0.7\textwidth]{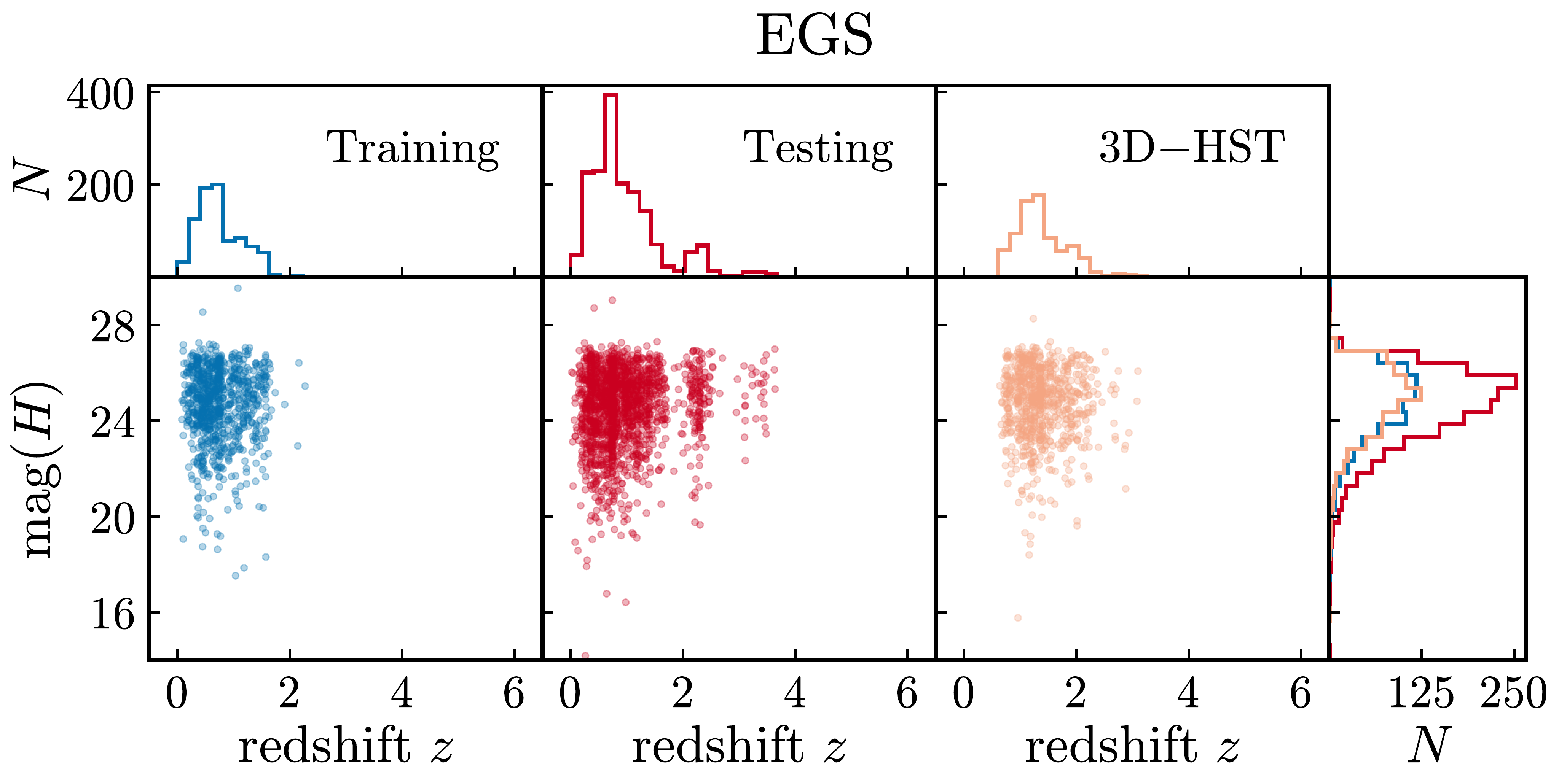}
      \\[0.5cm]
      \includegraphics[width=0.7\textwidth]{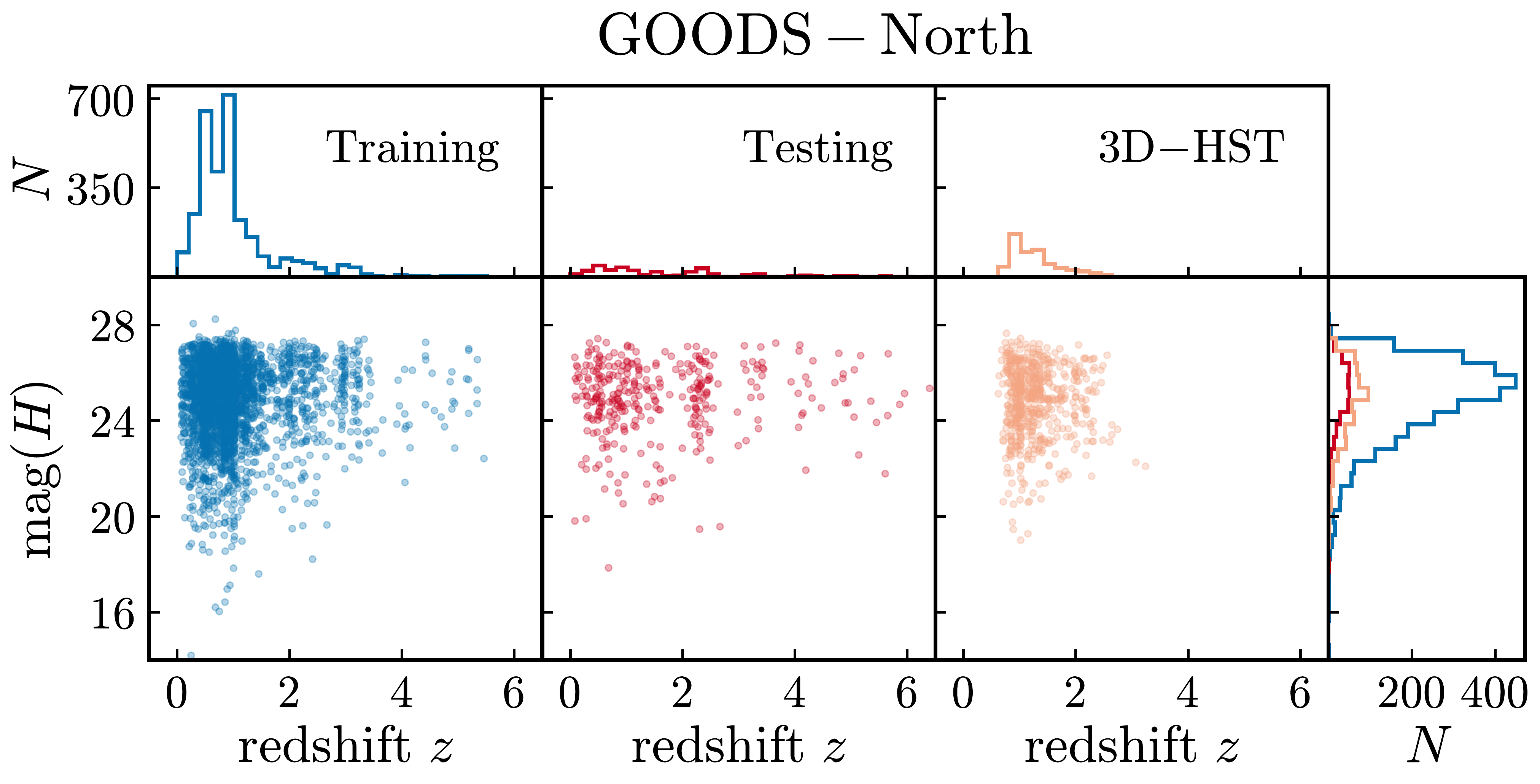}
  \end{center}
  \caption{$H$-band magnitude as a function of redshift for objects with spectroscopic redshifts (divided into training and testing sets) and 3D-HST grism redshifts for the COSMOS, EGS, and GOODS-N fields. The redshift and magnitude ranges of the testing and 3D-HST data sets strongly differ from the training set; as a result, they provide highly independent assessments of the quality of photo-$z$ PDFs, which are optimized based upon the training set.}
  \label{fig:specz_samples1}
\end{figure*}

\begin{figure*}
  \begin{center}
      \includegraphics[width=0.7\textwidth]{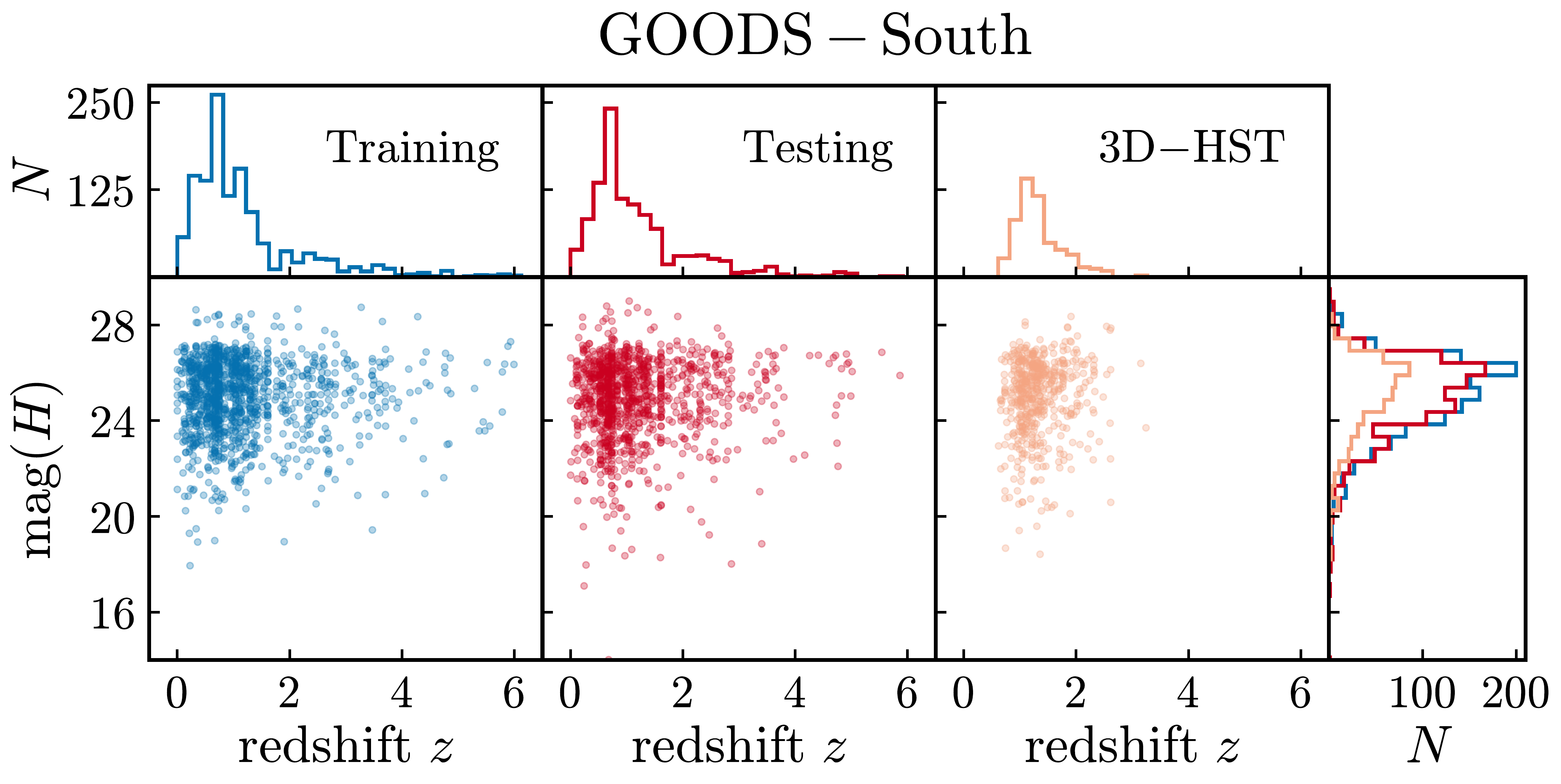}
      \\[0.5cm]
      \includegraphics[width=0.7\textwidth]{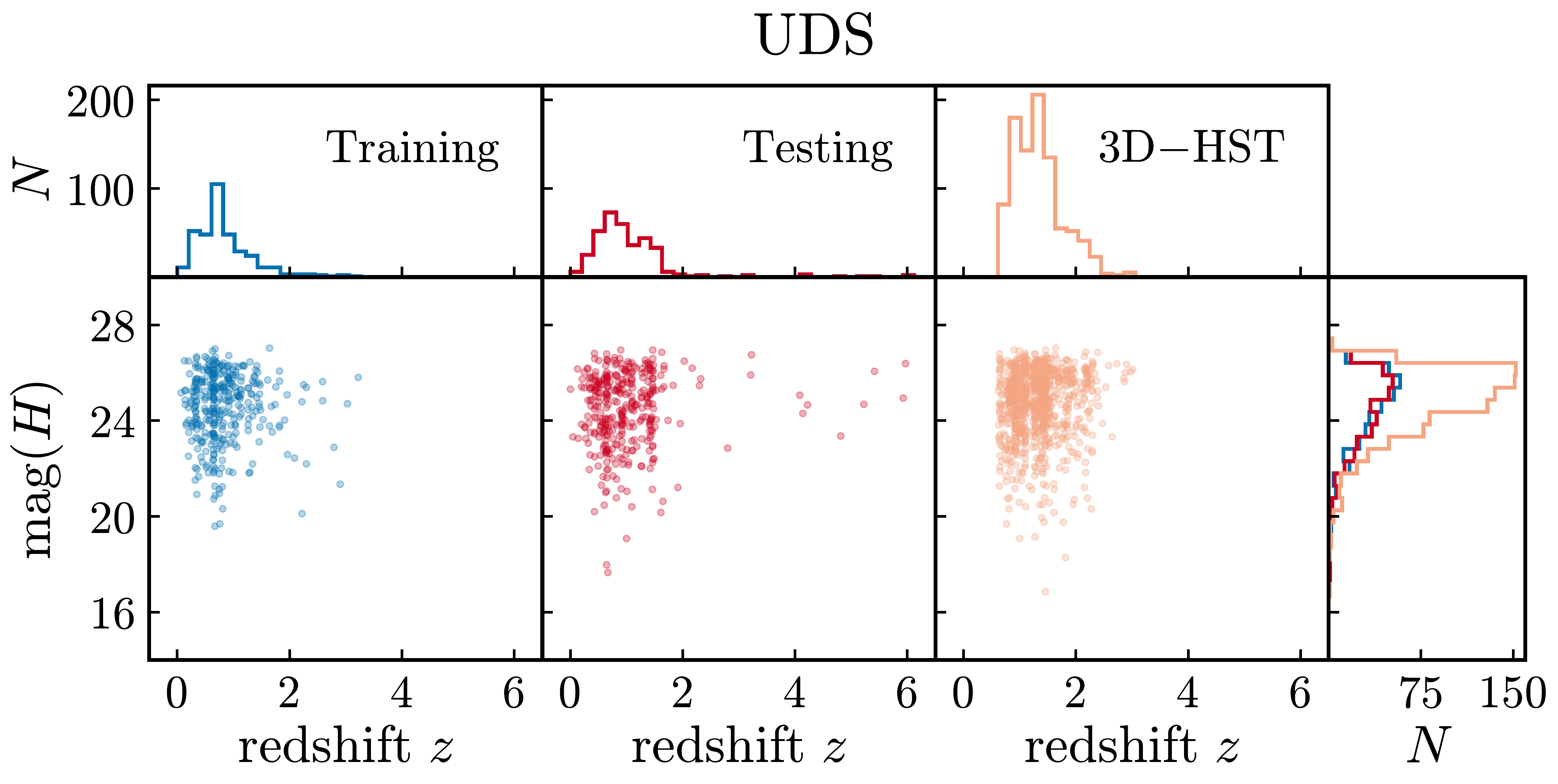}
  \end{center}
  \caption{Same as Figure \ref{fig:specz_samples1} except showing the GOODS-S and UDS fields.}
  \label{fig:specz_samples2}
\end{figure*}

\autoref{fig:specz_samples1} and \autoref{fig:specz_samples2} show the $H$-band magnitude as a function of redshift for the spectroscopic sample, as well as the marginal histograms for each of the five CANDELS fields separately.  \autoref{table:redshifts} gives details regarding the construction of these three sets, including references to the original catalogs and the specific cuts applied to select only the highest-quality, most secure redshifts.

Due to the order in which \speczs\ were obtained, the objects in the training sets used to tune the photometric redshift methods differ in both galaxy properties (such as brightness) and redshift coverage from the testing and 3D-HST sets, as illustrated in \autoref{fig:specz_samples1} and \autoref{fig:specz_samples2}.  This difference is greatest in the COSMOS and EGS fields.  As a result, the performance metrics derived using these independent data sets would be expected to be worse than would be measured using a random training/testing assignment from a common population. In fact, the objects for which we estimate \photozs\ extend to fainter magnitudes and at higher redshifts than the training sets, so this more pessimistic assessment is likely more realistic for assessing the performance for the broader population, though we cannot exclude the possibility that \photoz\ errors are worse in regimes for which we have no spectroscopy.


\section{Optimization Methods}
\label{sec:optimization}

The photo-$z$ PDFs produced by the different groups are not analogous to the true probability distributions that meet the statistical definition of a PDF (see e.g., \citealt{Dey_2021a}, \citealt{dey2022_calpit}).  Given the high likelihood that the spectroscopic redshifts used here are correct (>95\%), we can treat them to first order as representing the true redshifts of the objects observed.  If the statistical definition of a PDF is being obeyed, then we would expect 68\% and 95\% of them to fall in the 68\% and 95\% credible intervals of the photo-$z$ PDFs, respectively.

In an earlier test with CANDELS data, \cite{Dahlen_2013} found that while some codes had better results than others, none of them performed well when the coverage of credible intervals was assessed in this way. This indicates that the photo-$z$ {PDFs} must have substantial, qualitative problems. Simple examples of issues would be the presence of bias due to template mismatch, such that the PDFs are shifted to higher or lower redshifts than expected, or inaccuracies in photometric error models that caused the PDFs to be too wide or too narrow.

In this section, we will describe how we optimize the photo-$z$ PDFs produced by each group for each CANDELS field.  We correct for (1) systematic shifts in the PDFs and (2) systematically inaccurate PDF widths.

\subsection{$Q$--$Q$ Plot and Metrics}
\begin{figure}
  \begin{center}
    \includegraphics[width=0.47\textwidth]{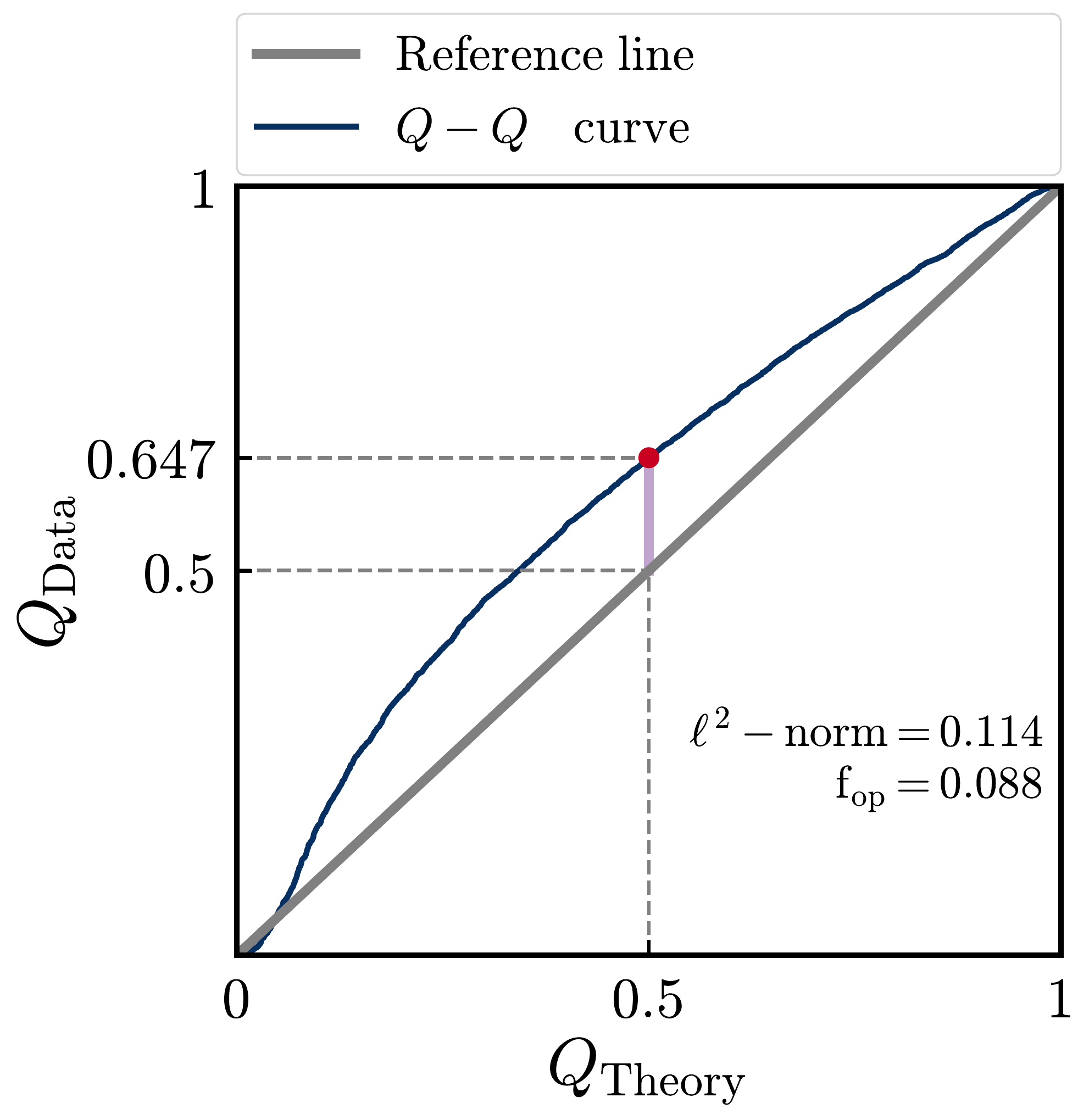}
  \end{center}
  \caption{An example Quantile--Quantile ($Q$--$Q$) plot constructed using a training set of \speczs\ and the corresponding set of \photoz\ PDFs from Finkelstein.  For a given \photoz\ CDF quantile ($Q_\mathrm{Theory}$), $Q_{\rm Data}$ is the fraction of objects for which the spectroscopic redshift is below the redshift corresponding to that \photoz\ quantile in the CDF for that particular object.  For instance, if the spectroscopic redshift is below the 50th percentile point in the \photoz\ CDF for 64.7\% of objects, then (0.5, 0.647) is a point (red circle) along the $Q$--$Q$ curve (blue line).  If photo-$z$ PDFs obey the statistical definition of a PDF, $Q$--$Q$ curves will lie along the solid gray reference line from (0, 0) to (1, 1). Their deviation is quantified by the square root of the sum of the vertical distance (pink line) at each point squared ($\ell^2$-norm). The $\ell^2$-norm is normalized by dividing by the square root of the number of objects. Additionally, the $f_\mathrm{op}$ gives the fraction of objects with spec-$z$'s that fall outside of the main region of the corresponding PDFs.}
\label{fig:QQ_example}
\end{figure}

\begin{figure*}
  \begin{center}
    \includegraphics[width=\textwidth]{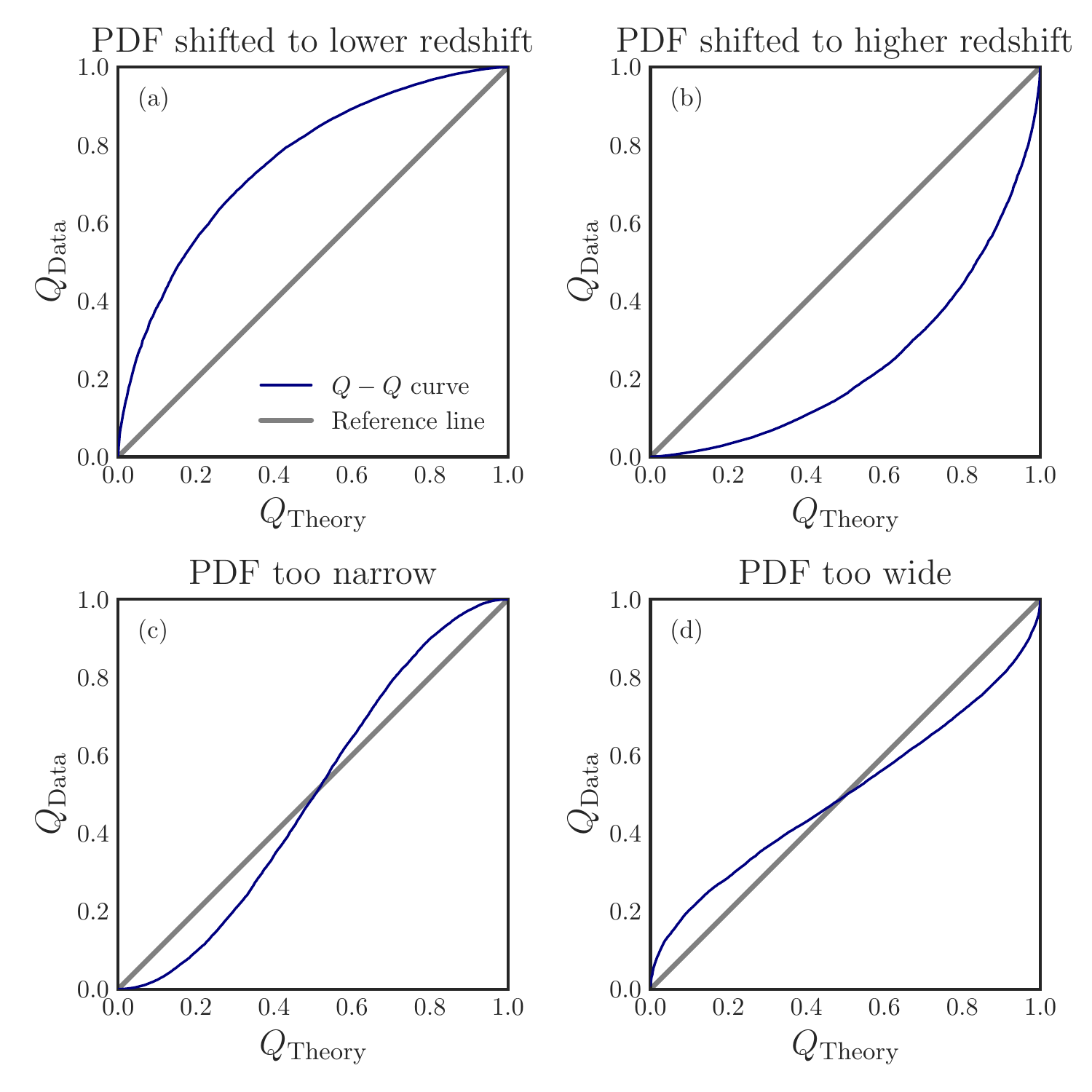}
  \end{center}
  \caption{Example $Q$--$Q$ plots showing the effects of \photoz\ PDFs that are systematically (a) shifted to lower redshift, (b) shifted to higher redshift, (c) too narrow, and (d) too wide.  Ideally, PDFs would lie along the gray diagonal reference line.}
\label{fig:QQ_example_grid}
\end{figure*}

In this work, we make use of a set of statistics related to the Quantile--Quantile ($Q$--$Q$) plot to optimize the recalibration of the photo-$z$ PDFs produced by each group.  While these statistics will prioritize adjustments that improve agreement with the statistical definition of a PDF, they should also deliver more accurate redshift point values and error estimates by reducing biases.  Figure \ref{fig:QQ_example} depicts the $Q$--$Q$ plot for a set of objects from the spec-$z$ training set with Finkelstein \photozs\ as an illustrative example.  This plot shows $Q_\mathrm{Data}$---the fraction of the time that the spectroscopic redshift is below the redshift corresponding to a given quantile in the \photoz\ cumulative distribution function (CDF)---as a function of the chosen value for the quantile, $Q_\mathrm{Theory}$.  By construction, the value of $Q_\mathrm{Theory}$ must range from 0 to 1, just as the CDF does (e.g., $Q_\mathrm{Theory}$ = 0.345 corresponds directly to the CDF value of 0.345).  In our example,  $Q_\mathrm{Data}(0.345)$ will be the fraction of the time that the spectroscopic redshift of an object is below the redshift where the CDF value for that object is $Q_\mathrm{Theory}$ = 0.345.
If the PDFs satisfy the statistical definition, $Q_\mathrm{Theory}$ will be the same as $Q_\mathrm{Data}$ (apart from small variations due to sampling error), so the $Q$--$Q$ plot will hew closely to the unit line between (0, 0) and (1, 1).  To help in interpreting the $Q$--$Q$ plots shown in this paper, Figure \ref{fig:QQ_example_grid} demonstrates how systematic shifts or incorrect widths of PDFs will manifest in them.

We use the normalized ${\ell^2}$-norm to quantify the deviation of the $Q$--$Q$ plot from the identity line to assess how close a set of PDFs is to the ideal case.  The quantity $\Delta Q = Q_\mathrm{Data} - Q_\mathrm{Theory}$, evaluated at a particular value of $Q_\mathrm{Data}$, will correspond to the gap between the $Q$--$Q$ curve for a given sample and the unity line (doing this as a function of $Q_\mathrm{Data}$ has computational advantages).  The ${\ell^2}$-norm will then correspond to the square root of the sum of this distance squared, $(\int (\Delta Q)^2 dQ_\mathrm{Data})^{1/2}$.  This distance will be zero in the ideal case.

For this analysis, we only include galaxies whose spectroscopic redshift lies between the 0.0001 and 0.9999 quantiles of their CDF in the construction of the $Q$--$Q$ plot and in the calculation of the ${\ell^2}$-norm.  Objects outside that range correspond to catastrophic outliers (as commonly occur in photometric redshift analyses) whose true redshift lies beyond the range that would be expected based on their estimated PDF.  We do not want their distribution to affect the recalibration of PDFs for the much more common nonoutlier objects.  Excluding them also makes our analysis considerably more robust to the possibility of erroneous spectroscopic redshifts. Hence, in the end, we calculate a normalized version of the $\ell^2$-norm by summing up the value of $(\Delta Q)^2$ at the $Q_{\rm Data}$ values corresponding to each object in the sample excluding those that were outside these CDF bounds, dividing by the total number of $(\Delta Q)^2$ values that were summed, and then taking the square root.

The second basic statistic we use to characterize the quality of PDFs from a given algorithm is  the fraction of objects for which their spectroscopic redshift has a CDF value less than 0.0001 or greater than 0.9999.  We label this quantity as  ${f_{\rm op}}$ (for ``fraction outside of PDF'').  This statistic roughly corresponds in its nature to the catastrophic outlier rate used to characterize the performance of point estimates from photometric redshift algorithms but is based on PDF outputs rather than scalar quantities.

\begin{figure*}
  \begin{center}
      \includegraphics[width = 0.45\textwidth]{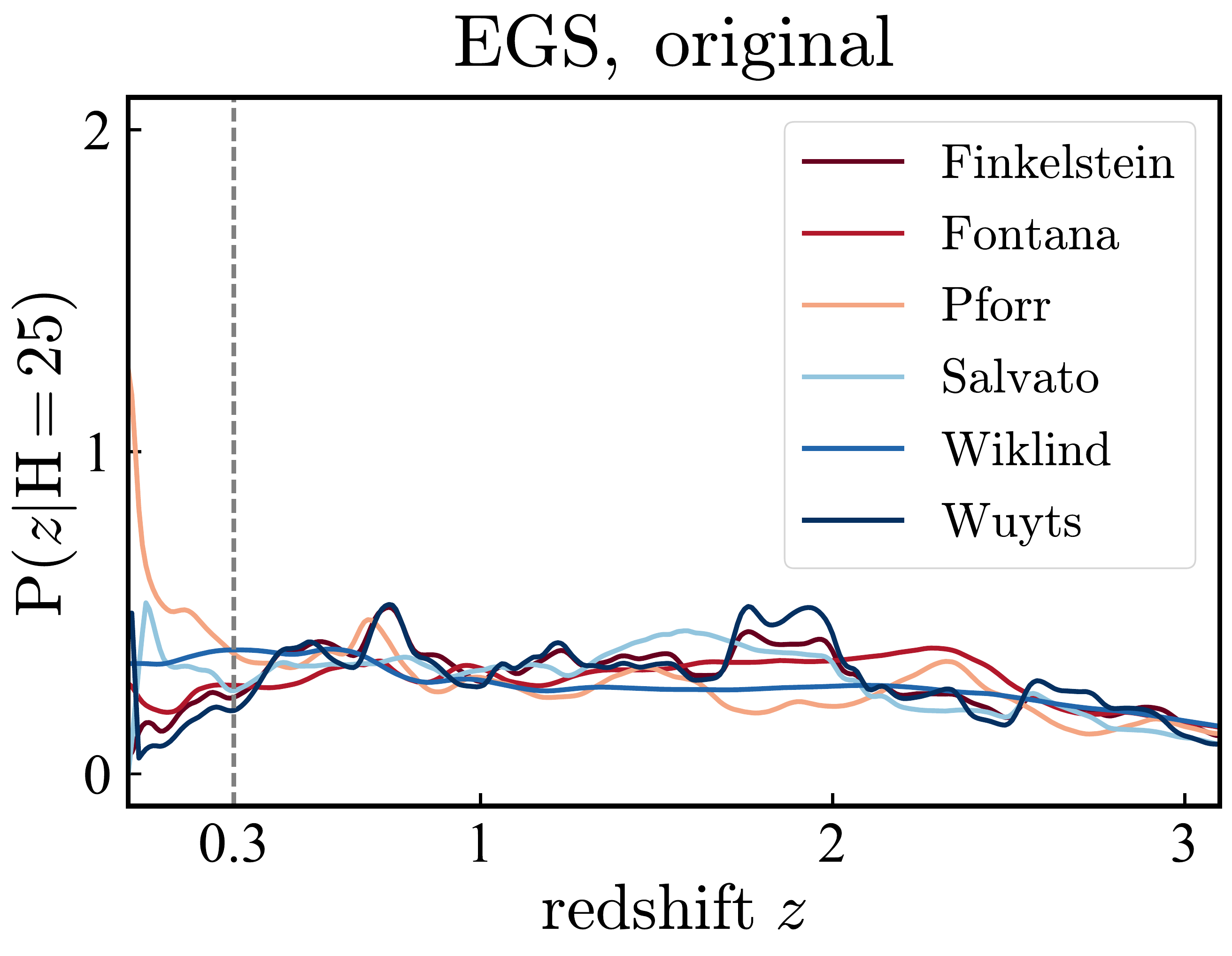}
      \hspace{0.5cm}
      \includegraphics[width = 0.45\textwidth]{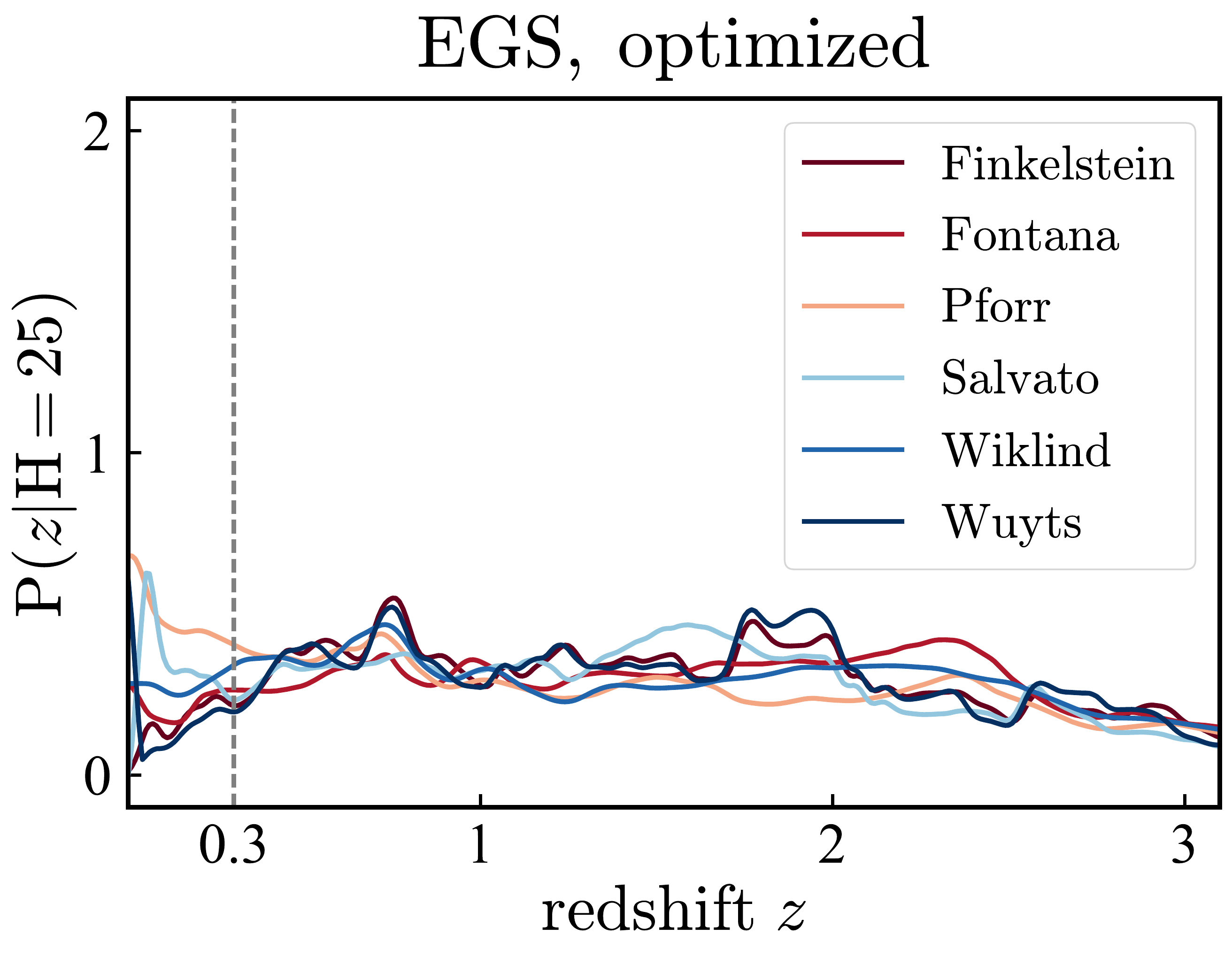}
  \end{center}
  \caption{\small Sum of the photo-$z$ PDFs for all objects in a magnitude bin of width 0.5 centered at $H = 25$ mag in the EGS field.  The curves correspond to the expectation value for the redshift distribution in this bin.  We exclude objects with missing PDFs for any of the groups from the sums to ensure that all curves are equivalent. The disagreement between groups is greatest at low redshifts (below $z = 0.3$, indicated by the gray dashed line).  Because of this divergence, we exclude objects with $z \leq 0.3$ from the calculation of $Q$--$Q$ statistics.  Although all codes deliver good performance when PDF peak redshifts are compared to spec-$z$'s, the aggregate predictions for broad galaxy samples differ significantly from code to code.}
\label{fig:All6_summed_h_egs}
\end{figure*}

To understand possible redshift- and magnitude-dependent biases among the PDFs generated by different participant codes, we analyze the normalized $P(z)$ in a fixed magnitude bin. In \autoref{fig:All6_summed_h_egs}, we show the set of the summed PDF curves for EGS objects in the magnitude bin centered at $H=25$ predicted by each group, both before and after the optimization procedures (described below) have been applied. Even after optimization, the predicted number of objects at low redshifts varies greatly from group to group. Results in this regime may have limited reliability.  In \autoref{fig:All6_summed_egs}, we show a set of images constructed from curves such as those shown in \autoref{fig:All6_summed_h_egs}.  These images show redshift as a function of magnitude (using the center of the magnitude bin used to construct the summed PDF curves).  The intensity of the color scale indicates the value of the summed PDF at a given magnitude and redshift (corresponding to the $y$-values in \autoref{fig:All6_summed_h_egs}).  Although all of the codes used to produce the photo-$z$ PDFs yield good results when evaluated with standard point statistics for objects with spectroscopic redshifts (as described below), and all are using the same photometry as inputs, the redshift distributions as a function of magnitude from each group differ in many details, even after optimizing the individual photo-$z$ PDFs such that they better obey the statistical definition.  We find that \photoz\ PDFs by different groups show divergent behavior at $z<0.3$. Therefore, we exclude these objects for our further analysis.

\begin{figure}
  \begin{center}
      \includegraphics[width = 0.45\textwidth]{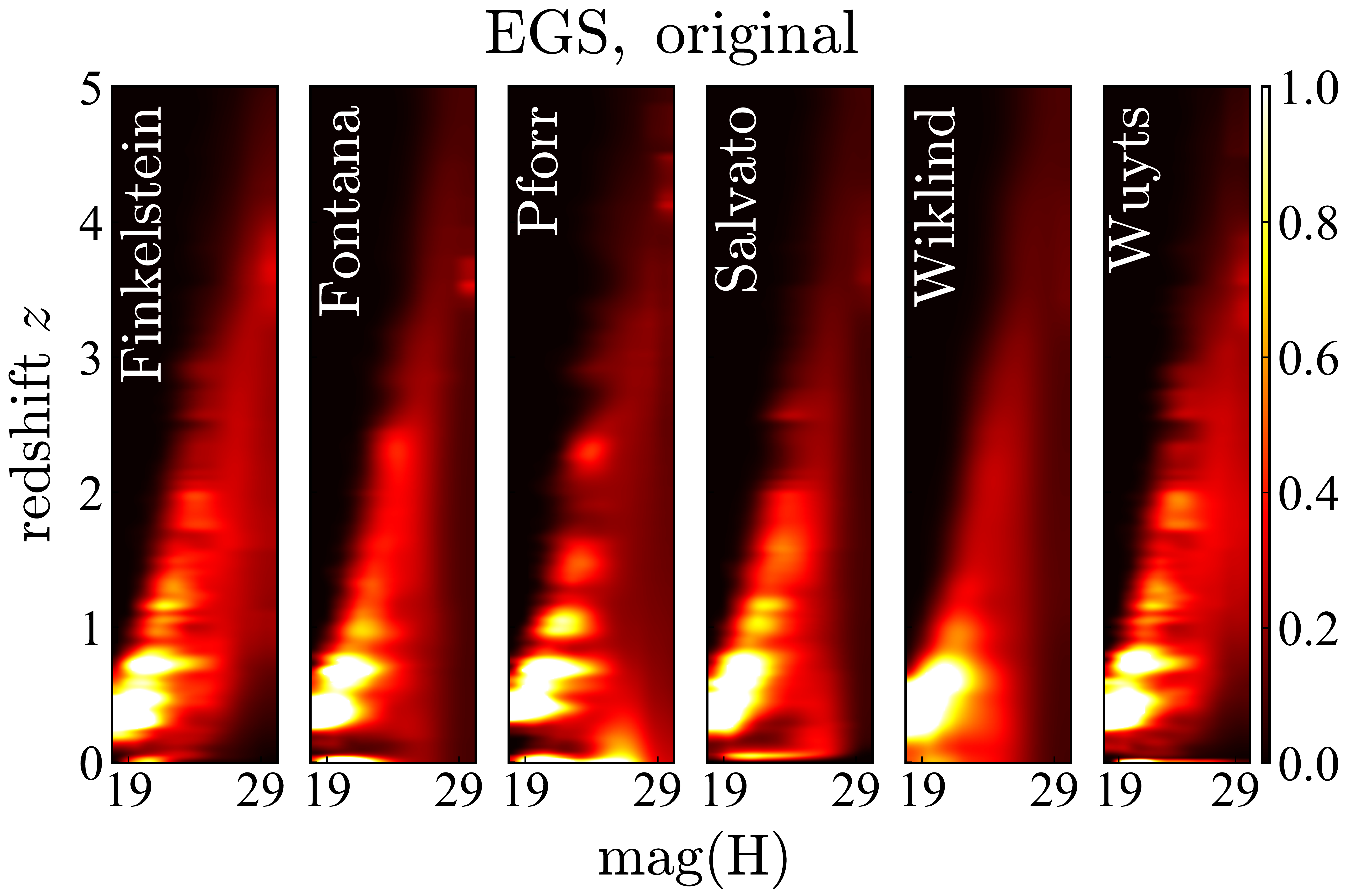}
      \\[0.5cm]
      \includegraphics[width = 0.45\textwidth]{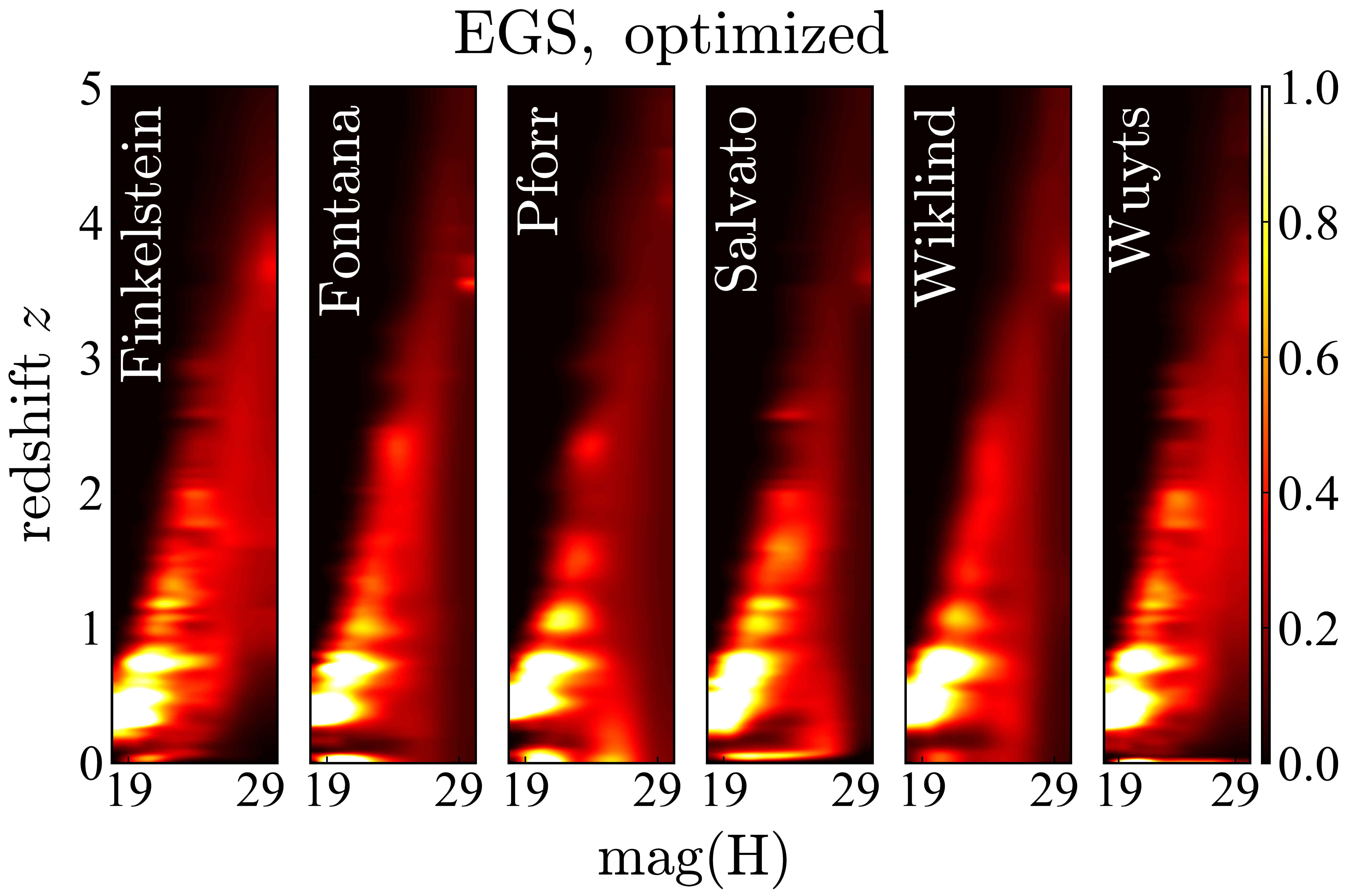}
  \end{center}
  \caption{\small Summed PDFs as a function of magnitude and redshift, before and after the photo-$z$ optimization procedures, for the EGS field. The color scale of the image indicates the summed and renormalized probability that an object in a particular magnitude bin is at the selected redshift, equivalent to the $y$-axis in \autoref{fig:All6_summed_h_egs}. Objects with missing $\mathrm{PDFs}$ for any of the groups are excluded from the sums. The shape and locations of the bright regions (which should correspond to redshifts where there is an excess of objects due to sample/cosmic variance), as well as other detailed features, differ significantly from group to group, even though all groups used identical photometry for the same set of objects to calculate the PDFs used here.}
\label{fig:All6_summed_egs}
\end{figure}

\subsection{Correcting for Global Shifts}

Using the $Q$--$Q$ statistics defined above, we can identify any aggregate bias in the photo-$z$ PDFs of a given group by applying negative or positive shifts in the redshift direction (${dz}$) to the PDFs. Our sign convention is such that negative ${dz}$ values correspond to shifts to the left on a plot of PDF($z$), such that the PDFs will peak at lower values of redshifts, and vice versa. We apply shifts corresponding to an array of 151 equally spaced values over the interval $[-0.5, 1.]$, and construct the $Q$--$Q$ curve in each case after removing objects with spec-$z$'s outside their PDFs. We then identify the shift value that minimizes the normalized ${\ell^2}$-norm; that is, the value that yields a $Q$--$Q$ curve as close as possible to the ideal unit line. To do this, we interpolate the normalized $\ell^2-{\rm norm}$ values using a quadratic univariate spline (using the {\tt scipy.interpolate.UnivariateSpline} routine with parameters $k = 2$ and $s = 1$.) The goal of this interpolation is for greater accuracy and smoother results.

\begin{figure*}
  \begin{center}
      \includegraphics[width=0.95\textwidth]{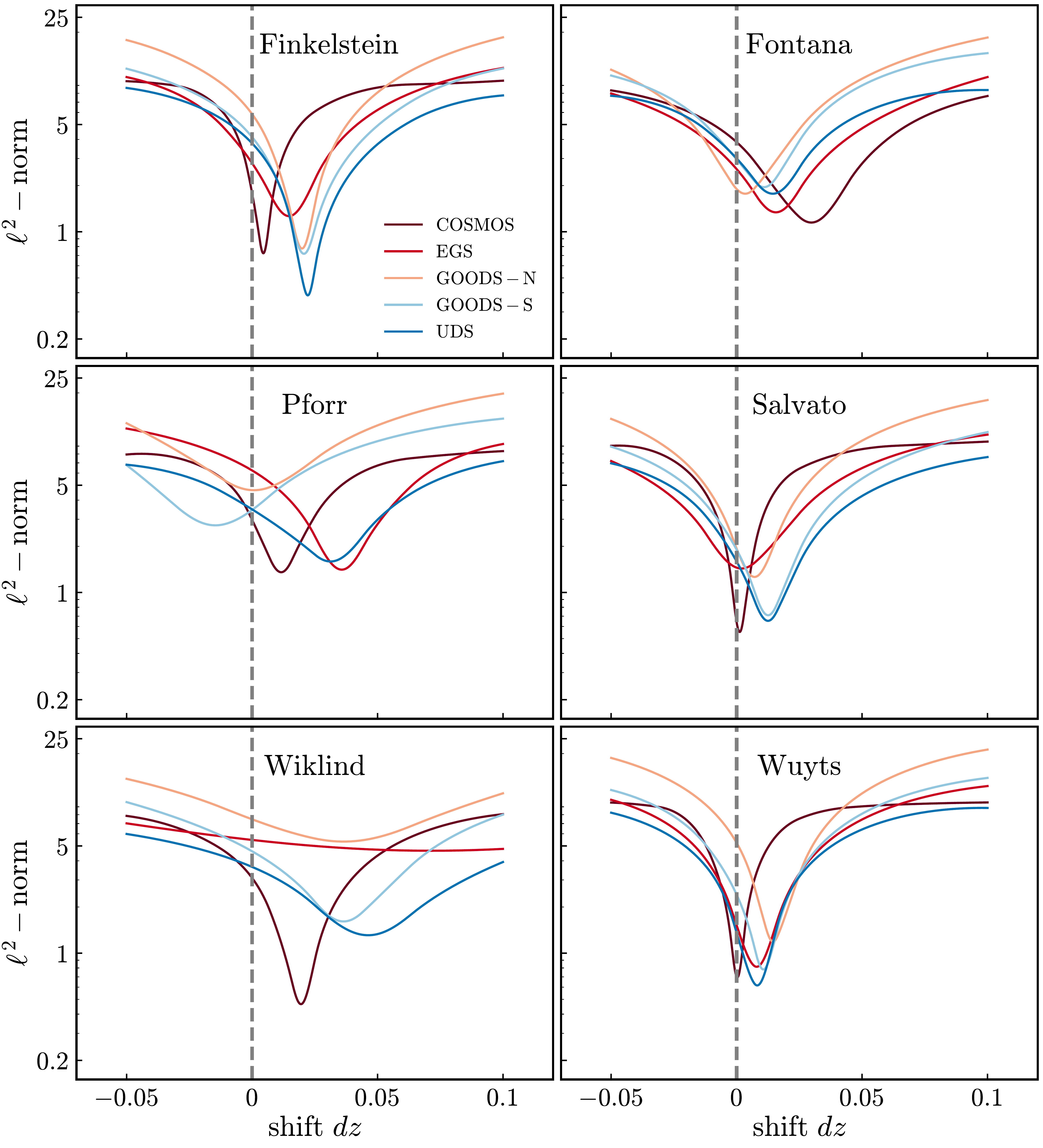}
  \end{center}
  \caption{Normalized $\mathrm{\ell^2}$-norm as a function of the shift ${dz}$ applied to the photo-$z$ $\mathrm{PDFs}$ for each of the five $\mathrm{CANDELS}$ fields, calculated using the spectroscopic redshift training set. The normalized $\mathrm{\ell^2}$-norm will have a smaller value when the PDFs more closely obey the statistical definition of a probability distribution; as a result, we wish to minimize this quantity.  The vertical gray dashed line indicates a zero shift, meaning that the calibration of the PDFs from a given group were accurate. A different shift seems to be needed for each group and each field, but nearly all of the shifts are positive. Consequently, the uncorrected photo-$z$ point estimates are biased low.}
\label{fig:L2norms_dz}
\end{figure*}

In \autoref{fig:L2norms_dz}, we present the dependence of the normalized ${\ell^2}$-norm on ${dz}$ for each group and CANDELS field.  Photo-$z$'s from some groups exhibit larger biases than others, with the bias varying from field to field; this may be due to the differences in the imaging bands included in each field and the target selection function of spectroscopic surveys carried out in those fields.  While the required shift for the COSMOS field is the smallest for four of the groups, we did not include all of the available medium and narrow band photometry for this field (that would lead to improved \photoz\ estimates) in the interest of uniformity across fields, so we expect that this feature may be due to the spectroscopy available in the COSMOS field.  With only one exception (Pforr results for the GOODS-South field) the shifts are all positive. That indicates that for almost all fields and groups, the photo-$z$ PDFs peak at lower values than they should, biasing point estimates (such as the redshift of the PDF peak) low. \autoref{table_dz} gives the optimal shift values that should be applied to the photo-$z$ PDFs from each group and each field to yield better performance in the $Q$--$Q$ plot.  We will assess the level of improvement achieved in both $Q$--$Q$ statistics and point estimates in sections to follow, using the independent testing set of spectroscopic redshifts as well as 3D-HST.

\begin{table*}
\caption{The optimal values of shifts ${dz}$ for the PDFs from each group for each field, determined by identifying the minima in the curves shown in \autoref{fig:L2norms_dz}. Most of the shifts are positive, indicating that  PDF$(z)$ curves need to be shifted to the right. Results from some groups exhibit larger biases than others, with Salvato and Wuyts needing the smallest shifts and Wiklind the largest, especially in the EGS field.}
\label{table_dz}
\begin{center}
\begin{tabular}{lrrrrrr}
%
\hline
\hline
$\mathrm{Field}$ &
$\mathrm{Finkelstein}$ &
$\mathrm{Fontana}$ &
$\mathrm{Pforr}$ &
$\mathrm{Salvato}$ &
$\mathrm{Wiklind}$ &
$\mathrm{Wuyts}$ \\
\hline
\hline
COSMOS  & 0.005 & 0.030 & 0.012    & 0.001 & 0.020 & 0.000 	\\
EGS 	& 0.015 & 0.016 & 0.036    & 0.002 & 0.071 & 0.008  \\
GOODS-N & 0.020 & 0.003 & 0.001    & 0.007 & 0.036 & 0.015  \\
GOODS-S & 0.021 & 0.011 & $-$0.015 & 0.013 & 0.036 & 0.011  \\
UDS 	& 0.022 & 0.014 & 0.031    & 0.013 & 0.046 & 0.008  \\
\hline
\hline
\end{tabular}
\end{center}
\end{table*}

\subsection{Correcting PDF Width}

To correct for either overestimation or underestimation of photo-$z$ errors, we stretch/sharpen the PDF widths.  We do this by raising a PDF to a power of $\alpha = \mathrm{1/\gamma}$ and then renormalizing the PDF to integrate to one.  For a Gaussian PDF, this is equivalent to multiplying the $\sigma$ parameter by $\gamma^{1/2}$, and so provides a way to correct for either overestimates or underestimates of errors; however, the procedure we use can be applied to any PDF whether it is Gaussian or not.  Optimal values of the parameter $\mathrm{\gamma}$ that are close to unity imply that little change is needed in the width of the PDFs.  Conversely, when the optimal values differ significantly from unity, then large differences in the widths of the PDFs are implied (corresponding to significantly overconfident or underconfident error models).

We determine optimal values of $\gamma$ after we apply the optimized shifts $dz$ that were determined by the preceding analysis. We consider values in the range $0 < \mathrm{\gamma} < 1$ as well as in the range $\mathrm{\gamma} \geq 1$, corresponding to $\mathrm{\alpha} > 1$ and $0 < \mathrm{\alpha} \leq 1$, respectively. The former range corresponds to making photo-$z$ PDFs sharper, since high peaks of the PDF become higher while low PDF values in the tails or valleys between peaks become smaller.  Conversely, $\gamma \geq 1$ stretches the photo-$z$ PDFs into broader shapes, since the contrast between the highest peaks and low values of the PDF is reduced.

For each group's PDFs in a particular field, we search a coarse array of $\gamma$ values spanning $[0.05, 7]$, construct the $Q$--$Q$ plot in each case, and evaluate the normalized $\mathrm{\ell^2}$-norm for each value of $\gamma$. We then use a finer grid of $\gamma$-values focused around the coarse-grid value that minimizes the $\mathrm{\ell^2}$-norm to obtain improved precision in the optimal value.

\begin{figure*}
  \begin{center}
      \includegraphics[width=0.95\textwidth]{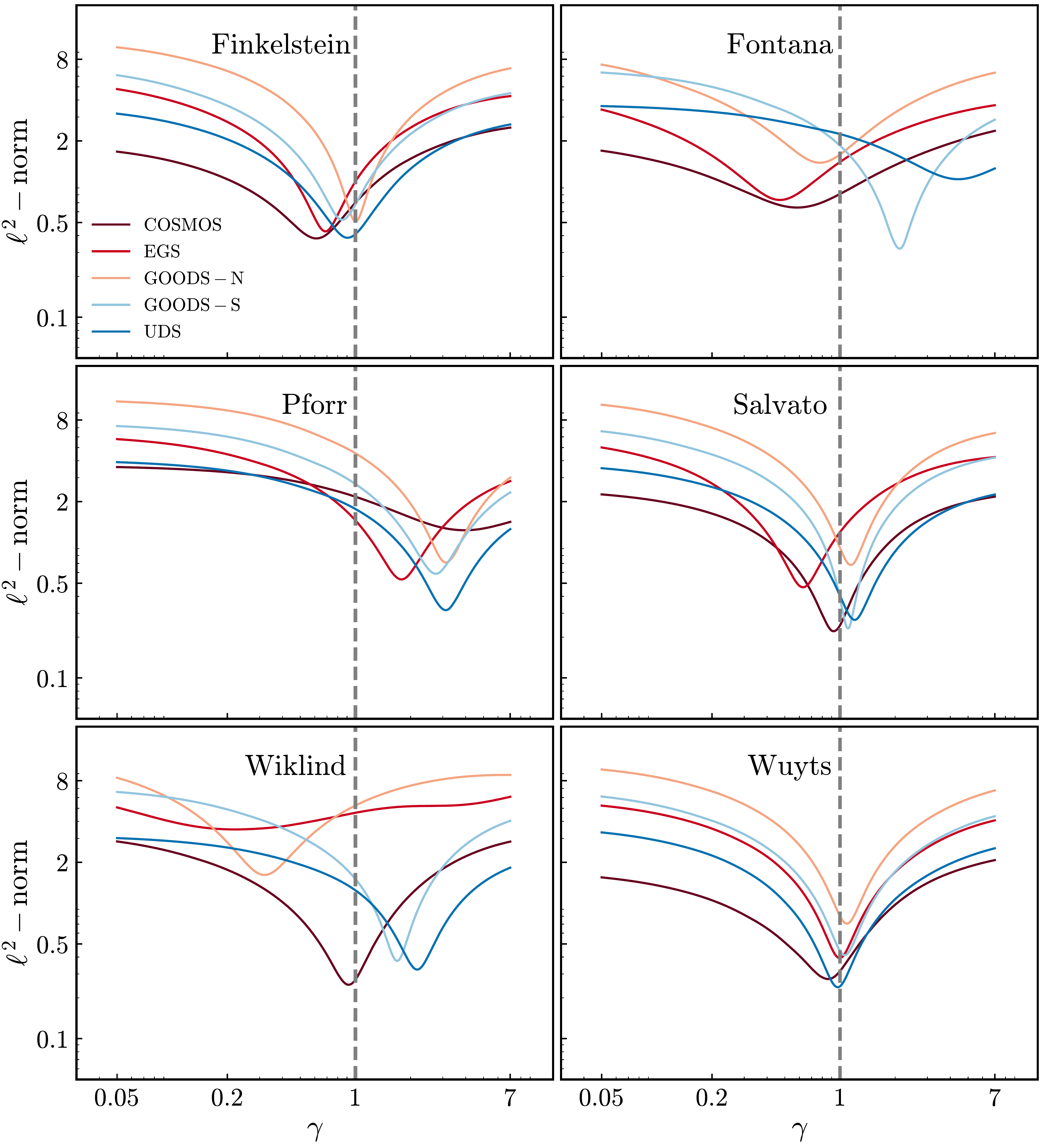}
  \end{center}
  \caption{Normalized $\mathrm{\ell^2}$-norm as a function of $\mathrm{\gamma}$, a scaling parameter used to improve the widths of the photo-$z$ {PDFs}. PDFs are raised to the power $\mathrm{1/\gamma}$ in this analysis, corresponding to altering the standard deviation parameter of a Gaussian distribution by a factor of $\sqrt{\gamma}$.  These plots were constructed using only the training set of spectroscopic redshifts in each of the five $\mathrm{CANDELS}$ fields, for each group's PDFs independently, after applying the shifts in redshift listed in \autoref{table_dz}. The vertical gray dashed line indicates where $\mathrm{\gamma = 1}$, corresponding to the case where no changes are applied to PDFs. Different values of the parameter $\mathrm{\gamma}$ are needed for each group's PDFs in each field, without any clear pattern.}
\label{fig:L2norms_gamma}
\end{figure*}

\autoref{fig:L2norms_gamma} shows the dependence of the $\mathrm{\ell^2}$-norm on $\mathrm{\gamma}$ for each group and $\mathrm{CANDELS}$ field.  The Wuyts PDFs are the best calibrated with all fields having best-fit $\gamma$-values near unity.  The Salvato PDFs show moderate scatter around $\gamma = 1$ with some fields having PDFs that are too narrow and some too wide. The Finkelstein and Pforr PDFs are consistently too wide and too narrow, respectively.  The Fontana and Wiklind PDFs demonstrate larger field-to-field scatter with some fields having PDFs that are too narrow and some too wide.  In \autoref{table_gamma}, we list the $\gamma$ values required to optimize the raw photo-$z$ PDFs such that they most closely match the statistical definition.

\begin{table*}
\caption{The optimal values of the PDF rescaling parameter $\gamma$ (used to improve the widths of the {PDFs}) for each participant group and each CANDELS field. While some groups' results yield $\gamma$ values that are close to unity, others have significantly lower or significantly larger values, implying substantial changes to the widths of their photo-$z$ PDFs.}
\label{table_gamma}
\begin{center}
\begin{tabular}{lcccccc}
\hline
\hline
 $\mathrm{Field}$ &
$\mathrm{Finkelstein}$ &
$\mathrm{Fontana}$ &
$\mathrm{Pforr}$ &
$\mathrm{Salvato}$ &
$\mathrm{Wiklind}$ &
$\mathrm{Wuyts}$ \\
\hline
\hline
COSMOS  & 0.611 & 0.582 & 3.959 & 0.922 & 0.916 & 0.859  \\
EGS     & 0.686 & 0.470 & 1.784 & 0.628 & 0.220 & 0.996  \\
GOODS-N & 1.003 & 0.777 & 3.118 & 1.145 & 0.315 & 1.090  \\
GOODS-S & 0.848 & 2.115 & 2.750 & 1.106 & 1.683 & 1.060  \\
UDS	    & 0.905 & 4.418 & 3.114 & 1.203 & 2.174 & 0.973  \\
\hline
\hline
\end{tabular}
\end{center}
\end{table*}


\subsection{Evaluation Using $Q$--$Q$ Statistics}
\label{sec:results_qq}

\begin{figure*}
  \begin{center}
    \includegraphics[width=0.7\textwidth]{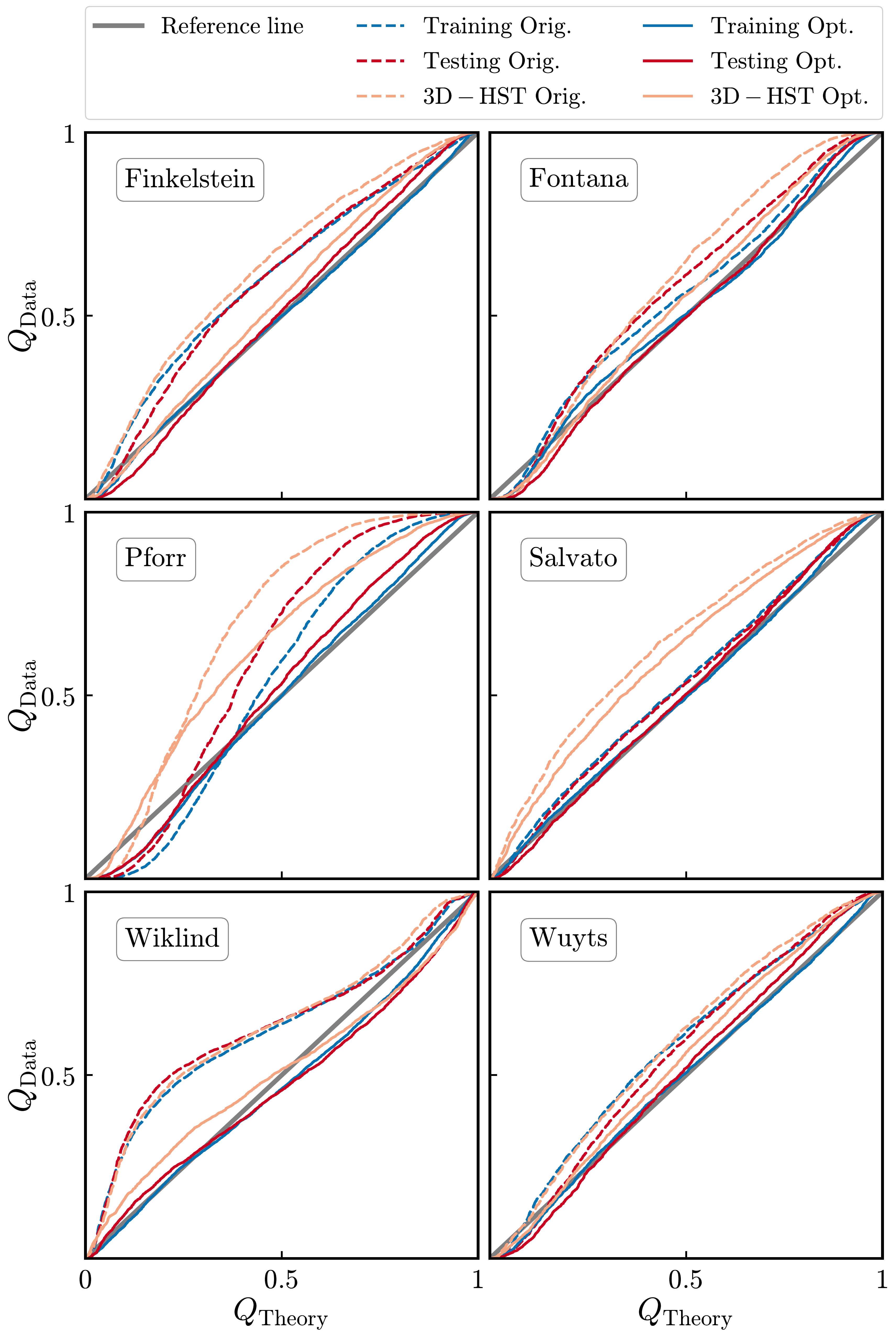}
  \end{center}
  \caption{Quantile--Quantile ($Q$--$Q$) plots for photo-$z$ PDFs both before (``original,'' dashed curves) and after (``optimized,'' solid curves) optimized transformations have been applied.  The $Q$--$Q$ plot shows the CDF value evaluated at the spectroscopic redshift of an object, with $Q_{\rm data}$ corresponding to a given quantile of the set of CDF values, $Q_{\rm theory}$. For instance, if the 30th percentile in the set of CDF($z_{\rm spec}$) values is 0.21, (0.3, 0.21) would be a point along the $Q$--$Q$ curve.  If photo-$z$ PDFs obey the statistical definition of a PDF, $Q$--$Q$ curves will lie along the unit line from (0, 0) to (1, 1). We evaluate the results from each group separately using the independent training and testing sets of spectroscopic redshifts as well as the 3D-HST grism redshifts; results from all five CANDELS fields are combined together here. It is clear that our optimization methods improve the results for each group and for every set of redshifts.  The normalized $\ell^2$-norm used to evaluate PDF accuracy corresponds to the RMS deviation between a $Q$--$Q$ curve and the diagonal in the $y$-direction in these plots.}
\label{fig:QQ_All6}
\end{figure*}

After identifying the optimal values of the $dz$ and $\gamma$ parameters using the training set of spectroscopic redshifts and their corresponding photo-$z$ PDFs, we can apply the corresponding transformations to all of the PDFs from a particular group and field. We can then construct $Q$--$Q$ plots and evaluate the overall normalized $\ell^2$-norm and $f_{\rm op}$ values for each group evaluated with the training data (which should have results that are biased low) as well as with the independent testing set of spectroscopic redshifts and the 3D-HST grism redshifts. Since some fields have relatively few testing redshifts available, we combine the objects from all five CANDELS fields to make the $Q$--$Q$ plots. We present the $Q$--$Q$ plots for each group in \autoref{fig:QQ_All6}. In each panel, dashed curves correspond to the $Q$--$Q$ plots for each spectroscopic data set before applying shift and power transformations, and solid curves show the $Q$--$Q$ plots after transformations. It is clear that all curves more closely follow the diagonal reference line after optimization.

\begin{figure}
  \begin{center}
      \includegraphics[width=0.47\textwidth]{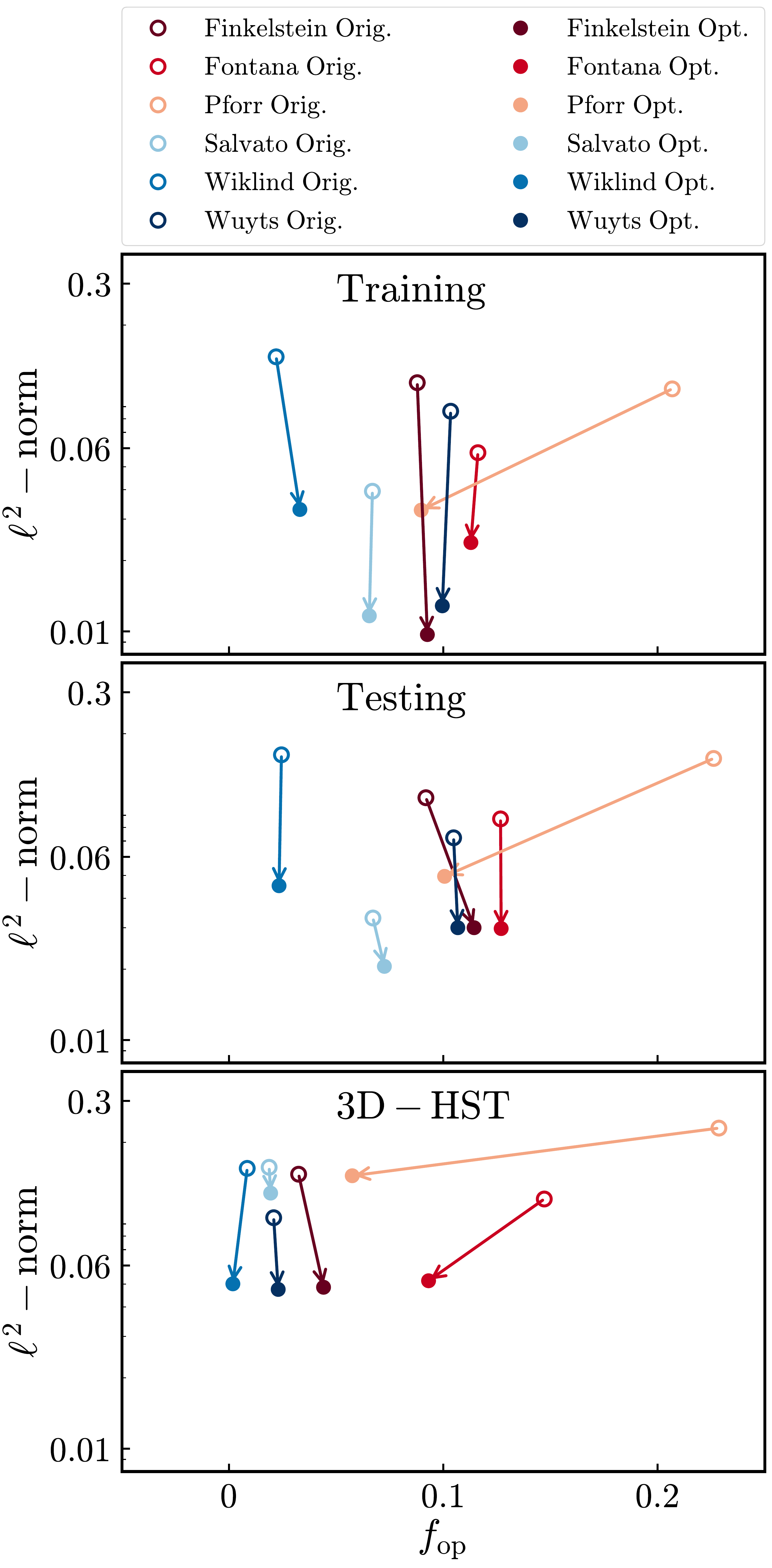}
  \end{center}
	\caption{The normalized $\ell^2$-norm as a function of the out-of-PDF fraction $f_{\rm op}$ for the PDFs from each group both before (open circles at arrow bases) and after (filled circles at arrow heads) optimization. The three panels show these statistics evaluated using the training set of spectroscopic redshifts (top); the testing set (middle); and 3D-HST grism redshifts (bottom). The $\ell^2$-norms are reduced in all cases; the $f_{\rm op}$ values are lower in most cases and only increase slightly otherwise. No single group produced superior results in all cases.}
\label{fig:All6_l2norm_fop}
\end{figure}

In \autoref{fig:All6_l2norm_fop}, we show the normalized $\ell^2$-norm and $f_{\rm op}$ values before and after optimization of the PDFs. Although in a few cases the $f_{\rm op}$ value becomes slightly larger after optimization, gains in the normalized $\ell^2$-norm are substantial and ubiquitous. This is to be expected as the optimized $\ell^2$-norm values are lowest for the training set, as by construction our optimization chooses the transformation parameter values that make that quantity as small as feasible. However, it is very encouraging that substantial improvements are also found for the testing and 3D-HST data sets, which have very different coverage in magnitude--redshift space from the training set.  We note that the Wiklind analysis produced significantly more objects with missing PDFs, which were not treated as out-of-PDF here.  Thus, the low values of $f_{\rm op}$ for that analysis should be considered lower limits on what the performance would be for the full set of objects.

\subsection{Evaluation Using Point Statistics}
\label{sec:point_stats}

Although our primary focus is on producing accurate photometric redshift probability distributions, we also can evaluate the performance of point (or summary) statistics for the photometric redshifts of CANDELS objects before and after optimization.  We focus on two different point statistics which are constructed from the photo-$z$ PDFs: the most probable redshift (which we label $z_{\rm peak}$, as it corresponds to the redshift where the highest peak of the photo-$z$ PDF occurs), and the probability-weighted expectation value of the redshift ($z_{\rm weight}$).  Following \citet{Dahlen_2013}, we compute this quantity using only the region surrounding the highest peak of the PDF (specifically, the redshift range within that peak where the PDF value remains above $0.05 \times \mathrm{PDF}(z_{\rm peak})$). To determine this range, we first linearly interpolate the PDF onto a grid of 64,001 redshifts in place of the original 1001, then use cubic spline interpolation on this finer grid to find the redshifts around $z_{\rm peak}$ where the PDF is equal to $0.05 \times \mathrm{PDF}(z_{\rm peak})$. If the PDF values are never lower than $0.05 \times \mathrm{PDF}(z_{\rm peak})$, then the bounds used to calculate $z_{\rm weight}$ are the same as the bounds of the grid on which the PDFs are defined.  In general, the \citet{Dahlen_2013} definition of $z_{\rm weight}$ yields better results than the overall expectation value of the redshift evaluated over the full PDF, as in cases where PDFs have multiple peaks, the overall expectation value will often lie between peaks in regions of negligible probability.

We use two quantities to evaluate the accuracy of these point statistics before and after optimization. The first is the normalized median absolute deviation ($\mathrm{\sigma_{NMAD}}$) of the differences between photo-$z$ point estimates and spectroscopic redshifts for the corresponding objects, defined by:

  \begin{equation}
    \sigma_{\text{NMAD}} = 1.48 \times \, \mathrm{median} \left( \frac{|\Delta \mathit{z}|}{1 + \mathit{z}_{\text{spec}}} \right) ,
  \end{equation}
where $\Delta z = z_{\rm phot} - z_{\rm spec}$ is the difference between a point estimate of the photometric redshift for an object and its spectroscopic redshift.
The normalizing factor of 1.48 in the definition of $\sigma_{\rm NMAD}$  causes the expectation value of the NMAD to equal the standard deviation for data that is drawn from a normal distribution.

In addition to the NMAD, we track the fraction of catastrophic outliers in a particular group's results, $f_{\rm co}$, as another useful statistic for characterizing photo-$z$ quality from each group. We define catastrophic outliers as those objects that fulfill the condition:
\begin{equation}
  \frac{ |\Delta z| }{1 + \mathit{z}_{\text{spec}}} > 0.15  .
\end{equation}
Here the limit $0.15$ is somewhat arbitrarily chosen; we will show below that typical photometric redshift errors for CANDELS objects with spectroscopic redshifts are $\sim 0.03(1+z)$, so the threshold of 0.15 approximately corresponds to  $\mathrm{5 \sigma}$ outliers.

We find that the NMAD is smaller when computed using $z_{\rm weight}$ rather than $z_{\rm peak}$; that is, $z_{\rm weight}$ provides a superior estimate of the redshift of an object. Therefore, we only show results for statistics computed using $z_{\rm weight}$ in the remainder of this paper, although we still include $z_{\rm peak}$ in our final photo-$z$ catalogs.

\begin{figure*}
  \begin{center}
    \includegraphics[height=0.85\textheight]{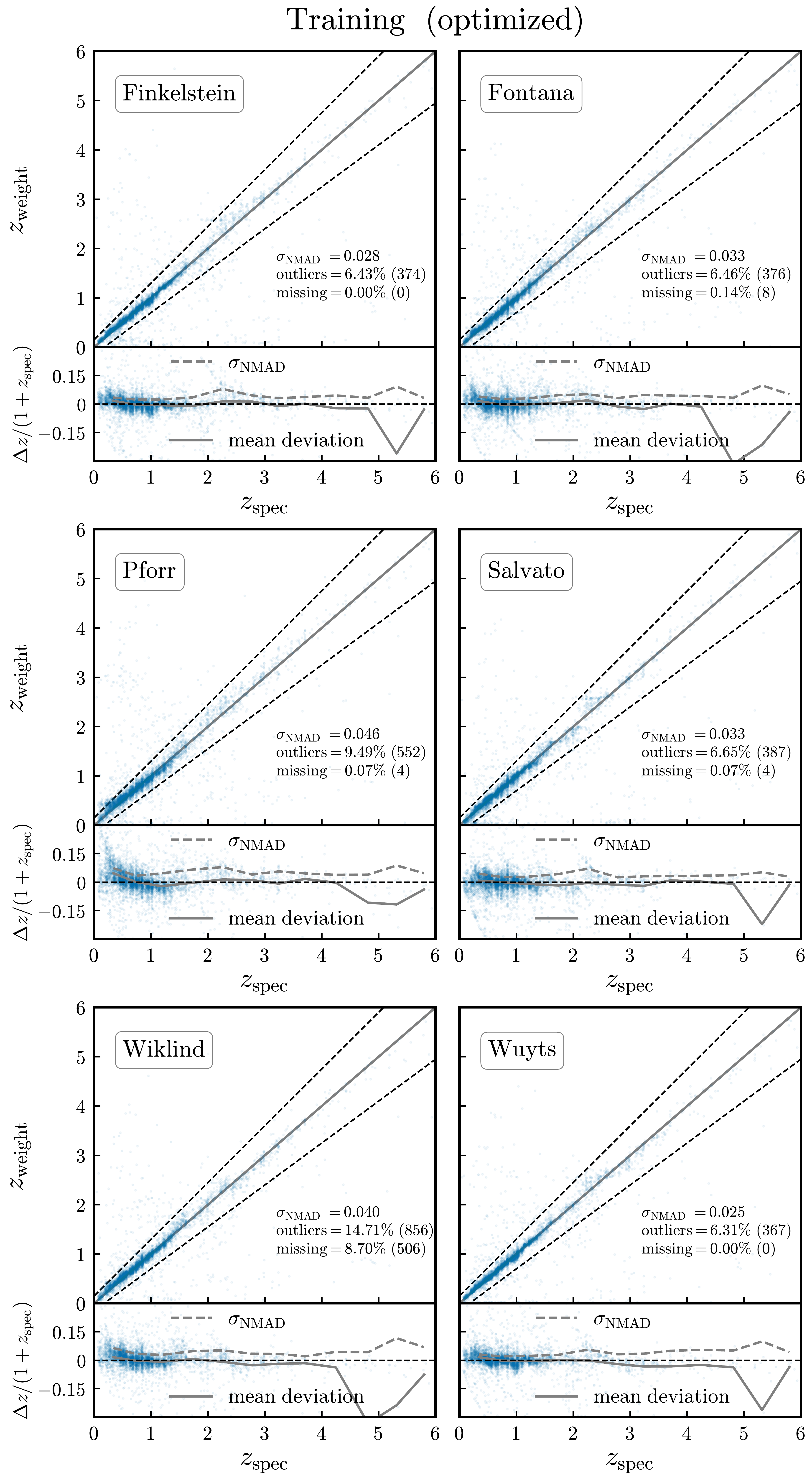}
  \end{center}
  \caption{Scatter plot of $z_{\rm weight}$ vs.\ $z_{\rm spec}$ (upper panels), using the optimized version of photo-$z$ PDFs for the calculation of $z_{\rm weight}$ from the six participating groups.  The $\sigma_{\rm NMAD}$, the outlier percentage, and the percentage of objects with missing photo-$z$ PDFs (and therefore missing $z_{\rm weight}$) are listed along with the actual number of objects (given in parentheses) for the latter two statistics.  Additionally, $\Delta z/({1 + z_{\rm spec}})$ vs.\ $z_{\rm spec}$ (lower panels) is presented below each scatter plot. The gray dashed and solid lines correspond to $\sigma_{\rm NMAD}$ and Hodges--Lehmann mean of $\Delta z/({1 + z_{\rm spec}})$, as a function of $z_{\rm spec}$. Bins of $0.5$ in $z_{\rm spec}$ have been used for the calculation of these two curves.}
\label{fig:All6_scatter}
\end{figure*}

\autoref{fig:All6_scatter} shows the scatter plots of $z_{\rm weight}$ vs.\ $z_{\rm spec}$ for all six groups, using only objects belonging to the training set of spec-$z$'s. The $z_{\rm weight}$ values are calculated using the optimized version of the photo-$z$ PDFs. In the same plots we present the $\sigma_{\rm NMAD}$, the percentages (and number of objects in parentheses) of catastrophic outliers, as well as the percentage (and number of objects in parentheses) of the objects that have missing photo-$z$ PDFs and therefore missing $z_{\rm weight}$ values. Below each scatter plot, we show the dependence of $\Delta z/({1 + z_{\rm spec}})$ on $z_{\rm spec}$, where $\Delta z = z_{\rm weight} - z_{\rm spec}$. In the same plots we also present the redshift dependence of $\sigma_{\rm NMAD} = 1.48 \times \Delta z/({1 + z_{\rm spec}})$ (dashed gray line), and the Hodges--Lehmann mean of $\Delta z/({1 + z_{\rm spec}})$ (solid gray line), where bins of $0.5$ have been used along $z_{\rm spec}$ for their construction. They generally do not seem to vary significantly with redshift, apart from the spikes observed at $z_{\rm spec} \sim 5.5$, where the number of objects becomes very low.

\begin{figure}
  \begin{center}
      \includegraphics[width=0.45\textwidth]{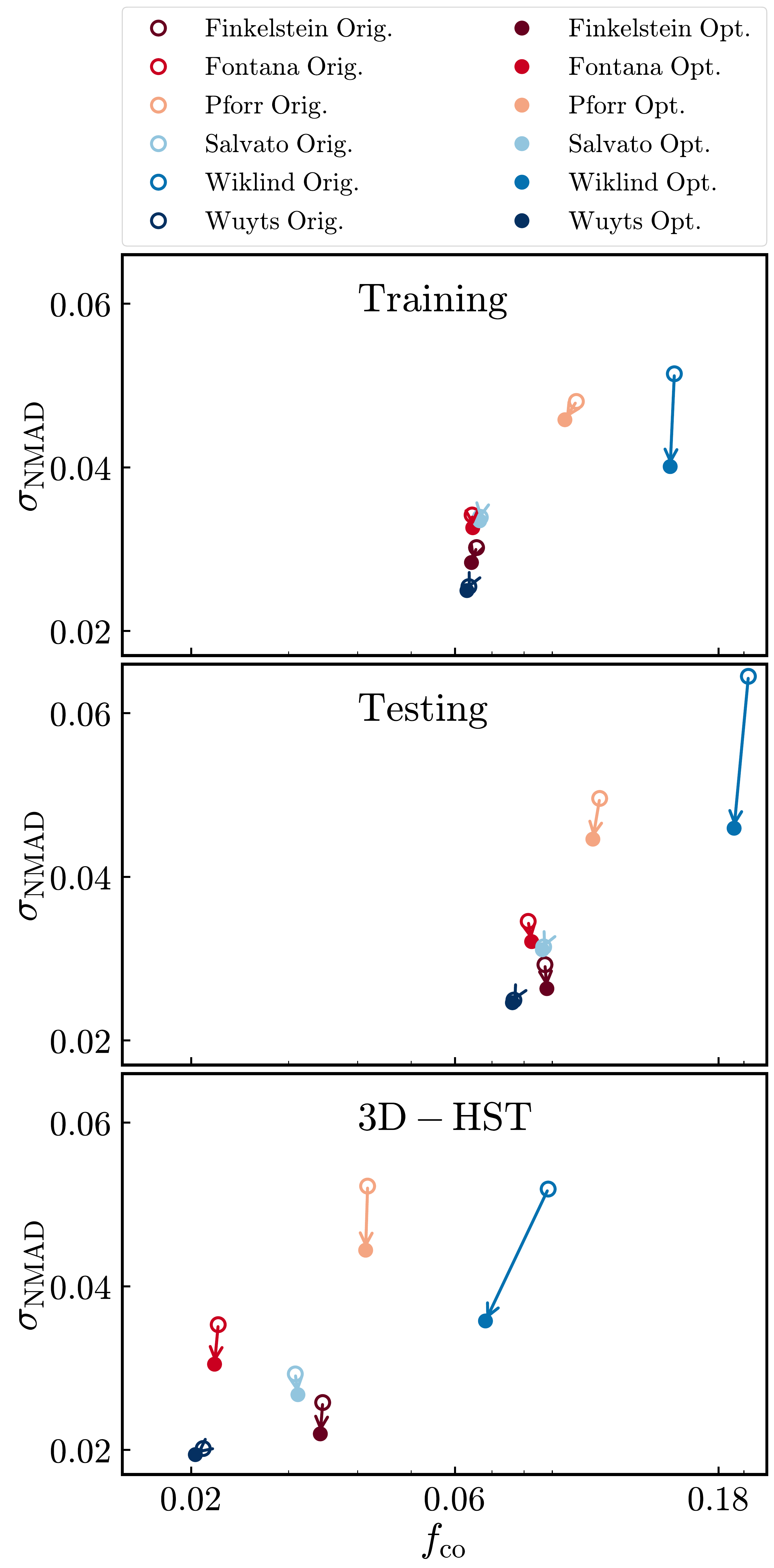}
  \end{center}
  \caption{The normalized median absolute deviation between photometric and spectroscopic redshifts, $\sigma_{\rm NMAD}$, as a function of the catastrophic outlier fraction, $f_{\rm co}$ from each group both before (open circles at arrow bases) and after (filled circles at arrow heads) optimization. The three panels show these statistics evaluated using the training set of spectroscopic redshift (top); the testing set (middle); and 3D-HST grism redshifts (bottom).  Results shown were calculated using the weighted-mean photo-$z$ peak redshifts ($z_{\rm weight}$) from each group.  Results from Wuyts outperform the other groups in all cases, especially when evaluated using the 3D-HST data set.  We caution that the 3D-HST redshift determination incorporates results from the EAZY code used by Finkelstein and Wuyts as part of the grism redshift determination process, so the level of agreement for them may be artificial.}
\label{fig:All6_zw_nmad_fo}
\end{figure}

\autoref{fig:All6_zw_nmad_fo} plots the $\sigma_{\rm NMAD}$ values and catastrophic outlier fractions for the photometric redshifts from each group, calculated using the PDFs both before and after optimization has been applied; results for each spectroscopic sample are shown in separate panels. In general, recalibrating the PDFs significantly improves the scatter between photo-$z$'s and spectroscopic redshifts, while improving or only negligibly degrading the outlier fraction.  The Wiklind analysis produced significantly more objects without any point estimates, which were treated as catastrophic outliers in this case.  Consequently, the high values of $f_{\rm co}$ for that analysis should be considered upper limits on what the performance would be for the full set.

For the objects belonging to the training or testing sets, the outlier fraction is similar from all of the different groups (roughly $8\%$), while it ranges from $2\%$ to $5\%$ for objects with 3D-HST grism redshifts.  There are two possible explanations for this discrepancy.  One is that a significant ($\sim 5\%$) fraction of the objects in the training and testing samples have incorrect redshifts assigned to them (potentially due to incorrect matches between the spectroscopic target and CANDELS galaxy).  The other possibility is that the smaller outlier fraction for 3D-HST is artificially low due to the grism redshift fits incorporating photo-$z$ information, though we limited to objects whose spectrum drove the grism-$z$ fit to mitigate this possibility.

Finally, we compute the fraction of objects that are both out-of-PDF and catastrophic outliers (${f_{\rm op \, \& \, co}}$).  These objects are the most problematic because they have both a poorly estimated photo-$z$ and poorly characterized photo-$z$ uncertainty.  Generally, around 3\% of objects fit into this category across codes (excluding the Wiklind results that should be viewed as lower limits due to their high missing PDF rate driving a low out-of-PDF fraction).

\section{Methods for Combining Probability Density Functions from Multiple Codes}
\label{sec:combining_pdfs}

Previous works have found that combining results from multiple photometric redshift codes can yield superior performance over individual codes when testing with spec-$z$'s. \citet{Dahlen_2013}, \citet{Castellano_2016}, and \citet{Merlin_2021} found that the median of the photo-$z$'s from all the groups contributing measurements yielded a lower scatter than any single code. Since we are not only interested in point values (such as $z_{\rm weight}$ or $z_{\rm peak}$) but also desire accurate photo-$z$ PDFs (or, equivalently, accurate error estimates), we require methods that can combine multiple PDFs. In this paper, we utilize two different methods for combining PDFs: the HB approach first presented in \citet{Dahlen_2013}, which produces an entirely new {PDF} constructed based on the input PDFs (see also \citealt{duncan2018a}, \citealt{duncan2018b}, and \citealt{hatfield2022}), and the minimum \fdiv\ method (mFD), which selects one of the input PDFs as being the most representative for a given object (much as the median of a set of numbers is the element of that set which approximates its central value).


\subsection{Hierarchical Bayesian Combination of Photometric Redshift PDFs}
\label{sec:hierarchical_bayesian}

The HB approach introduced by \citet{Dahlen_2013} combines PDF results from multiple codes based upon the assumptions that estimated PDFs for a given object are not entirely statistically independent (since they are based upon the same photometry) and that some may be inaccurate.  This basic framework has a number of antecedents in the literature \citep{Press_1997, Newman_1999,Lang_Hogg_2012}.

If we assume that the PDF from a given group is either informative about the true redshift of an object (``good'') or uninformative (``bad''), the posterior probability for the redshift of an object from the $i$th code, $P_i(z)$, can be described as:
\begin{equation}
 P_i(z) = P_i(z | {\rm bad}) P({\rm bad}) + P_i(z | {\rm good}) [1- P({\rm bad})],
\end{equation}
where $P_i(z | {\rm bad})$ is the PDF resulting from an uninformative measurement (which we take here to be a uniform probability distribution from $z=0$ to $z=10$); $P({\rm bad})$ is the probability a randomly selected object has an uninformative measurement; and $P_i(z | {\rm good})$ is the redshift PDF for a given object predicted by the $i$th photo-$z$ code. We assume that $P({\rm bad})$ is the same for all objects, and hence it will be equal to the fraction of all PDF measurements that are uninformative, which we label as $f_{\rm bad}$.  Therefore, for a given value of $f_{\rm bad}$, the posterior PDF is given by:
\begin{equation}
 P_i(z, f_{\rm bad}) = P_i(z | {\rm bad}) f_{\rm bad} + P_i(z | {\rm good}) (1- f_{\rm bad}).
\end{equation}
For a given object, the Bayesian posterior combining the information of each (participant) PDF is given by:
\begin{align}
  P(z, f_{\text{bad}})
  = \prod_i{ P_i(z, f_{\text{bad}})^{1/\alpha} },
\end{align}
\noindent where we introduce a parameter $\alpha$ that provides a correction for the covariance between the different PDFs.  If the PDFs are all statistically independent, the probabilities from each one would multiply, so $\alpha = 1$.  In contrast, if they were completely covariant, we would need $\alpha$ to match the number of PDFs being combined, $n_p$, so that after multiplication and exponentiation $P(z, f_{\text{bad}})$ would match the result from a single PDF.

Based on tests with the CANDELS data, we find that the optimal value of $\alpha$ is best described by the equation:
\begin{align}
  \alpha = 1 + (n_p - 1) \times 1.1/4 ,
\end{align}
\noindent which yields $\alpha = 1$ for $n_p = 1$ and $\alpha = 2.1$ for $n_p = 5$ (matching the results from \citealt{Dahlen_2013} when $n_p = 5$). This implies that the covariance between results from different participating groups is nonnegligible (since $\alpha > 1$ is optimal in general) but also far from complete (since $\alpha < n_p$).

We then marginalize over the fraction of bad measurements to obtain a PDF for redshift alone:
\begin{align}
  P(z)
  = \int_0^1{ P(z, f_{\text{bad}}) \, d f_{\text{bad}} },
\end{align}
\noindent assuming a flat prior distribution for $f_{\text{bad}}$ over $[0, 1]$.

The result of this procedure is a PDF that matches the PDFs from each code when they agree but will include extra probability at the redshifts predicted by each individual PDF when they disagree.  Hence, HB photo-$z$ constraints are appropriately degraded when the PDFs disagree, in a manner that is agnostic about which PDFs are most accurate.


\subsection{Minimum \fdiv\ Approach}
\label{sec:fdiv}

\begin{figure}
  \centering
    \includegraphics[width = 0.45\textwidth]{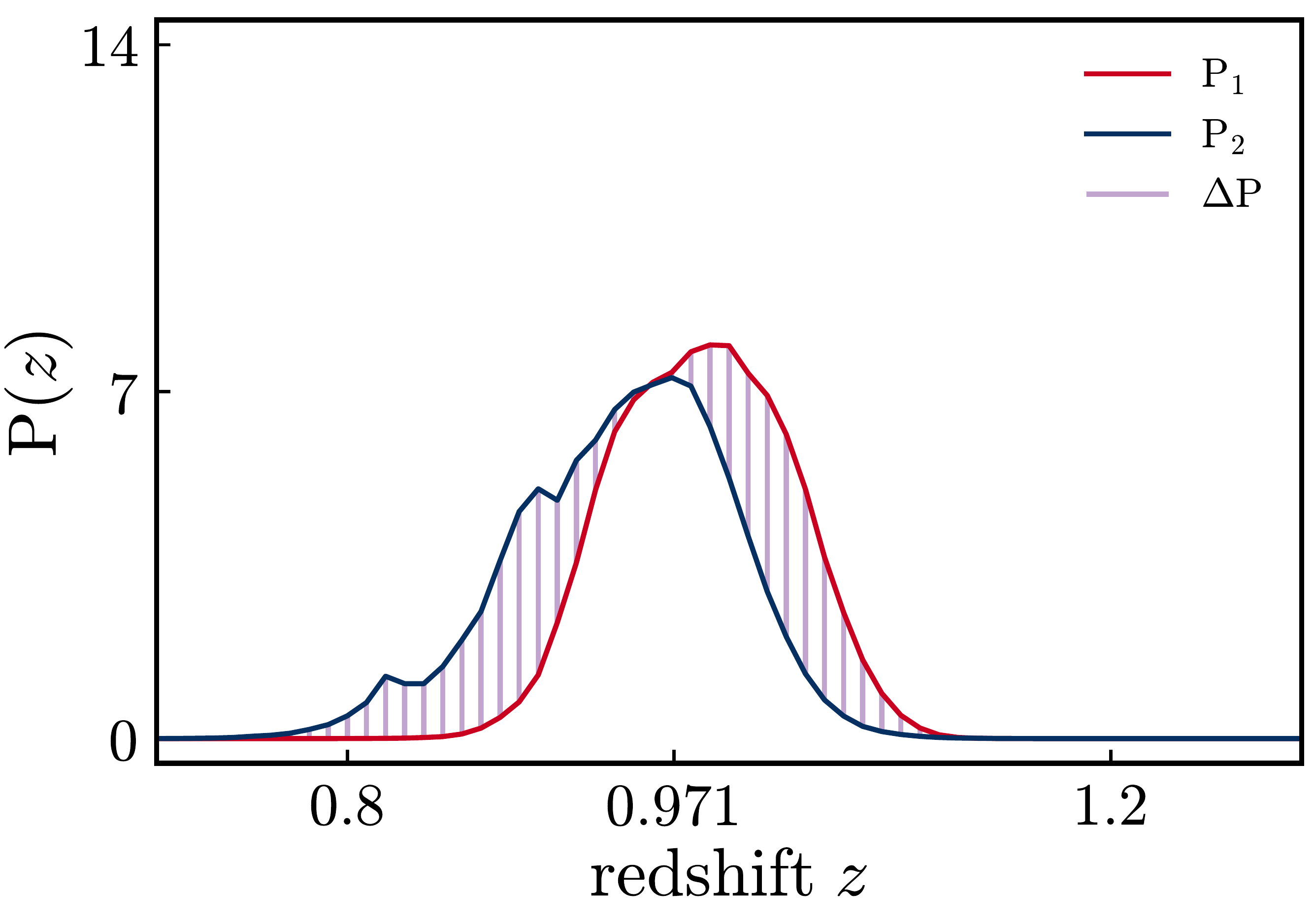}
    \caption{An example illustration of the calculation of the \fdiv\ distance measures between two \photoz\ PDFs, $P_1 (z)$ and $P_2 (z)$, represented by the red and blue curves. The vertical purple hatching indicates the vertical difference between them at each point where they are defined, $\Delta P (z_i) = P_1 (z_i) - P_2 (z_i)$.  We compute the \fdiv\ between the two curves in two ways: with an $\ell^1$ distance (i.e., the sum of the absolute values of the differences) and an $\ell^2$ distance (i.e., the square root of the sum of the squared differences).}
\label{fig:fd}
\end{figure}

In addition to the hierarchical Bayesian method of combining PDFs, we also test a new approach of choosing the single PDF that minimizes a measure of the total distance to the other PDFs; for the distance calculations, we consider two measures that fall in the general class of \fdiv\ measures.  In this paper, we consider \fdivs\ calculated either by summing the absolute values of the differences between two PDFs (an $\ell^1$ distance) or by taking the square root of the sum of the squares of the differences (an $\ell^2$ distance) as measured at many equally spaced $z$-values:

\begin{subequations}
  \begin{align}
    FD_{\text{abs}} & = \sum_i{ |P_1(z_i) - P_2(z_i)| } \quad \text{and}\\
    FD_{\text{sqr}} & = \left\{ \sum_i{ \left[ P_1(z_i) - P_2(z_i) \right]^2 } \right\}^{1/2},
  \end{align}
\end{subequations}
where $z_i$ indicates the $i$th redshift point at which the distances are evaluated, and we use $\mathrm{FD}_{\text{abs}}$ and $\mathrm{FD}_{\text{sqr}}$ to label the $\ell^1$ and $\ell^2$ versions of the \fdiv, respectively.  For two given curves $P_1(z)$ and $P_2(z)$ that are defined at the same set of discrete points, the \fdiv\ is found by combining the difference between the values of the curves at each ordinate point, as illustrated in \autoref{fig:fd}.  In these calculations, we perform a discrete sum over redshift points, which should closely approximate the integral calculated using a continuous PDF.

In our case, six groups have provided photo-$z$ PDFs for each object in the CANDELS photometric catalogs.  We then calculate the sum of the \fdiv\ between each PDF and each of the other five results, yielding the total distance of that curve from the rest (in the case where the $\ell^2$ distance is used, we sum in quadrature). We repeat this procedure for each PDF for a given object, and identify the one for which the total distance to the other PDFs is lowest; that is, the one with the mFD from the other PDFs.  This curve will be the most similar to all the rest, and therefore provides a reasonable way to summarize the ensemble of PDFs.  This construction is analogous to the use of the median value of an array as a summary statistic; the median minimizes the sum of the absolute values of the deviations from all points in an array.

The $\ell^1$ distance ($\mathrm{FD}_\mathrm{abs}$) will be less affected by the largest excursions between two curves than the $\ell^2$ distance ($\mathrm{FD}_\mathrm{sqr}$) because the latter metric squares the differences before summation, providing more weight to larger deviations. As a result, the minimum \fdiv\ curve selected using an $\ell^1$ distance metric, which we label as ``mFDa'' as it is based upon the sum of absolute values, is less sensitive to outlier curves than when we use an $\ell^2$ metric, which we label as ``mFDs'' as it is based upon the sum of squared differences. The situation is analogous to the reason why the median statistic is more robust to outliers in a data array than the mean.


\subsection{Comparing Combination Methods}
\label{sec:comparing_methods}

\begin{figure}
  \centering
    \includegraphics[width = 0.45\textwidth]{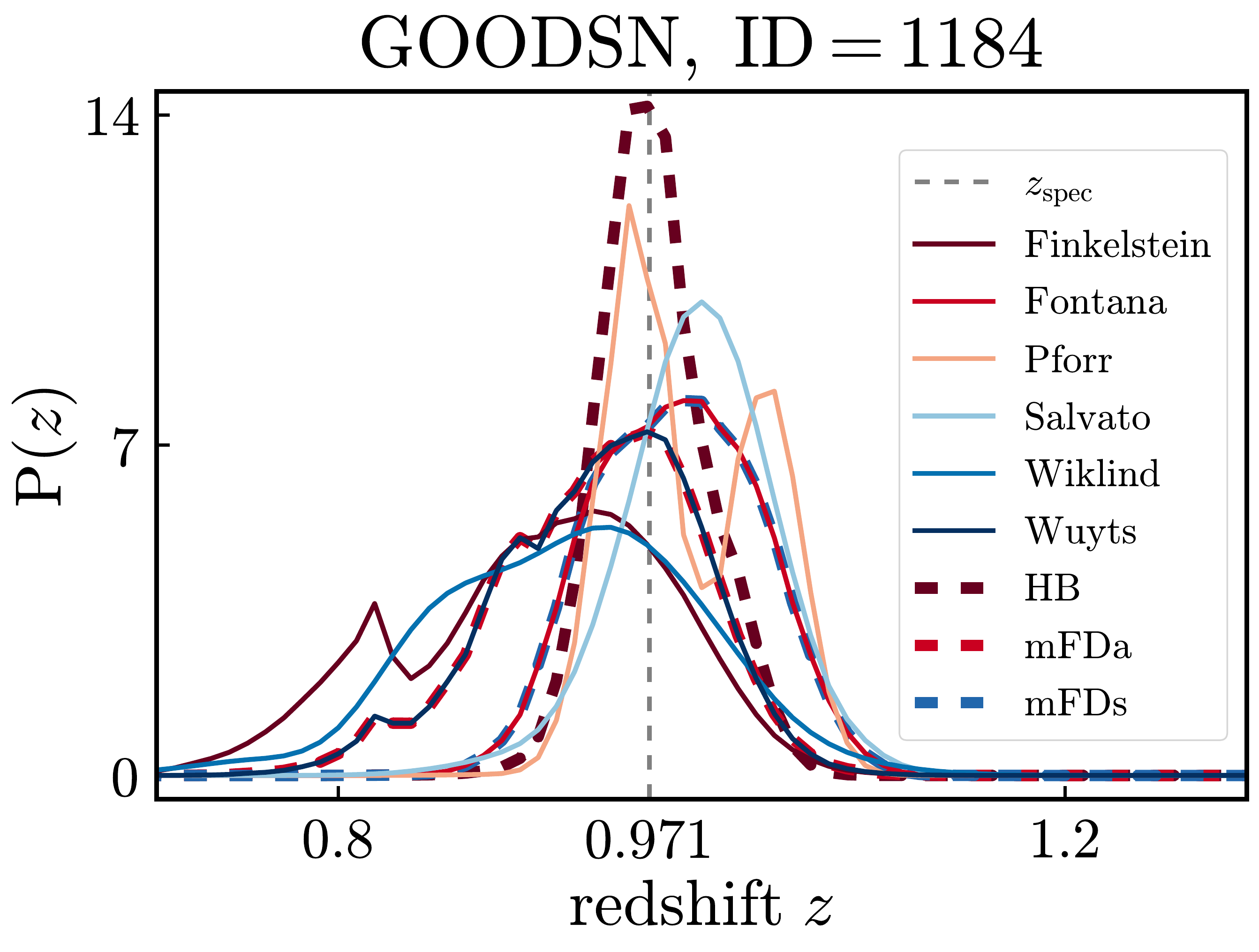}
    \caption{Optimized PDFs from all six participant photo-$z$ groups, as well as the results from three PDF combination methods, for a CANDELS galaxy in the GOODS-North field (ID = 1184).  The HB (dark brown dashed curve) method produces a new PDF by combining information from each input probability distribution and accounting for the possibility that some are inaccurate.  The minimum \fdiv\ method identifies the PDF that has the smallest total \fdiv\ from the other curves. When the sum of absolute values of differences is used as a distance metric, the minimum \fdiv\ curve corresponds to the Salvato PDF for this object (mFDa, red dashed curve), whereas when the square root of the sum of the squares of separations is used to compute distances the Fontana curve is selected (mFDs, blue dashed curve).  The vertical gray dashed line indicates the galaxy's spectroscopic redshift, $z_{\rm spec} = 0.971$, which is consistent with all PDFs shown.}
\label{fig:All_pdfs}
\end{figure}

\autoref{fig:All_pdfs} illustrates the results of the hierarchical Bayesian and minimum \fdiv\ approaches for a galaxy in the GOODS-North field. In this figure, we present the {PDFs} from the six different groups, as well as the results of each combination method.  Whereas the hierarchical Bayesian method produces a completely new photo-$z$ {PDF} that is distinct from all of the input curves, the minimum \fdiv\ method selects one of the original PDFs as being the best representative for a given object.  For this object, the mFDa and mFDs algorithms select different PDFs, but it is more common that they agree with each other than not.

For each of the three methods of combining photometric redshift PDFs applied here, we also explore how the quality of results changes when only a subset of the groups' PDFs is used as inputs.  We note that some groups' PDFs performed significantly better than others when evaluated using the $\mathrm{\ell^2}$-norm, $f_{\rm op}$, $\sigma_{\rm NMAD}$, or $f_{\rm co}$ for both the testing set of spectroscopic redshifts and the 3D-HST grism redshifts. Given the possibility that some groups' PDFs are more useful for computing combined distributions than others, we have computed PDFs using the HB and mFD methods using only subsets of groups that yielded the lowest $\sigma_{\rm NMAD}$ and $f_{\rm co}$ values (using the same set of groups' \photozs\ for all objects). We use the label ``6'' for when all six groups are used, the label ``5'' for when the five best groups are used, the label ``4'' for when the four best groups are used, and finally the label ``3'' for when only the three best groups are used; hence, results labeled HB4 correspond to a hierarchical Bayesian combination of PDFs from the four best-performing codes.

\begin{figure}
  \begin{center}
      \includegraphics[width=0.47\textwidth]{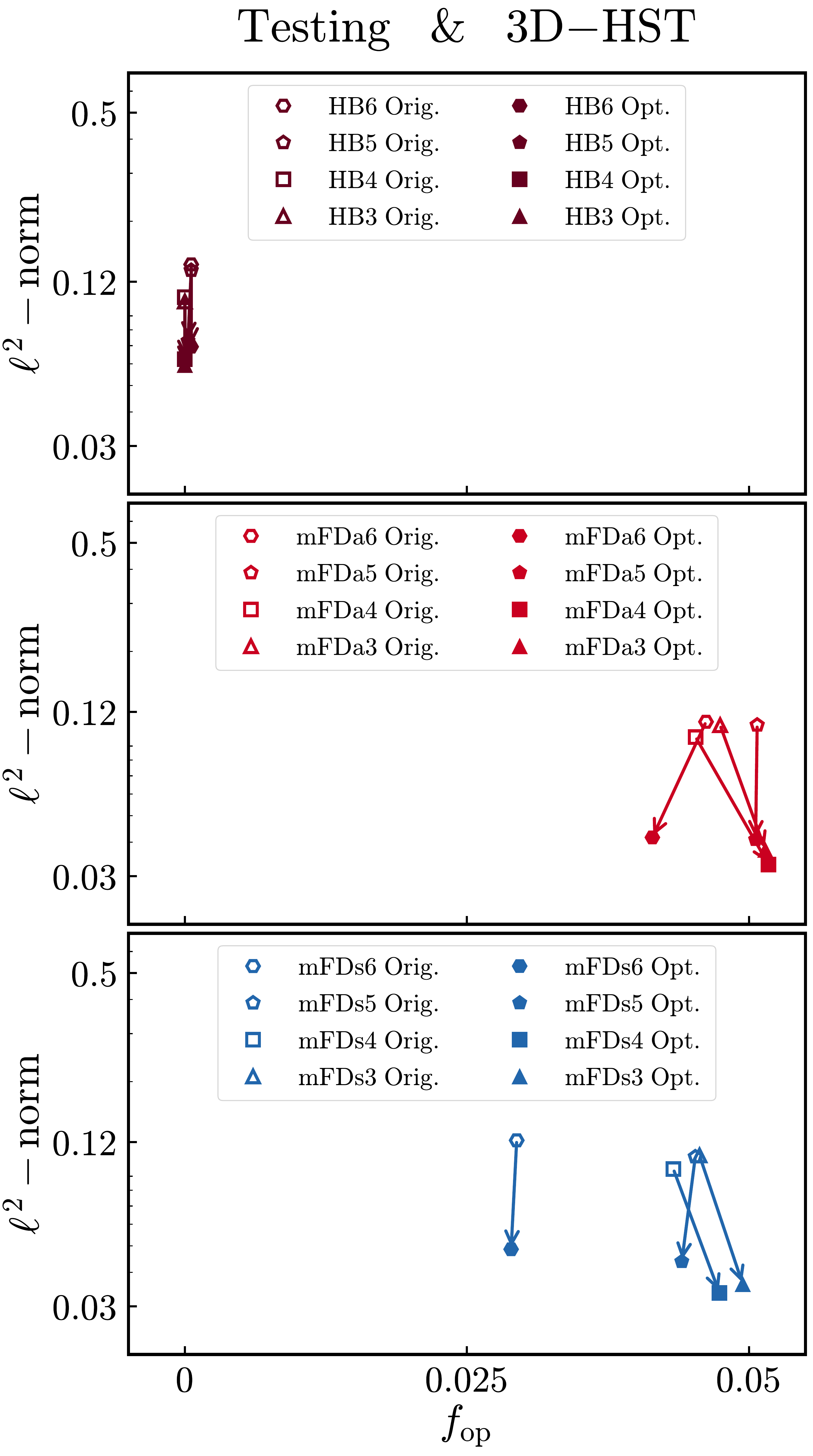}
  \end{center}
  \caption{Average values of the normalized $\mathrm{\ell^2}$-norm and out-of-PDF fraction $f_{\rm op}$ for each PDF combination method considered here, combining results from the testing and 3D-HST redshifts across all CANDELS fields. We show values both without (open symbols at arrow bases) and with (filled symbols at arrow heads) optimization of the PDFs from each group in advance of combination.  The panels differ in the method of combination considered; here ``HB'' indicates hierarchical Bayesian combination (top), ``mFDa'' indicates the minimum \fdiv\ curve selected using an $\ell^1$ (sum of absolute values) metric (middle); and ``mFDs'' indicates the minimum \fdiv\ curve selected using an $\ell^2$ (sum of squares) metric (bottom).  The number following the combination type indicates the number of PDFs combined; e.g., in the HB3 case, we combine the PDFs from the best three groups (ranked according to their performance at point value metrics).  Within each type of combination, the lowest values for the $\mathrm{\ell^2}$-norm are obtained using the optimized PDFs for $\mathrm{HB3}$, $\mathrm{mFDa4}$, and $\mathrm{mFDs4}$, respectively.}
\label{fig:All_comb_l2norm_fop}
\end{figure}

In \autoref{fig:All_comb_l2norm_fop}, we plot results for the $\mathrm{\ell^2}$-norm and $f_{\rm op}$ for the HB, mFDa, and mFDs methods with the 3, 4, 5, and 6 best-performing codes both before and after PDF optimization, averaging the results for these statistics from the testing set of spectroscopic redshifts and the 3D-HST set of grism redshifts.  For the HB combination, HB3 produces the lowest $\mathrm{\ell^2}$-norm.  Note that $f_{\rm op}$ is zero for the HB combined PDFs by construction, as the uniform probability distribution used to model uninformative measurements yields a small amount of probability at all redshifts in the final HB combination.  For the minimum \fdiv\ combinations, mFDa4 and mFDs4 have the lowest $\mathrm{\ell^2}$-norm values and are our preferred combinations even though mFDa6 and mFDs6 have lower $f_{\rm op}$ values.  For all three combination methods, combinations that did not use all six codes produced the lowest $\mathrm{\ell^2}$-norm values.

\begin{figure}
  \begin{center}
      \includegraphics[width=0.47\textwidth]{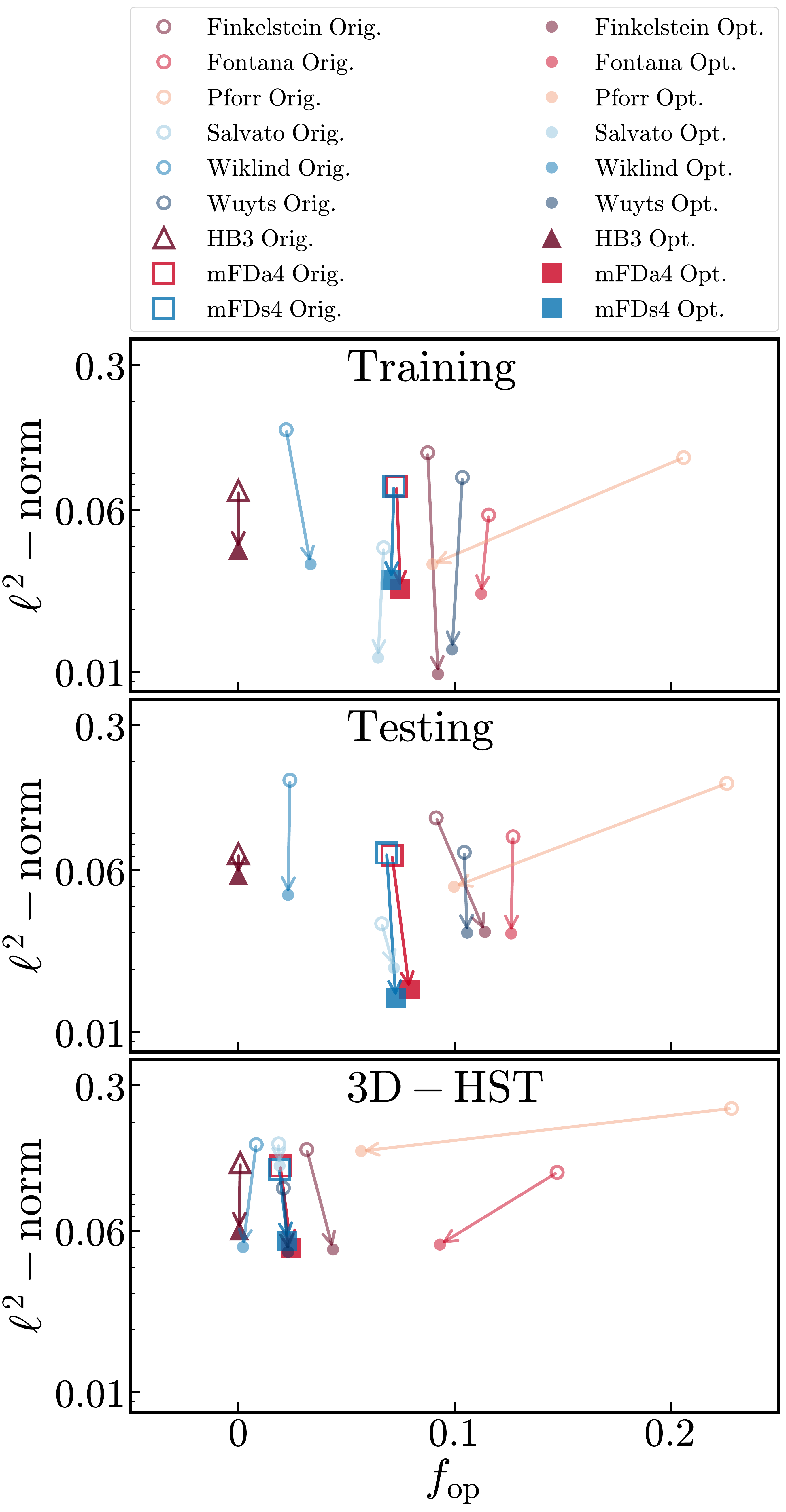}
  \end{center}
  \caption{Normalized $\mathrm{\ell^2}$-norm and $f_{\rm op}$ for the lowest-norm PDF combinations of each type---$\mathrm{HB3}$, $\mathrm{mFDa4}$, and $\mathrm{mFDs4}$---with results from individual groups shown for comparison (same as Figure \ref{fig:All6_l2norm_fop}). We show results for the training and testing sets of spectroscopic redshifts and the 3D-HST grism redshifts separately in each panel but combine objects from all CANDELS fields. $\mathrm{mFDa4}$ and $\mathrm{mFDs4}$ produce similarly excellent results for all three data sets, give lower $\mathrm{\ell^2}$-norm values than HB3 in every case, and outperform the individual groups on the testing set.  We prefer the $\mathrm{mFDa4}$ combination because it gives the lowest $\mathrm{\ell^2}$-norm value averaged across all three data sets, indicating that the minimum \fdiv\ curve constructed from the four highest-quality PDFs comes closest to meeting the statistical definition of a probability density function for the actual redshift of a galaxy.}
\label{fig:Best_comb_l2norm_fop}
\end{figure}

\autoref{fig:Best_comb_l2norm_fop} takes the best-performing subset for each combination method (i.e., HB3, mFDa4, and mFDs4) and compares them for the training set, spectroscopic testing set, and 3D-HST testing set.  It also shows the results from the individual groups for comparison.  The mFDa4 and mFDs4 combinations perform similarly well and are clearly better than HB3 (and the individual groups) on the spectroscopic testing set.  We prefer the mFDa4 combination because it has the lowest $\mathrm{\ell^2}$-norm averaged across all three data sets.  The mFDa4 combination has slightly higher $f_{\rm op}$ values than the mFDs4 combination for all three data sets.  We conclude that the minimum \fdiv\ (with the absolute value as the distance metric) PDF selected from the four highest-quality results comes closest to meeting the statistical definition of a PDF for the actual redshift of a galaxy, and therefore is the preferred combination technique for CANDELS when a consensus PDF is desired.

\begin{figure}
  \begin{center}
  	\includegraphics[width=0.47\textwidth]{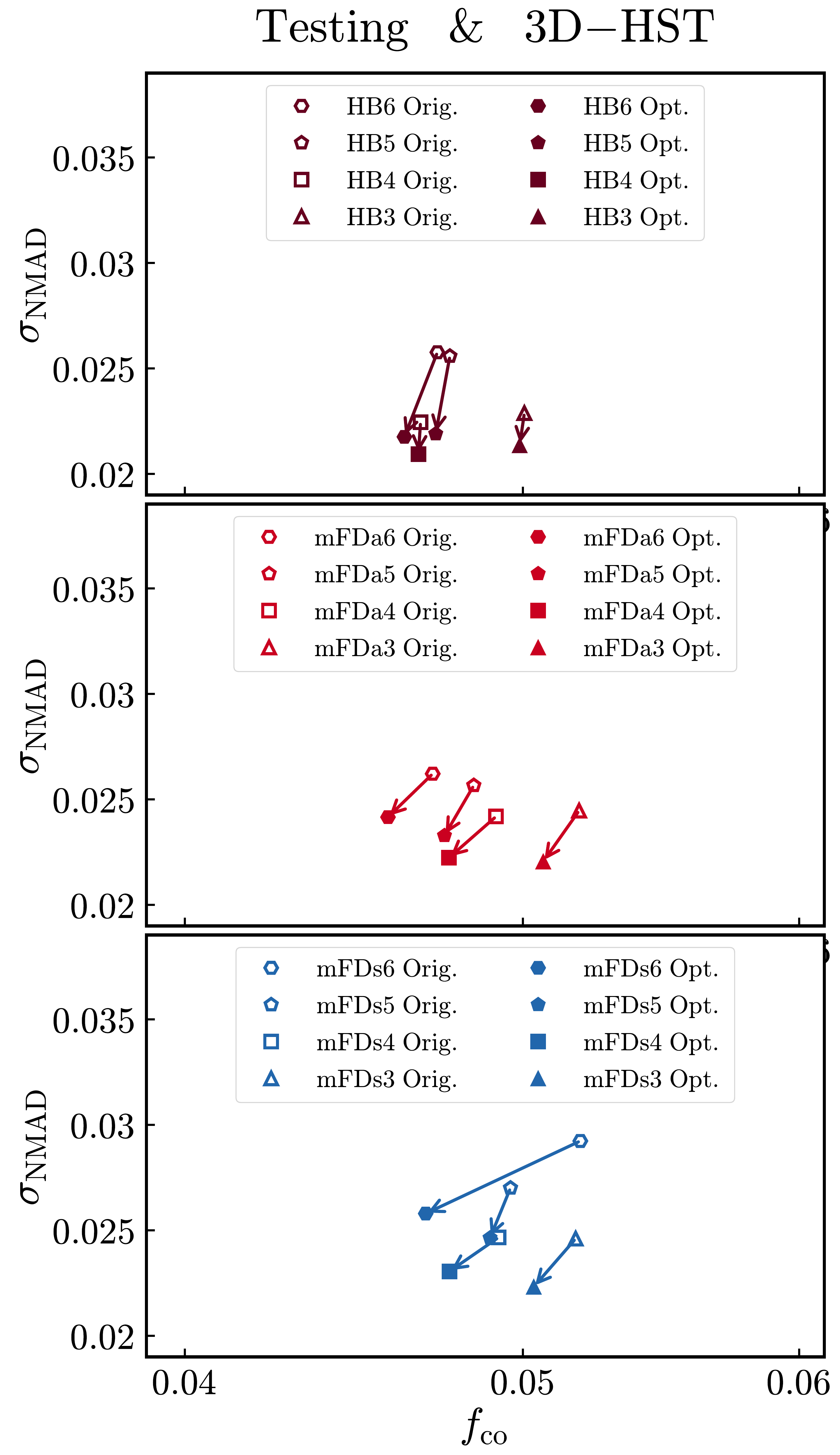}
  \end{center}
  \caption{Comparison of \photoz\ point statistics ($z_{\rm weight}$) from different combination methods using the NMAD ($\sigma_{\rm NMAD}$) and outlier fraction ($f_{\rm co}$) statistics both before (open symbols at arrow bases) and after (filled symbols at arrow heads) optimization.  The results shown are for the testing and 3D-HST redshifts across all five CANDELS fields.  The HB4, mFDa4, and mFDs3 combination methods yielded the lowest NMAD values, and hence the most accurate point estimates when evaluated with these samples.
 \vspace{2cm}
  }
\label{fig:All_comb_zw_nmad_fo}
\end{figure}

\begin{figure}
  \begin{center}
      \includegraphics[width=0.47\textwidth]{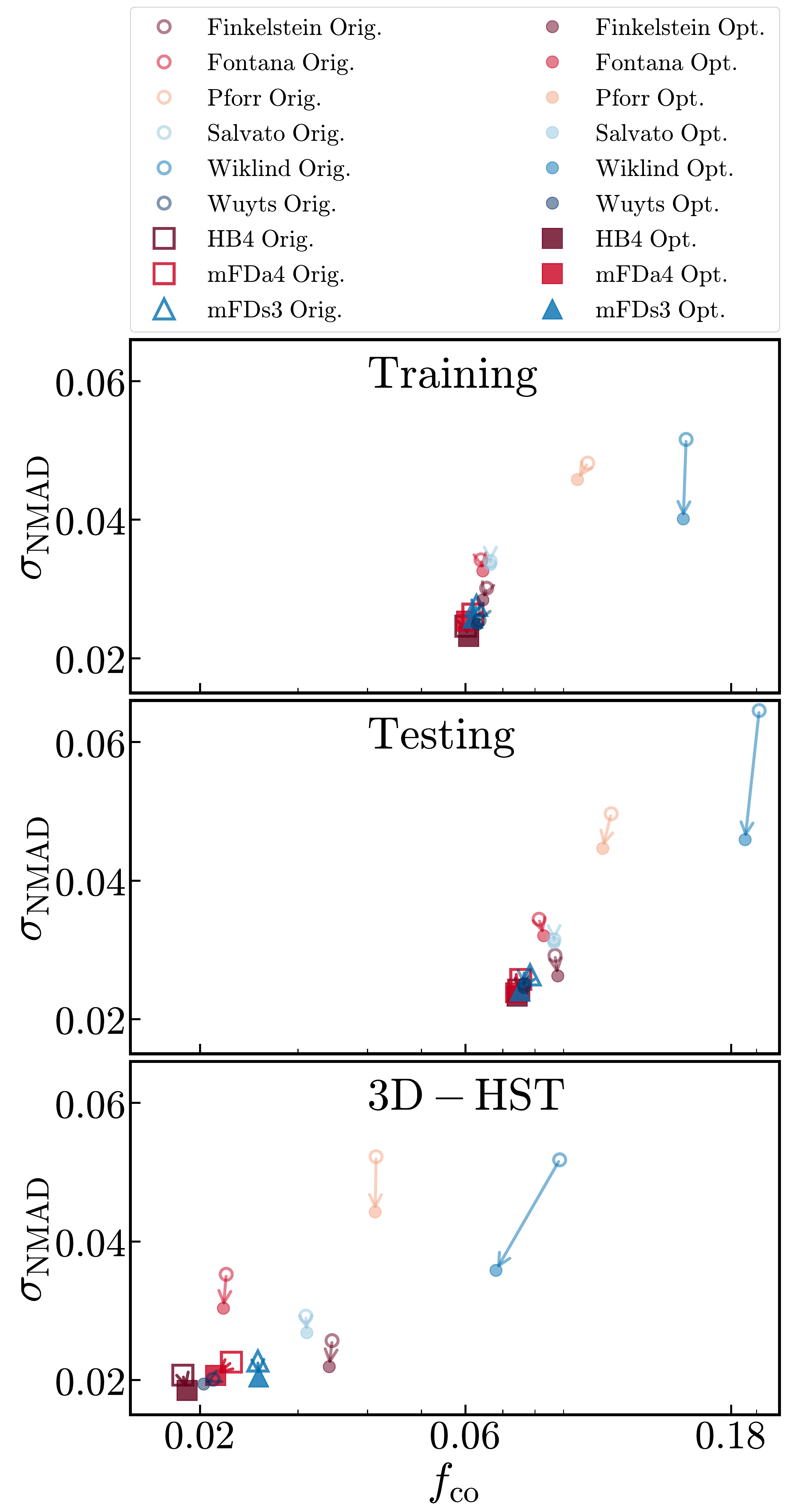}
  \end{center}
  \caption{Comparison of the best \photoz\ point statistic results from each PDF combination method---$\mathrm{HB4}$, $\mathrm{mFDa4}$, and $\mathrm{mFDa3}$---using $\sigma_{\rm NMAD}$ and $f_{\rm co}$ both before (open symbols at arrow bases) and after (filled symbols at arrow heads) optimization.  We also show results from Figure \ref{fig:All6_zw_nmad_fo} for the individual groups. The three panels show these statistics evaluated using the training set of spectroscopic redshift (top); the testing set (middle); and 3D-HST grism redshifts (bottom).  In every case, the hierarchical Bayesian combination of the best four PDFs, $\mathrm{HB4}$, gives the lowest $\sigma_{\rm NMAD}$ values for each set and outlier fractions comparable to or lower than other methods, making this the combination that provides the most accurate point estimates.  That being said, the improvement over the Wuyts results is slight.}
\label{fig:Best_comb_zw_nmad_fo}
\end{figure}

Although mFDa4 is our preferred combination method for PDFs, there are cases when the accuracy of point measures of the redshift is more desirable than the accuracy of the PDF. We therefore evaluate the accuracy of $z_{\rm weight}$ estimates from the same set of combination methods and subsets of the input PDFs considered above, using the $\sigma_{\rm NMAD}$ and $f_{\rm co}$ statistics to evaluate them. We present the averaged results from the testing set of spectroscopic redshifts and the 3D-HST grism redshifts in \autoref{fig:All_comb_zw_nmad_fo}, combining objects from all CANDELS fields. Among the hierarchical Bayesian combinations considered, HB4 has the lowest $\sigma_{\rm NMAD}$ value; mFDa4 and mFDs3 yield the lowest scatters among the minimum \fdiv\ combinations. We compare the results for  these three best cases against each other and the individual groups using each set of redshifts separately in \autoref{fig:Best_comb_zw_nmad_fo}. In every case, the hierarchical Bayesian combination of the best four PDFs, HB4, yields the lowest $\sigma_{\rm NMAD}$; it is therefore the combination method that provides the most accurate point values of photo-$z$'s, with $\sigma_z \sim 0.02 (1+z)$ or better for all three spectroscopic samples.  In every case it outperforms the best individual group results, though the improvement over the best individual group (Wuyts) is small.

Table \ref{table:test} summarizes the performance of the optimized photo-$z$ PDFs from each group and the PDFs produced by the best combination methods of each type (HB, minimum \fdiv\ computed with an $\ell^1$ metric [mFDa], and minimum \fdiv\ computed with an $\ell^2$ metric [mFDs]) for each of the three spectroscopic data sets (training, testing, or 3D-HST) across the five CANDELS fields.  The method yielding the best performance for a given statistic is highlighted in blue, and the combination method with the best performance on average is indicated in red.  The quantities used to evaluate the quality of photo-$z$ PDFs are the normalized $\ell^2$-norm between the $Q$--$Q$ curve for a given set of PDFs and the unity line ($\ell^2$-norm) and the fraction of spectroscopic redshifts that lie outside the photo-$z$ PDF for their corresponding object (${f_{\rm op}}$; which are zero by construction for the hierarchical Bayesian combination method).  To test the performance of each group and combination method for the weighted-mean peak redshift, $z_{\rm weight}$, we use the normalized median absolute deviation ($\mathrm{\sigma_{NMAD}}$) and the $\Delta z / (1+z) > 0.15$ outlier fraction ${f_{\rm co}}$. Finally, we list the fraction of objects for which a PDF file was not provided by a given group as ${f_{\rm missing}}$ (fraction of missing files). Combinations of different numbers of PDFs (either the best three or the best four) yielded the best results for PDF statistics (where HB3, mFDa4, and mFDs4 proved superior), which differed from the best ones for point statistics (where HB4, mFDa4, and mFDs3 were preferred).

%
\begin{deluxetable*}{cc  cccccc  ccc}
\label{table:test}
\tablecaption{Table of quantities used to assess the quality of the optimized photometric redshift PDFs and their point statistics.  All values multiplied by 100. The lowest average result among the combination methods is shown in red, and the lowest value in a given row is shown in blue (unless that entry is also the lowest average result for the combination methods).  We note that the Wiklind analysis has significantly higher fractions of objects with missing photo-$z$ PDFs and point estimates than the other codes, so its summary statistics are sensitive to the treatment of these objects.  If objects with missing PDFs are counted as out-of-pdf, then Wiklind would no longer produce the lowest $f_{\rm op}$ and \fopco\ for all three data sets.  In fact, it would have the highest $f_{\rm op}$ for the training and testing sets (and an unremarkable $f_{\rm op}$ for the 3D-HST set).  On the other hand, if objects with a missing point estimate are not counted as catastrophic outliers, then Wiklind would produce nearly the best $f_{\rm co}$ for the training and testing sets (and an unremarkable $f_{\rm co}$ for the 3D-HST set).
}

%
\tablehead{
\colhead{Quantity} &
\colhead{Set} &
\colhead{Finkelstein} &
\colhead{Fontana} &
\colhead{Pforr} &
\colhead{Salvato} &
\colhead{Wiklind} &
\colhead{Wuyts} &
\colhead{HB3} &
\colhead{mFDa4} &
\colhead{mFDs4}
}
%
\startdata
%
\multirow{5}{*}{$\mathbf{\mathrm{\ell^2-norm}}$} &
$\mathrm{Training}$ &
{\color{blue} $\mathbf{0.97}$}  &
$2.38$  &
$3.27$  &
$1.16$  &
$3.29$  &
$1.28$  &
$3.95$  &
$2.52$  &
$2.80$
\\
&
$\mathrm{Testing}$ &
$3.00$  &
$2.97$  &
$4.96$  &
$2.06$  &
$4.52$  &
$3.00$  &
$5.81$  &
$1.61$  &
{\color{blue} $\mathbf{1.46}$}
\\
&
$\mathrm{3D{-}HST}$ &
$4.85$  &
$5.17$  &
$14.4$  &
$12.2$  &
$5.01$  &
{\color{blue} $\mathbf{4.75}$}  &
$6.08$  &
$5.01$  &
$5.27$
\\[0.05cm]
\cmidrule(r){2-11}
%
&
$\mathrm{Average}$ &
{\color{blue} $\mathbf{2.94}$}  &
$3.51$  &
$7.55$  &
$5.13$  &
$4.28$  &
$3.01$  &
$5.28$  &
{\color{red} $\mathbf{3.05}$}  &
$3.18$
\\[0.05cm]
%
\hline
%
\multirow{5}{*}{${f_{\rm op}}$} &
$\mathrm{Training}$ &
$9.25$  &
$11.3$  &
$8.97$  &
$6.54$  &
{\color{blue} $\mathbf{3.30}$}  &
$9.95$  &
$0$  &
$7.43$  &
$7.06$
\\
&
$\mathrm{Testing}$ &
$11.4$  &
$12.7$  &
$10.1$  &
$7.25$  &
{\color{blue} $\mathbf{2.33}$}  &
$10.7$  &
$0$  &
$7.95$  &
$7.22$
\\
&
$\mathrm{3D{-}HST}$ &
$4.41$  &
$9.31$  &
$5.75$  &
$1.94$  &
{\color{blue} $\mathbf{0.18}$}  &
$2.29$  &
$0$  &
$2.40$  &
$2.26$
\\[0.05cm]
\cmidrule(r){2-11}
%
&
$\mathrm{Average}$ &
$8.36$  &
$11.1$  &
$8.26$  &
$5.24$  &
{\color{blue} $\mathbf{1.94}$}  &
$7.64$  &
$0$  &
$5.92$  &
{\color{red} $\mathbf{5.51}$}
\\[0.05cm]
%
%
\hline
%
\multirow{5}{*}{$\mathrm{\sigma_{NMAD}}$} &
$\mathrm{Training}$ &
$2.84$  &
$3.26$  &
$4.58$  &
$3.35$  &
$4.01$  &
$2.41$  &
{\color{blue} $\mathbf{2.32}$}  &
$2.53$  &
$2.58$
\\
&
$\mathrm{Testing}$ &
$2.63$  &
$3.21$  &
$4.46$  &
$3.11$  &
$4.59$  &
$2.46$  &
{\color{blue} $\mathbf{2.33}$}  &
$2.38$  &
$2.41$
\\
&
$\mathrm{3D{-}HST}$ &
$2.20$  &
$3.05$  &
$4.44$  &
$2.68$  &
$3.58$  &
$1.94$  &
{\color{blue} $\mathbf{1.85}$}  &
$2.07$  &
$2.05$
\\[0.05cm]
\cmidrule(r){2-11}
%
&
$\mathrm{Average}$ &
$2.56$  &
$3.17$  &
$4.49$  &
$3.04$  &
$4.06$  &
$2.27$  &
{\color{red} $\mathbf{2.17}$}  &
$2.33$  &
$2.35$
\\[0.05cm]
%
\hline
%
\multirow{5}{*}{${f_{\rm co}}$} &
$\mathrm{Training}$ &
$6.43$  &
$6.46$  &
$9.49$  &
$6.65$  &
$14.7$  &
$6.31$  &
$6.08$  &
{\color{blue} $\mathbf{6.05}$}  &
$6.17$
\\
&
$\mathrm{Testing}$ &
$8.80$  &
$8.26$  &
$10.4$  &
$8.64$  &
$19.2$  &
$7.62$  &
$7.44$  &
{\color{blue} $\mathbf{7.39}$}  &
$7.53$
\\
&
$\mathrm{3D{-}HST}$ &
$3.42$  &
$2.20$  &
$4.14$  &
$3.12$  &
$6.81$  &
$2.03$  &
{\color{blue} $\mathbf{1.90}$}  &
$2.14$  &
$2.54$
\\[0.05cm]
\cmidrule(r){2-11}
%
&
$\mathrm{Average}$ &
$6.22$  &
$5.64$  &
$7.99$  &
$6.13$  &
$13.6$  &
$5.32$  &
{\color{red} $\mathbf{5.14}$}  &
$5.19$  &
$5.41$
\\[0.05cm]
\hline
\multirow{5}{*}{${f_{\rm op \, \& \, co}}$} &
$\mathrm{Training}$ &
$5.14$  &
$4.81$  &
$2.59$  &
$3.57$  &
{\color{blue} $\mathbf{2.37}$}  &
$5.28$  &
$-$ &
$4.83$  &
$-$
\\
&
$\mathrm{Testing}$ &
$6.12$  &
$4.28$  &
$5.13$  &
$3.74$  &
{\color{blue} $\mathbf{1.76}$}  &
$5.81$  &
$-$ &
$5.15$  &
$-$
\\
&
$\mathrm{3D{-}HST}$ &
$1.79$  &
$1.63$  &
$2.22$  &
$0.89$  &
{\color{blue} $\mathbf{0.27}$}  &
$1.01$  &
$-$ &
$0.93$  &
$-$
\\[0.05cm]
\cmidrule(r){2-11}
%
%
&
$\mathrm{Average}$ &
$3.95$  &
$2.96$  &
$3.67$  &
$2.32$  &
{\color{blue} $\mathbf{1.02}$}  &
$3.41$  &
$-$ &
$3.04$  &
$-$
\\[0.05cm]
%
\hline
\multirow{3}{*}{${f_{\rm missing}}$} &
$\mathrm{Training}$ &
$0$  &
$0.11$  &
$0.057$  &
$0.057$  &
$7.51$  &
$0$ &
$-$ &
$-$ &
$-$
\\
&
$\mathrm{Testing}$ &
$0$  &
$0.18$  &
$0.33$  &
$0.076$  &
$9.59$  &
$0$  &
$-$ &
$-$ &
$-$
\\
&
$\mathrm{3D{-}HST}$ &
$0$  &
$0.030$  &
$0.12$  &
$0.030$  &
$4.51$  &
$0$  &
$-$ &
$-$ &
$-$
\enddata
\end{deluxetable*}
%
%


The Finkelstein group's PDFs yielded the smallest average normalized $\mathrm{\ell^2}$-norm, while the Wiklind PDFs had the smallest out-of-PDF fraction $f_{\rm op}$ (caused by objects with missing PDFs not being considered as out-of-PDF).  If objects with missing PDFs are considered out-of-PDF, then Salvato's PDFs would have had the smallest $f_{\rm op}$ of the individual codes.  However, the mFDa4 combination method yielded only marginally larger $\mathrm{\ell^2}$-norm than Finkelstein while having a smaller $f_{\rm op}$; we therefore recommend the use of this combination if the most accurate PDFs are desired.

Whereas individual groups' results yielded the best performance for some PDF quality statistics, the best point statistics were obtained using combinations of multiple photo-$z$ PDFs.  The HB4 (hierarchical Bayesian combination of the four best PDFs) combination yielded both the smallest average NMAD (0.022) and the lowest average outlier rate (5.1\%).  However, it is worth noting that the results from the Wuyts group taken on their own were almost as accurate.

The most dramatic failures of the photo-$z$ codes are objects that are both out-of-PDF and catastrophic outlier point estimates.  Generally, there is some overlap between the out-of-PDF objects and catastrophic outlier objects; but there are also a considerable number of objects that are either one or the other but not both.  We note that calculation of $f_{\rm op}$, $f_{\rm co}$, and \fopco\ for the Wiklind analysis in particular is sensitive to the treatment of missing PDFs/photo-$z$'s due to the significantly higher ${f_{\rm missing}}$ for Wiklind compared to the other codes.


\section{CANDELS Photometric Redshift Catalogs}
\label{sec:catalogs}

\begin{figure*}
  \begin{center}
    \includegraphics[width=0.95\textwidth]{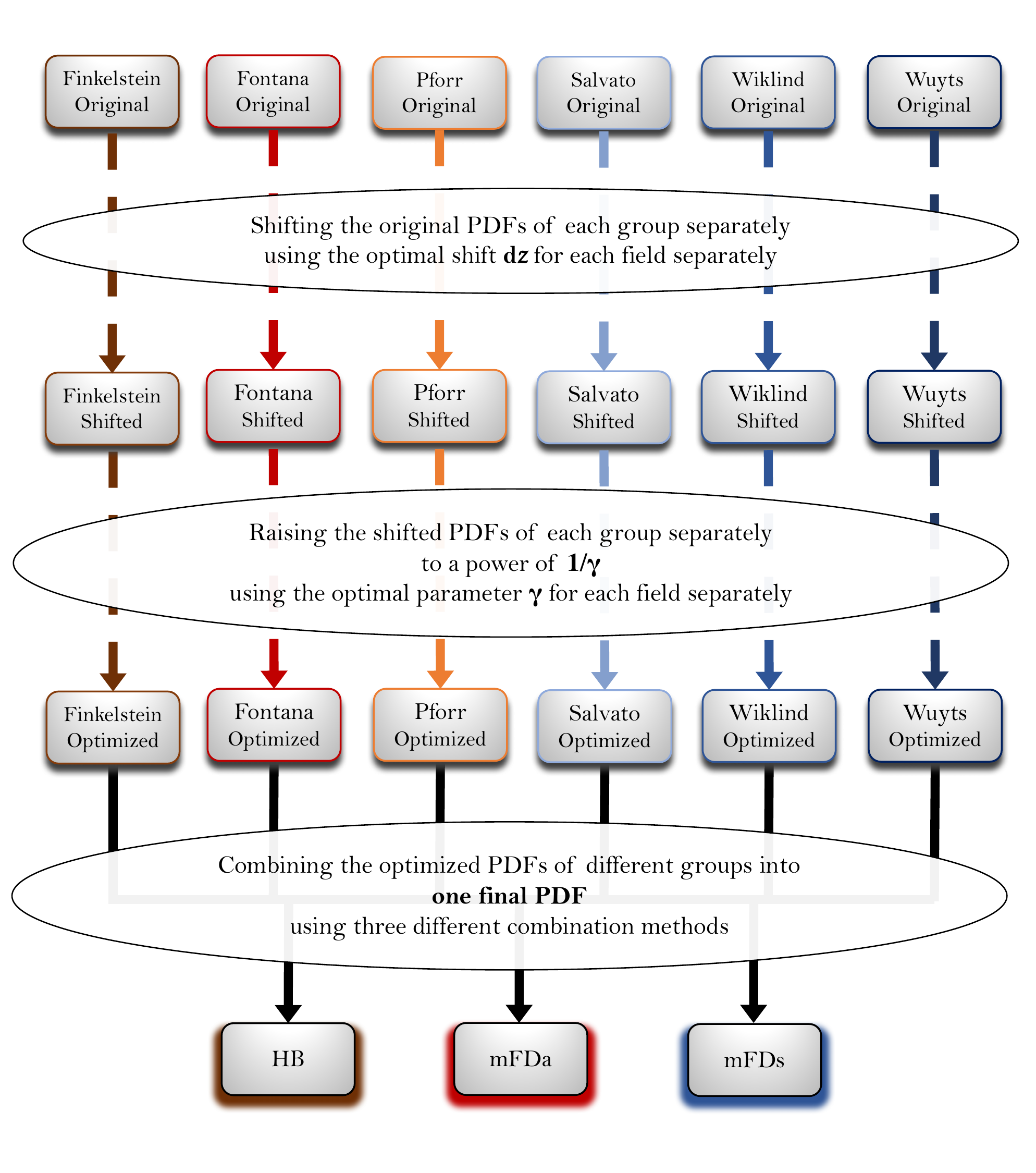}
  \end{center}
  \caption{Diagram of the optimization procedure for obtaining the final products of photo-$z$ PDFs, starting from the initially provided PDFs. First the PDFs are shifted, then raised to a power, resulting in optimized PDFs. Then the PDFs from different groups are combined using three different combination methods into final PDFs that can be used for science.}
\label{fig:procedure}
\end{figure*}

\autoref{fig:procedure} summarizes the procedure from the initial photo-$z$ PDFs obtained from the six different groups (``Original'') to the ``Shifted'' and ``Optimized'' versions of the individual photo-$z$ PDFs and then to the combined photo-$z$ PDFs used to generate the final photometric redshift catalogs.  We provide tabulated ``Original,'' ``Optimized,'' and combined photo-$z$ PDFs and point estimates for all objects in the five CANDELS fields.

First, we provide photo-$z$ PDFs in a separate file for every CANDELS object (with the object identifier specified in the filename, e.g., \texttt{ALL\_OPTIMIZED\_PDFS\_GOODSN\_ID00001.pzd}). Each file has columns specifying the redshift; the PDF provided by each group after the optimization procedure from Section \ref{sec:optimization} has been applied; and the PDFs resulting from the two best combination methods, $\mathrm{HB4}$ and $\mathrm{mFDa4}$ (hence the term \texttt{ALL} in the filename). The PDFs cover the redshift interval $[0, 10]$ with a step size of $\Delta z = 0.01$. The details of the columns included in these files are presented in \autoref{table:PDFs}. We also provide files with the original photo-$z$ PDFs as provided by the six different groups before applying the optimization method. The format of these files is exactly the same as the one with the optimized PDFs, while the file name in this case is \texttt{ALL\_ORIGINAL\_PDFS\_GOODSN\_ID00001.pzd}.

Additionally, we provide summary catalogs of objects in each CANDELS field.  These catalogs contain photo-$z$ point statistics constructed from the optimized PDFs and the best combinations, as well as spectroscopic and/or grism redshifts where available. These catalogs include estimates of the $1\sigma$ and $2\sigma$ credible intervals for the photometric redshift constructed from the PDFs.  The columns of these files are described in detail in \autoref{table:summary_catalogs}.  The catalogs are available at \catalogurl.

%
\begin{table*}
\begin{center}
\begin{tabular}{ll}
\specialrule{2.pt}{1pt}{1pt}
$\mathrm{Column}$ &
$\mathrm{Description}$
\\
\specialrule{2.pt}{1pt}{1pt}
$\#1 \,\, z$ &
Redshift values for the grid on which PDFs are tabulated
\\
$\#2 \,\, \mathrm{Finkelstein}$ &
Probability Density Function (PDF) from Finkelstein
\\
$\#3 \,\, \mathrm{Fontana}$ &
PDF from Fontana
\\
$\#4 \,\, \mathrm{Pforr}$ &
PDF from Pforr
\\
$\#5 \,\, \mathrm{Salvato}$ &
PDF from Salvato
\\
$\#6 \,\, \mathrm{Wiklind}$ &
PDF from Wiklind
\\
$\#7 \,\, \mathrm{Wuyts}$ &
PDF from Wuyts
\\
$\#8 \,\, \mathrm{HB4}$ &
PDF from Hierarchical Bayesian combination, constructed using the PDFs from the best-performing four groups
\\
$\#9 \,\, \mathrm{mFDa4}$ &
PDF from the minimum Fr\'{e}chet Distance combination (computed with $\ell^1$ distance metric), constructed using
\\
 & the PDFs from the best-performing four groups
\\
\\
\specialrule{2.pt}{1pt}{1pt}
\end{tabular}
\caption{Detailed description of the files (e.g., \texttt{ALL\_OPTIMIZED\_PDFS\_GOODSN\_ID00001.pzd}) containing the PDFs from each participant as well as the two best combination methods. The number of models used in the combination methods is reported in the header, as well as the value of the parameter $\alpha$ from the Hierarchical Bayesian method. Note that while the four best participants are included in the evaluation of the combination methods, one or more participants might have missing PDFs for a given object, therefore the total number of PDFs used is not always 4. Both the original and optimized versions of PDFs are provided in separate files.}
\label{table:PDFs}
\end{center}
\end{table*}

%
%

%

\begin{table*}
\begin{center}
\begin{tabular}{ll}
%
\specialrule{2.pt}{1pt}{1pt}
$\mathrm{Column}$ &
$\mathrm{Description}$
\\
\specialrule{2.pt}{1pt}{1pt}
$\# \,\, 1 \,\, \mathrm{file}$ &
Name of the PDF file used to estimate photometric point values.
\\
$\# \,\, 2 \,\, \mathrm{ID}$ &
CANDELS ID of the object as used in the photometric catalogs.
\\
$\# \,\, 3 \,\, \mathrm{RA}$ &
Right Ascension of object (from photometric catalog).
\\
$\# \,\, 4 \,\, \mathrm{DEC}$ &
Declination of object (from photometric catalog).
\\
$\# \,\, 5 \,\, \mathrm{z\_best}$ &
Best redshift value which can be spectroscopic, grism, or photometric.
\\
$\# \,\, 6 \,\, \mathrm{z\_best\_type}$ &
Type of photometric redshift: s = spec-z, g = grism-z, p = photo-z.
\\
$\# \,\, 7 \,\, \mathrm{z\_spec}$ &
Spectroscopic redshift if available.
\\
$\# \,\, 8 \,\, \mathrm{z\_spec\_ref}$ &
Reference of catalog from which the spectroscopic redshift is obtained.
\\
$\# \,\, 9 \,\, \mathrm{z\_grism}$ &
3D-HST grism redshift of object if available
\\
$\# \,\, 10 \,\, \mathrm{mFDa4\_z\_peak}$ &
Peak value of mFDa4 PDF
\\
$\# \,\, 11 \,\, \mathrm{mFDa4\_z\_weight}$ &
Weighted average value of mFDa4 PDF
\\
$\# \,\, 12 \,\, \mathrm{mFDa4\_z683\_low}$ &
Lower boundary of 68.3\% ($1\sigma$) credible interval of mFDa4 PDF
\\
$\# \,\, 13 \,\, \mathrm{mFDa4\_z683\_high}$ &
Higher boundary of 68.3\% ($1\sigma$) credible interval of mFDa4 PDF
\\
$\# \,\, 14 \,\, \mathrm{mFDa4\_z954\_low}$ &
Lower boundary of 95.4\% ($2\sigma$) credible interval of mFDa4 PDF
\\
$\# \,\, 15 \,\, \mathrm{mFDa4\_z954\_high}$ &
Higher boundary of 95.4\% ($2\sigma$) credible interval of mFDa4 PDF
\\
$\# \,\, 16 \,\, \mathrm{HB4\_z\_peak}$ &
Peak value of HB4 PDF
\\
\hspace{1cm} \vdots
\\
$\# \,\, 22 \,\, \mathrm{Finkelstein\_z\_peak}$ &
Peak value of Finkelstein PDF
\\
\hspace{1cm} \vdots
\\
$\# \,\, 28 \,\, \mathrm{Fontana\_z\_peak}$ &
Peak value of Fontana PDF
\\
\hspace{1cm} \vdots
\\
$\# \,\, 34 \,\, \mathrm{Pforr\_z\_peak}$ &
Peak value of Pforr PDF
\\
\hspace{1cm} \vdots
\\
$\# \,\, 40 \,\, \mathrm{Salvato\_z\_peak}$ &
Peak value of Salvato PDF
\\
\hspace{1cm} \vdots
\\
$\# \,\, 46 \,\, \mathrm{Wiklind\_z\_peak}$ &
Peak value of Wiklind PDF
\\
\hspace{1cm} \vdots
\\
$\# \,\, 52 \,\, \mathrm{Wuyts\_z\_peak}$ &
Peak value of Wuyts PDF
\\
\hspace{1cm} \vdots
\\
\specialrule{2.pt}{1pt}{1pt}
\end{tabular}
\caption{Detailed description of the columns of the CANDELS photometric redshift catalogs (e.g., \texttt{zcat\_EGS\_v2.0.cat}), which provide point statistics constructed from the optimized photometric PDFs, as well as spectroscopic and/or grism redshifts where available, for all objects in each CANDELS field. Each CANDELS object corresponds to one row in the catalog.  Statistics based upon the optimized PDFs from all six groups, as well as the two best combination methods, $\mathrm{mFDa4}$ and $\mathrm{HB4}$, are provided within the catalog.  The full set of statistics tabulated for the minimum \fdiv\ (mFDa4) PDF are detailed.  Corresponding statistics are provided for each groups' results are included in the catalog, with the column identifier only differing in its prefix (i.e., HB4, Finkelstein, Fontana, Pforr, Salvato, Wiklind, or Wuyts) from the column identifier for mFDa4.}
\label{table:summary_catalogs}
\end{center}
\end{table*}




\section{Summary and Discussion}
\label{sec:summary}

In this paper, we developed a technique to measure the \photoz\ PDFs and point estimates for galaxies in the CANDELS field using the final photometric catalogs.  Using this technique, we present photometric redshifts for over 150,000 galaxies.  We began with probability density functions measured by six groups within the collaboration by applying a variety of template-based methods to the same photometric catalogs.  We determined the optimal shift and stretching/sharpening parameters for the PDFs from each group using statistics based upon the $Q$--$Q$ plot, which we measured using the same training set of spectroscopic redshifts provided to each group to tune their photo-$z$ algorithms.

In tests with a training set of spectroscopic redshifts and with independent sets of spectroscopic and grism redshifts, the optimized PDFs much more closely match the statistical definition of a probability density function than those originally provided by each group, with the normalized $\ell^2$-norm statistic (a measure of the accuracy of the photo-$z$ CDF) improving by more than a factor of 2 in some cases.
Point estimates of the redshift (e.g., $z_{\rm weight}$) derived from the optimized photo-$z$ PDFs also exhibit significantly smaller scatter (as measured by the normalized median absolute deviation) and smaller or negligibly worse catastrophic outlier rates, in the best cases yielding photo-$z$ errors of $\sim 0.02(1+z)$.

After optimizing the results from individual groups, we have explored the gains from three different methods of combining the six PDFs available for each object: the HB method described in \citet{Dahlen_2013} as well as two techniques introduced here that identify the PDF with the minimum \fdiv\ based on the sum of absolute values ($\mathrm{mFDa}$) and squared differences ($\mathrm{mFDs}$). We construct new {PDFs} by applying each method to subsets of the six results for each object. Comparing them to each other with the same statistics used to assess individual groups' results shows that combining the PDFs from the four best-performing groups produced the best results. The hierarchical Bayesian method yielded the lowest scatter in point statistics, while the minimum \fdiv\ curve computed with an $\ell^1$ metric ($\mathrm{mFDa}$) had the lowest $\mathrm{\ell^2}$-norm values, indicating that it provides the most accurate PDFs, and hence the most accurate credible intervals as well.

The optimized methods were then used to estimate \photoz\ PDFs for galaxies in CANDELS.  We constructed publicly available catalogs of optimized PDFs and photometric redshift summary statistics for all objects from the CANDELS photometric catalogs used to calculate the \photozs.  Instructions on how best to use these catalogs are as follows:

\begin{itemize}

\item In general, results from different photometric redshift codes are sufficiently different from each other (as can be seen most clearly in \autoref{fig:All6_summed_egs}) that we recommend performing an analysis multiple times using photo-$z$ point estimates or PDFs from different groups each time to ensure that conclusions are robust to these variations.

\item Different columns of the summary catalog are better for different purposes.  For instance, if one wants the best estimate for the redshift of an individual object (where uniformity does not matter), the {\tt z\_best} value from the catalog (which is determined from the combined data set of spectroscopic redshifts, 3D-HST grism redshifts, and mFDa4 photometric redshifts) would be most appropriate.  If instead the smallest-scatter estimator of redshift for a uniform sample is needed, {\tt HB4\_z\_weight} (the $z_{\rm weight}$ value computed from the hierarchical Bayesian combination of the four best PDFs for each object) would be most appropriate.  This photo-$z$ point estimate yielded $\sigma_{\rm NMAD} = 0.0227/0.0189$ and $|\Delta z/(1+z)| > 0.15$ outlier fraction $= 0.067/0.019$ for the testing \speczs\ and 3D-HST \grismzs, respectively.  We note that redshift uncertainties will never be uniform across galaxy samples---even when restricting uniformly to using only photo-$z$'s---because different objects have different signal-to-noise ratios and different SED shapes with more or less pronounced breaks.

\item The mFDa4 (minimum \fdiv\ curve constructed from the best four PDFs from individual groups) yielded PDFs that best meet the statistical definition of a PDF.  As a consequence, this is the preferred set of PDFs to use when the accuracy of credible intervals on redshifts is desired.  Correspondingly, the {\tt mFDa4\_z\_weight} column of the summary table will have the best-characterized error estimates associated with it.

\end{itemize}

The photometric redshift catalogs presented here represent the culmination of a considerable amount of effort by the CANDELS collaboration to obtain a broad range of imaging data, measure uniform photometry with TFIT, and calculate photometric redshifts.  They represent a public legacy of the survey that should contribute to a wide variety of science in the future, such as the estimation of stellar masses of galaxies.


\vspace{5mm}

\noindent We would like to thank Janine Pforr for performing the \texttt{HyperZ} fits.  We also wish to acknowledge helpful discussions with Larry Wasserman, Ann Lee, Peter Freeman, and the International Computational Astrostatistics Group at Carnegie Mellon University; Rongpu Zhou; members of the CANDELS Multiwavelength Catalog Working Group and the LSST Dark Energy Science Collaboration Photometric Redshift Working Group.  We appreciate the careful reading and thoughtful suggestions by the referee and the AAS Journals statistician.  This work is based on observations taken by the CANDELS Multi-Cycle Treasury Program with the NASA/ESA HST and was supported by HST program No. GO-12060.  Support for Program No. GO-12060 was provided by NASA through a grant from the Space Telescope Science Institute, which is operated by the Association of Universities for Research in Astronomy, Incorporated, under NASA contract NAS5-26555. D.C.K.\ acknowledges support from NSF grant AST-1615730.  This research has made use of NASA's Astrophysics Data System.

%

\vspace{5mm}
\facilities{HST(WFC3 and ACS)}





\appendix

\bigskip
\bigskip
\section{Spectroscopic and Grism Redshifts}

For training and testing the photo-$z$ codes, we compiled spectroscopic and grism redshifts for numerous sources. Table \ref{table:redshifts} lists the original sources, the relevant CANDELS field, and any quality cuts applied.

%
%

\begin{longrotatetable}
\begin{deluxetable*}{l ccccc l}
\tablecaption{Details of the spectroscopic redshifts and grism redshifts used in this study. For each data set used we provide the name of the survey or instrument used, where applicable; a reference for the source catalog; the number of redshifts provided in each CANDELS field; and any cuts applied in order to restrict to the most robust redshifts.}
%
%
\tablehead{
\colhead{Survey / Instrument (Reference)} &
\multicolumn{5}{ c }{Number Of Redshifts in each field} &
\colhead{Cuts applied}
\\ \cmidrule(r){2-6}
\colhead{} &
\colhead{COSMOS} &
\colhead{EGS} &
\colhead{GOODS-N} &
\colhead{GOODS-S} &
\colhead{UDS} &
\colhead{}
}


%
%
%
\startdata
$\mathrm{Training, \quad COSMOS}$ (Private Communication: B.~Mobasher) &
$\mathrm{370}$ &
$\mathrm{-}$ &
$\mathrm{-}$ &
$\mathrm{-}$ &
$\mathrm{-}$ &
$\mathrm{\mathit{z} > 0, \,\, flag = 1}$
\\
$\mathrm{Training, \quad EGS}$ (DEEP3, \citealt{DEEP3_1, DEEP3_2, Zhou_2019})&
$\mathrm{-}$ &
$\mathrm{840}$ &
$\mathrm{-}$ &
$\mathrm{-}$ &
$\mathrm{-}$ &
$\mathrm{\mathit{z} > 0, \,\, flag = none}$
\\
$\mathrm{Training, \quad GOODS-N}$ (Private Communication: B.~Mobasher) &
$\mathrm{-}$ &
$\mathrm{-}$ &
$\mathrm{2994}$ &
$\mathrm{-}$ &
$\mathrm{-}$ &
$\mathrm{\mathit{z} > 0, \,\, flag \geq 3}$
\\
$\mathrm{Training, \quad GOODS-S}$ (Private Communication: B.~Mobasher) &
$\mathrm{-}$ &
$\mathrm{-}$ &
$\mathrm{-}$ &
$\mathrm{1249}$ &
$\mathrm{-}$ &
$\mathrm{\mathit{z} > 0, \,\, GRISM\_FLAG = 0}$\footnote{Flag to avoid \grismzs\ and only keep \speczs.}
\\
$\mathrm{Training, \quad UDS}$ (Private Communication: B.~Mobasher) &
$\mathrm{-}$ &
$\mathrm{-}$ &
$\mathrm{-}$ &
$\mathrm{-}$ &
$\mathrm{354}$ &
$\mathrm{\mathit{z} > 0, \,\, flag = 1}$
\\
\\
\hline
\\
%
%
$\mathrm{3D{-}HST}$ (\citealt{3DHST}) &
$\mathrm{566}$ &
$\mathrm{771}$ &
$\mathrm{579}$ &
$\mathrm{523}$ &
$\mathrm{928}$ &
$\mathrm{\mathit{z}\_max\_grism > 0.6}$
\\
 &
 &
 &
 &
 &
 &
$\mathrm{use\_zgrism = 1, \,\, use\_phot = 1}$
\\
 &
 &
 &
 &
 &
 &
$\mathrm{flag1 = 0, \,\, flag2 = 0}$
\\
 &
 &
 &
 &
 &
 &
$\mathrm{z\_best\_s \neq 0, \,\, z\_spec \leq 0}$
\\
 &
 &
 &
 &
 &
 &
$\mathrm{z\_phot\_u68 - z\_phot\_l68 > 0}$
\\
 &
 &
 &
 &
 &
 &
$\mathrm{z\_grism\_u68 - z\_grism\_l68 < 0.01}$
\\
 &
 &
 &
 &
 &
 &
$\mathrm{\frac{z\_grism\_u68 - z\_grism\_l68}{z\_phot\_u68 - z\_phot\_l68} < 0.1}$
\\
 &
 &
 &
 &
 &
 &
$\mathrm{z\_max\_grism > z\_phot\_l95}$
\\
 &
 &
 &
 &
 &
 &
$\mathrm{z\_max\_grism < z\_phot\_u95}$
\\
\\
\hline
\\
%
%
\hspace{1cm} \underline{$\mathrm{MULTIPLE \quad FIELDS}$} &
& & & & &
\\[0.2cm]
$\mathrm{DEIMOS}$ (\citealt{DEIMOS}, Private Communication: B.~Mobasher) &
$\mathrm{172}$ &
$\mathrm{-}$ &
$\mathrm{70}$ &
$\mathrm{39}$ &
$\mathrm{179}$ &
$\mathrm{\mathit{z} > 0, \,\, flag = 1}$
\\
\\
%
%
$\mathrm{MOSDEF}$ (\citealt{MOSDEF}) &
$\mathrm{189}$ &
$\mathrm{268}$ &
$\mathrm{73}$ &
$\mathrm{10}$ &
$\mathrm{26}$ &
$\mathrm{\mathit{z} > 0, \,\, flag > 0}$
\\
\\
%
%
$\mathrm{MOSFIRE}$ (\citealt{MOSFIRE_1, MOSFIRE_2}) &
$\mathrm{-}$ &
$\mathrm{-}$ &
$\mathrm{22}$ &
$\mathrm{75}$ &
$\mathrm{-}$ &
$\mathrm{\mathit{z} > 0, \,\, flag \geq 3}$
\\
\\
\hline
\\
%
%
\hspace{1cm} \underline{$\mathrm{COSMOS \quad FIELD \quad ONLY}$} &
& & & & &
\\[0.2cm]
$\mathrm{FMOS}$ (\citealt{FMOS}) &
$\mathrm{9}$ &
$\mathrm{-}$ &
$\mathrm{-}$ &
$\mathrm{-}$ &
$\mathrm{-}$ &
$\mathrm{\mathit{z} > 0, \,\, flag > 3}$
\\
$\mathrm{FORS2}$ (Private Communication: J.~Comparat) &
$\mathrm{33}$ &
$\mathrm{-}$ &
$\mathrm{-}$ &
$\mathrm{-}$ &
$\mathrm{-}$ &
$\mathrm{\mathit{z} > 0, \,\, flag > 3}$
\\
$\mathrm{Gemini{-}S}$ (Private Communication: M.~Balogh) &
$\mathrm{16}$ &
$\mathrm{-}$ &
$\mathrm{-}$ &
$\mathrm{-}$ &
$\mathrm{-}$ &
$\mathrm{\mathit{z} > 0, \,\, flag > 3}$
\\
$\mathrm{PRIMUS}$ (\citealt{PRIMUS}) &
$\mathrm{232}$ &
$\mathrm{-}$ &
$\mathrm{-}$ &
$\mathrm{-}$ &
$\mathrm{-}$ &
$\mathrm{\mathit{z} > 0, \,\, flag > 3}$
\\
$\mathrm{VUDS}$ (\citealt{VUDS_2015}) &
$\mathrm{101}$ &
$\mathrm{-}$ &
$\mathrm{-}$ &
$\mathrm{-}$ &
$\mathrm{-}$ &
$\mathrm{\mathit{z} > 0, \,\, flag = 3 \,\, and \,\, 4}$
\\
$\mathrm{WFC3}$ (\citealt{WFC3}, grism-$z$'s) &
$\mathrm{12}$ &
$\mathrm{-}$ &
$\mathrm{-}$ &
$\mathrm{-}$ &
$\mathrm{-}$ &
$\mathrm{\mathit{z} > 0, \,\, flag > 3}$
\\
$\mathrm{zBRIGHT}$ (\citealt{zBRIGHT}) &
$\mathrm{2}$ &
$\mathrm{-}$ &
$\mathrm{-}$ &
$\mathrm{-}$ &
$\mathrm{-}$ &
$\mathrm{\mathit{z} > 0, \,\, flag = 1.5, 2.5, 9.3, 9.5}$
\\
$\mathrm{}$ &
$\mathrm{}$ &
$\mathrm{}$ &
$\mathrm{}$ &
$\mathrm{}$ &
$\mathrm{}$ &
$\mathrm{\quad \quad \quad \quad \quad \,\, 3.x, 4.x, 13.x, 14.x}$
\\
$\mathrm{}$ &
$\mathrm{}$ &
$\mathrm{}$ &
$\mathrm{}$ &
$\mathrm{}$ &
$\mathrm{}$ &
$\mathrm{\quad \quad \quad \quad \quad \,\, secondary \,\, targets}$
\\
$\mathrm{zCOSMOS}$ (\citealt{zCOSMOS}) &
$\mathrm{7}$ &
$\mathrm{-}$ &
$\mathrm{-}$ &
$\mathrm{-}$ &
$\mathrm{-}$ &
$\mathrm{\mathit{z} > 0, \,\, flag = 1.5, 2.5, 9.3, 9.5}$
\\
$\mathrm{}$ &
$\mathrm{}$ &
$\mathrm{}$ &
$\mathrm{}$ &
$\mathrm{}$ &
$\mathrm{}$ &
$\mathrm{\quad \quad \quad \quad \quad \,\, 3.x, 4.x, 13.x, 14.x}$
\\
$\mathrm{}$ &
$\mathrm{}$ &
$\mathrm{}$ &
$\mathrm{}$ &
$\mathrm{}$ &
$\mathrm{}$ &
$\mathrm{\quad \quad \quad \quad \quad \,\, secondary \,\, targets}$
\\
\\
\hline
\\
%
%
\hspace{1cm} \underline{$\mathrm{EGS \quad FIELD \quad ONLY}$} &
& & & & &
\\[0.2cm]
$\mathrm{DEEP2}$ (\citealt{DEEP2_1}; \citealt{DEEP2_2}; \citealt{DEEP2_3}) &
$\mathrm{-}$ &
$\mathrm{1432}$ &
$\mathrm{-}$ &
$\mathrm{-}$ &
$\mathrm{-}$ &
$\mathrm{\mathit{z} > 0, \,\, flag \geq 3}$
\\
\\
\hline
\\
%
%
\hspace{1cm} \underline{$\mathrm{GOODS-N \quad FIELD \quad ONLY}$} &
& & & & &
\\[0.2cm]
$\mathrm{Barger}$ (24 $\mu$m, ACS, K, UV, \citealt{Barger}, and references therein) &
$\mathrm{-}$ &
$\mathrm{-}$ &
$\mathrm{80}$ &
$\mathrm{-}$ &
$\mathrm{-}$ &
$\mathrm{\mathit{z} > 0, \,\, flag = 4 \, or \, A}$
\\
$\mathrm{DEEP3}$ (\citealt{DEEP3_1, Zhou_2019}) &
$\mathrm{-}$ &
$\mathrm{-}$ &
$\mathrm{3}$ &
$\mathrm{-}$ &
$\mathrm{-}$ &
$\mathrm{\mathit{z} > 0, \,\, flag \geq 3}$
\\
$\mathrm{Pirzkal}$ (\citealt{Pirzkal_2013}, grism-$z$'s) &
$\mathrm{-}$ &
$\mathrm{-}$ &
$\mathrm{81}$ &
$\mathrm{-}$ &
$\mathrm{-}$ &
$\mathrm{\mathit{z} > 0, \,\, flag \geq 3}$
\\
$\mathrm{Reddy\_2006}$ (\citealt{Reddy_2006}) &
$\mathrm{-}$ &
$\mathrm{-}$ &
$\mathrm{2}$ &
$\mathrm{-}$ &
$\mathrm{-}$ &
$\mathrm{\mathit{z} > 0, \,\, flag \geq 3}$
\\
$\mathrm{Wirth\_2004}$ \citealt{Wirth_2004} &
$\mathrm{-}$ &
$\mathrm{-}$ &
$\mathrm{3}$ &
$\mathrm{-}$ &
$\mathrm{-}$ &
$\mathrm{\mathit{z} > 0, \,\, flag \geq 3}$
\\
\\
%
%
\hspace{1cm} \underline{$\mathrm{GOODS-S \quad FIELD \quad ONLY}$} &
& & & & &
\\[0.2cm]
$\mathrm{ACES}$ (\citealt{ACES_2012}) &
$\mathrm{-}$ &
$\mathrm{-}$ &
$\mathrm{-}$ &
$\mathrm{103}$ &
$\mathrm{-}$ &
$\mathrm{\mathit{z} > 0, \,\, flag = 0}$
\\
$\mathrm{COMBO{-}17}$ (\citealt{COMBO-17}) &
$\mathrm{-}$ &
$\mathrm{-}$ &
$\mathrm{-}$ &
$\mathrm{6}$ &
$\mathrm{-}$ &
$\mathrm{\mathit{z} > 0, \,\, flag = 0}$
\\
$\mathrm{CXO{-}CDFS}$ (\citealt{CXO-CDFS}) &
$\mathrm{-}$ &
$\mathrm{-}$ &
$\mathrm{-}$ &
$\mathrm{77}$ &
$\mathrm{-}$ &
$\mathrm{\mathit{z} > 0, \,\, flag = 0}$
\\
$\mathrm{FW\_7}$ (\citealt{FW_7}) &
$\mathrm{-}$ &
$\mathrm{-}$ &
$\mathrm{-}$ &
$\mathrm{7}$ &
$\mathrm{-}$ &
$\mathrm{\mathit{z} > 0, \,\, flag = 0}$
\\
$\mathrm{FW\_10}$ (\citealt{FW_10}) &
$\mathrm{-}$ &
$\mathrm{-}$ &
$\mathrm{-}$ &
$\mathrm{3}$ &
$\mathrm{-}$ &
$\mathrm{\mathit{z} > 0, \,\, flag = 0}$
\\
$\mathrm{FW\_13}$ (\citealt{Wuyts_2009}) &
$\mathrm{-}$ &
$\mathrm{-}$ &
$\mathrm{-}$ &
$\mathrm{2}$ &
$\mathrm{-}$ &
$\mathrm{\mathit{z} > 0, \,\, flag = 0}$
\\
$\mathrm{FW\_14}$ (\citealt{FW_14}) &
$\mathrm{-}$ &
$\mathrm{-}$ &
$\mathrm{-}$ &
$\mathrm{1}$ &
$\mathrm{-}$ &
$\mathrm{\mathit{z} > 0, \,\, flag = 0}$
\\
$\mathrm{FW\_16}$ (Private Communication: J.-S.~Huang) &
$\mathrm{-}$ &
$\mathrm{-}$ &
$\mathrm{-}$ &
$\mathrm{2}$ &
$\mathrm{-}$ &
$\mathrm{\mathit{z} > 0, \,\, flag = 0}$
\\
$\mathrm{K20}$ (\citealt{K20}) &
$\mathrm{-}$ &
$\mathrm{-}$ &
$\mathrm{-}$ &
$\mathrm{99}$ &
$\mathrm{-}$ &
$\mathrm{\mathit{z} > 0, \,\, flag = 0}$
\\
\citealt{Kurk_2013} &
$\mathrm{-}$ &
$\mathrm{-}$ &
$\mathrm{-}$ &
$\mathrm{49}$ &
$\mathrm{-}$ &
$\mathrm{\mathit{z} > 0, \,\, flag = 0}$
\\
\citealt{Morris_2015} &
$\mathrm{-}$ &
$\mathrm{-}$ &
$\mathrm{-}$ &
$\mathrm{54}$ &
$\mathrm{-}$ &
$\mathrm{\mathit{z} > 0, \,\, flag = 0}$
\\
$\mathrm{VIMOS\_10\_MR\/LR}$ (\citealt{VIMOS}) &
$\mathrm{-}$ &
$\mathrm{-}$ &
$\mathrm{-}$ &
$\mathrm{196}$ &
$\mathrm{-}$ &
$\mathrm{\mathit{z} > 0, \,\, flag = 0}$
\\
$\mathrm{VLT\_2008}$ (\citealt{VLT_2008}) &
$\mathrm{-}$ &
$\mathrm{-}$ &
$\mathrm{-}$ &
$\mathrm{144}$ &
$\mathrm{-}$ &
$\mathrm{\mathit{z} > 0, \,\, flag = 0}$
\\
$\mathrm{VLT\_IMAG}$ (\citealt{VLT_IMAG}) &
$\mathrm{-}$ &
$\mathrm{-}$ &
$\mathrm{-}$ &
$\mathrm{5}$ &
$\mathrm{-}$ &
$\mathrm{\mathit{z} > 0, \,\, flag = 0}$
\\
$\mathrm{VLT\_LBGs}$ (\citealt{VLT_LBGs}) &
$\mathrm{-}$ &
$\mathrm{-}$ &
$\mathrm{-}$ &
$\mathrm{1}$ &
$\mathrm{-}$ &
$\mathrm{\mathit{z} > 0, \,\, flag = 0}$
\\
$\mathrm{VUDS\_2015}$ (\citealt{VUDS_2015}) &
$\mathrm{-}$ &
$\mathrm{-}$ &
$\mathrm{-}$ &
$\mathrm{77}$ &
$\mathrm{-}$ &
$\mathrm{\mathit{z} > 0, \,\, flag = 0}$
\\
$\mathrm{VVDS \,\, (Final}$, \citealt{VVDS})&
$\mathrm{-}$ &
$\mathrm{-}$ &
$\mathrm{-}$ &
$\mathrm{116}$ &
$\mathrm{-}$ &
$\mathrm{\mathit{z} > 0, \,\, flag = 0}$
\\
$\mathrm{Wuyts\_2009}$ (\citealt{Wuyts_2009}) &
$\mathrm{-}$ &
$\mathrm{-}$ &
$\mathrm{-}$ &
$\mathrm{3}$ &
$\mathrm{-}$ &
$\mathrm{\mathit{z} > 0, \,\, flag = 0}$
\\
$\mathrm{Xray\_1}$ (\citealt{CXO-CDFS}) &
$\mathrm{-}$ &
$\mathrm{-}$ &
$\mathrm{-}$ &
$\mathrm{7}$ &
$\mathrm{-}$ &
$\mathrm{\mathit{z} > 0, \,\, flag = 0}$
\\
$\mathrm{Xray\_2}$ (\citealt{VLT_2008}) &
$\mathrm{-}$ &
$\mathrm{-}$ &
$\mathrm{-}$ &
$\mathrm{1}$ &
$\mathrm{-}$ &
$\mathrm{\mathit{z} > 0, \,\, flag = 0}$
\\
$\mathrm{Xray\_6}$ (\citealt{VIMOS/GOODS}) &
$\mathrm{-}$ &
$\mathrm{-}$ &
$\mathrm{-}$ &
$\mathrm{1}$ &
$\mathrm{-}$ &
$\mathrm{\mathit{z} > 0, \,\, flag = 0}$
\\
$\mathrm{Xray\_12}$ (\citealt{VIMOS}) &
$\mathrm{-}$ &
$\mathrm{-}$ &
$\mathrm{-}$ &
$\mathrm{1}$ &
$\mathrm{-}$ &
$\mathrm{\mathit{z} > 0, \,\, flag = 0}$
\\
$\mathrm{Xue\_4}$ (\citealt{K20}) &
$\mathrm{-}$ &
$\mathrm{-}$ &
$\mathrm{-}$ &
$\mathrm{2}$ &
$\mathrm{-}$ &
$\mathrm{\mathit{z} > 0, \,\, flag = 0}$
\\
$\mathrm{Xue\_5}$ (\citealt{VLT_IMAG}) &
$\mathrm{-}$ &
$\mathrm{-}$ &
$\mathrm{-}$ &
$\mathrm{2}$ &
$\mathrm{-}$ &
$\mathrm{\mathit{z} > 0, \,\, flag = 0}$
\\
$\mathrm{Xue\_6}$ (\citealt{VLT_2008}) &
$\mathrm{-}$ &
$\mathrm{-}$ &
$\mathrm{-}$ &
$\mathrm{2}$ &
$\mathrm{-}$ &
$\mathrm{\mathit{z} > 0, \,\, flag = 0}$
\\
$\mathrm{Xue\_7}$ (\citealt{VIMOS/GOODS}) &
$\mathrm{-}$ &
$\mathrm{-}$ &
$\mathrm{-}$ &
$\mathrm{2}$ &
$\mathrm{-}$ &
$\mathrm{\mathit{z} > 0, \,\, flag = 0}$
\\
$\mathrm{Xue\_8}$ (\citealt{Xue_8}) &
$\mathrm{-}$ &
$\mathrm{-}$ &
$\mathrm{-}$ &
$\mathrm{1}$ &
$\mathrm{-}$ &
$\mathrm{\mathit{z} > 0, \,\, flag = 0}$
\\
$\mathrm{Xue\_9}$ (\citealt{VIMOS}) &
$\mathrm{-}$ &
$\mathrm{-}$ &
$\mathrm{-}$ &
$\mathrm{1}$ &
$\mathrm{-}$ &
$\mathrm{\mathit{z} > 0, \,\, flag = 0}$
\\
$\mathrm{Xue\_10}$ (\citealt{Xue_10}) &
$\mathrm{-}$ &
$\mathrm{-}$ &
$\mathrm{-}$ &
$\mathrm{2}$ &
$\mathrm{-}$ &
$\mathrm{\mathit{z} > 0, \,\, flag = 0}$
\\
\\
\hline
\\
%
%
\hspace{1cm} \underline{$\mathrm{UDS \quad FIELD \quad ONLY}$} &
& & & & &
\\[0.2cm]
\citealt{Cooper_UDS} &
$\mathrm{-}$ &
$\mathrm{-}$ &
$\mathrm{-}$ &
$\mathrm{-}$ &
$\mathrm{61}$ &
$\mathrm{\mathit{z} > 0, \,\, flag > 3}$
\\
$\mathrm{Akiyama}$ (\citealt{Akiyama_2015}) &
$\mathrm{-}$ &
$\mathrm{-}$ &
$\mathrm{-}$ &
$\mathrm{-}$ &
$\mathrm{23}$ &
$\mathrm{\mathit{z} > 0, \,\, flag = A}$
\\
$\mathrm{UDSz}$ (\citealt{Simpson_1}; \citealt{Simpson_2}) &
$\mathrm{-}$ &
$\mathrm{-}$ &
$\mathrm{-}$ &
$\mathrm{-}$ &
$\mathrm{63}$ &
$\mathrm{\mathit{z} > 0, \,\, flag = 4 \, \, or \,\, A}$
\\
\\
\enddata
%
%
\label{table:redshifts}
\end{deluxetable*}
\end{longrotatetable}




\software{
\texttt{EAZY} \citep{EAZY},
\texttt{HyperZ} \citep{HyperZ_2000},
\texttt{LePhare} \citep{LePHARE_2011},
\texttt{WikZ} \citep{WikZ_2008},
\texttt{zphot} \citep{Giallongo_1998, Fontana_2000}
}

\bibliography{thesis, thesis_new}{}
\bibliographystyle{aasjournal}



\end{document}